\documentclass{aa}

\usepackage{graphicx}
\usepackage{epsfig}

\usepackage{txfonts}
\usepackage[colorlinks=true,linkcolor=blue,citecolor=blue,urlcolor=blue]{hyperref}%
\begin{document}

\title{Spectroscopic classification of a complete sample of astrometrically-selected quasar candidates using {\em Gaia} DR2}
\titlerunning{Astrometric selection of quasars using {\em Gaia} DR2}

\author{
K.~E.~Heintz\inst{1}, J.~P.~U.~Fynbo\inst{2,3}, S.~J.~Geier\inst{4,5}, P.~M{\o}ller\inst{6}, J.-K.~Krogager\inst{7} C.~Konstantopoulou\inst{8}, A.~de~Burgos\inst{8,9}, L.~Christensen\inst{3}, C.~L.~Steinhardt\inst{2,3}, B.~Milvang-Jensen\inst{2,3}, P.~Jakobsson\inst{1}, E. H{\o}g\inst{3}, B.~E.~H.~K.~Arvedlund\inst{3}, C.~R.~Christiansen\inst{3}, T.~B.~Hansen\inst{3}, P.~D.~Henriksen\inst{3}, K.~B.~Kuszon\inst{3}, I.~B.~McKenzie\inst{3}, K.~A.~Mosekj{\ae}r\inst{3}, M.~F.~K.~Paulsen\inst{3}, M.~N.~Sukstorf\inst{3}, S.~N.~Wilson\inst{3}, S.~K.~K.~{\O}rgaard\inst{3}
}
\institute{
Centre for Astrophysics and Cosmology, Science Institute, University of Iceland, Dunhagi 5, 107 Reykjav\'ik, Iceland \\
\email{keh14@hi.is}
\and
Cosmic DAWN Center
\and
Niels Bohr Institute, University of Copenhagen, Jagtvej 128, 2100 Copenhagen \O, Denmark
\and
Instituto de Astrof{\'i}sica de Canarias, V{\'i}a L{\'a}ctea, s/n, 38205, La Laguna, Tenerife, Spain
\and
Gran Telescopio Canarias (GRANTECAN), 38205 San Crist{\'o}bal de La Laguna, Tenerife, Spain
\and
European Southern Observatory, Karl-Schwarzschildstrasse 2, D-85748 Garching bei M\"unchen, Germany
\and
Institut d'Astrophysique de Paris, CNRS-SU, UMR7095, 98bis bd Arago, 75014 Paris, France
\and
Nordic Optical Telescope, Rambla Jos\'{e} Ana Fern\'{a}ndez P\'{e}rez, 7, ES-38711 Bren\~{a} Baja, Spain
\and
Universidad de La Laguna, Dpto. Astrof\'isica, E-38206 La Laguna, Tenerife, Spain
}
\authorrunning{K.~E.~Heintz et al.}

\abstract{
Here we explore the efficiency and fidelity of a purely astrometric selection of quasars as point sources with zero proper motions in the {\it Gaia} data release 2 (DR2). We have built a complete candidate sample including 104 {\it Gaia}-DR2 point sources, which are brighter than 20th magnitude in the {\it Gaia} $G$-band within one degree of the north Galactic pole (NGP); all of them have proper motions that are consistent with zero within 2$\sigma$ uncertainty. In addition to pre-existing spectra, we have secured long-slit spectroscopy of all the remaining candidates and find that all 104 stationary point sources in the field can be classified as either quasars (63) or stars (41).
One of the new quasars that we discover is particularly interesting as the line-of-sight to it passes through the disc of a foreground ($z=0.022$) galaxy, which imprints both \ion{Na}{D} absorption and dust extinction on the quasar spectrum.
The selection efficiency of the zero-proper-motion criterion at high Galactic latitudes is thus $\approx 60\%$. Based on this complete quasar sample, we examine the basic properties of the underlying quasar population within the imposed limiting magnitude. We find that the surface density of quasars is 20 deg$^{-2}$ (at $G<20$\,mag), the redshift distribution peaks at $z\sim 1.5$, and only eight systems ($13^{+5}_{-3}\%$) show significant dust reddening. We then explore the selection efficiency of commonly used optical, near-, and mid-infrared quasar identification techniques and find that they are all complete at the $85-90\%$ level compared to the astrometric selection. Finally, we discuss how the astrometric selection can be improved to an efficiency of $\approx 70\%$ by including an additional cut requiring parallaxes of the candidates to be consistent with zero within 2$\sigma$. The selection efficiency will further increase with the release of a future, more sensitive astrometric measurement from the {\it Gaia} mission. This type of selection, which is purely based on the astrometry of the quasar candidates, is unbiased in terms of colours and emission mechanisms of the quasars and thus provides the most complete census of the quasar population within the limiting magnitude of {\it Gaia}.
}


\maketitle

\section{Introduction}

Since the discovery of active galactic nuclei (AGN) in the middle of the last century
\citep{Seyfert1943,BM1954,Schmidt1963,Sandage1965}, the richness of the phenomenon has
gradually become evident to us. It now appears well-established that AGN are caused by accretion
onto supermassive black holes \citep{LyndenBell1969,Rees1984}. Most recently, it has also 
become clear that at least all massive galaxies harbour super-massive black holes in
their centres and hence one or more episodes in their past must have harboured 
AGN\@. The feedback from the AGN onto the gas in their host galaxies may be a crucial 
aspect in what regulates star-formation in massive galaxies and for turning galaxies
quiescent \citep{Sanders88,Granato2001,Ferrarese2002}. 

The brightest AGN, the quasars, were quickly found to also be powerful probes of the 
interstellar and intergalactic medium at high redshifts. Quasar absorption line
studies allow for the measurement of properties that are very difficult to get access 
to by other means \citep[e.g.][]{Weymann1981,Wolfe2005}.

As AGN and quasars are key factors for galaxy evolution directly as essential 
ingredients in the galaxy formation recipe and indirectly as probes of intervening material, for example, of chemical evolution \citep[][]{Smith1979,Pettini1994,Lu1996,Zafar2014,Berg2016,Krogager2019},
it is very important to get a complete census of AGN, in general, and of quasars in particular. 
The early work
of \citet{Sandage1965} already demonstrated that the original radio-selection of quasars 
only reveals a minority of the underlying population (the radio-loud quasars). It is thus clear 
that selection effects play a crucial role in the study of AGN\@. To accommodate this, a wide range of selection techniques have subsequently been developed, thereby further expanding the AGN taxonomy \citep[e.g.,][]{Mickaelian2015}. 

In this work we explore the use of the astrometric measurements from the {\it Gaia} mission
as a selection method for AGN and more specifically quasars. Due to their enormous distances, the proper motion of quasars on the 
sky has to be extremely small. A quasar at redshift 1, moving with the speed of light perpendicular to the line-of-sight, still has a proper motion of less than 0.1 mas yr$^{-1}$ \citep[the accuracy of {\it Gaia} at $G\approx 20$\,mag is 1.2 mas yr$^{-1}$;][]{Lindegren18}.
The use of astrometry in the selection of quasars was first discussed by \citet{Koo1986}. 
The great appeal of a purely astrometric selection is that the resulting selection function is 
very simple. There is of course a flux limit, which is set by the need to secure an accurate
proper motion measurement, but there is no colour sensitivity. This allows us to identify locations of quasars independent of their emission mechanisms (except for the requirement that the source is bright enough for a precise astrometric measurement), which are otherwise essential for other selection methods.

In a preliminary study, \citet{Heintz2015} explored the extent to which a purely astrometric selection of 
quasars as stationary sources in the {\it Gaia} survey would be feasible. The crucial problem 
is to not be overwhelmed by stellar sources with no apparent proper motion. We found that in large parts of the sky, well away from the Galactic disc, such a purely astrometric selection is in fact feasible. Based on the {\it Gaia} Universe model snapshot \citep[GUMS,][]{Robin2012}, we found that when targeting regions at Galactic latitudes $|b|>30^\mathrm{o}$, the ratio of quasars to apparently stationary stars should be above 
50\%; furthermore, when observing towards the Galactic poles, the fraction of quasars is expected to increase to $\approx 80\%$. After the release of the {\it Gaia} Data Release 2 \citep[DR2;][]{GaiaDR2}, we defined a pilot study \citep[][hereafter, paper I]{Heintz2018}, targeting a circular region within one degree
from the north Galactic pole (NGP), centred at RA = $12:51:26.00$, Dec = $+27:07:42.0$ (J\,2000). In this 
region we found 104 point sources (as classified from the SDSS imaging), which are brighter than 20th magnitude in the Gaia $G$-band. Of these, 34 were previously spectroscopically confirmed by other surveys to be quasars and four to be stars. 

The objective of this work is to explore the nature of the remaining 66
point sources in this sample with the goal of classifying them as stars, quasars, or potentially another class of objects. One of the main goals of our survey is to determine how many red quasars may have been overlooked with the commonly used selection methods \citep{Webster1995,Richards2003,Glikman2012,Fynbo2013}. In Sect.~\ref{observations} we describe our spectroscopic observations
of these 66 sources in addition to two previously known quasars for which we could not locate spectra in the literature. In Sect.~\ref{results} we present source classifications and our primary results. In Sect.~\ref{sec:disc} we discuss the implications of our results on the general question of quasar populations and quasar selection and in Sect.~\ref{sec:conc} we conclude on our work and discuss future outlooks.

\section{Observations}
\label{observations}

Spectroscopic observations of the 68 targets were secured using low-resolution long-slit spectroscopy since our main objective is simply to classify the targets and not to derive detailed properties (e.g. abundances or precise stellar parameters). We used the following four different instruments for this work: the Nordic Optical Telescope (NOT) equipped with the Alhambra Faint Object Spectrograph and Camera (AlFOSC), the Gran Telescopio Canarias (GTC) equipped with the Optical System for Imaging and low-Intermediate-Resolution Integrated Spectroscopy (OSIRIS) instrument, the William Herschel Telescope (WHT) equipped with
Intermediate-dispersion Spectrograph and Imaging System (ISIS), and the Telescopio Nazionale Galileo (TNG) equipped with the Device Optimized for the LOw RESolution (DOLORES) spectrograph. 

In Table~\ref{grisms}, information about the grisms used in the observations and the nominal spectral resolutions are provided. The spectra were secured on a wide range of dates between February 2019 and April 2020, under a range of observing conditions. As the primary purpose of the observations  is to classify the targets as either stars or quasars, high signal-to-noise is not required. We are thus still able to use spectra obtained under poor conditions as long as we can robustly conclude as to the nature of the targets. For some of the targets, we re-observed with higher resolution grisms in order to further examine specific aspects of the targets.
The full log of observations is presented in Table~\ref{tab:log}.

\begin{table}[!t]
\caption{Properties of the grisms used in the survey. The listed resolutions are for a 1.0 arcsec slit. For some of the observations, we used a slightly wider slit and hence the resolution is correspondingly lower.}
\begin{tabular}{lrrc}
\noalign{\smallskip} \hline \hline \noalign{\smallskip}
Instrument & Grism & Resolution & Wavelength range \\
     &       &            &    (\AA)  \\
\hline
AlFOSC &     4    & 360 & 3800--9200 \\
OSIRIS & 1000B    & 600 & 3750--7850 \\
OSIRIS & 2500R    & 1500 & 5575 - 7685 \\
OSIRIS & 2500I    & 1500 & 7350--10000 \\
ISIS   & R158B    & 600 & 3300--5400 \\
ISIS   & R600R    & 3500 & 5800-7200 \\
DOLORES & LR-B    & 600 & 3800--7800 \\
\noalign{\smallskip} \hline \noalign{\smallskip}
\end{tabular}
\label{grisms}
\end{table}

The spectroscopic data from all of the instruments were reduced using standard procedures mostly in IRAF\,\footnote{IRAF is distributed by the National Optical Astronomy Observatory, which is operated by the Association of Universities for
Research in Astronomy (AURA) under a cooperative agreement with the
National Science Foundation.}, but also using similar procedures written in {\tt Python}\footnote{https://github.com/keheintz/PyReduc}. The spectra were flux calibrated using 
observations of spectro-photometric standard stars, typically observed on the same nights as the science targets. 

\section{Results}
\label{results}

\begin{figure*}[!t]
\centering
    \includegraphics[height=8.0cm]{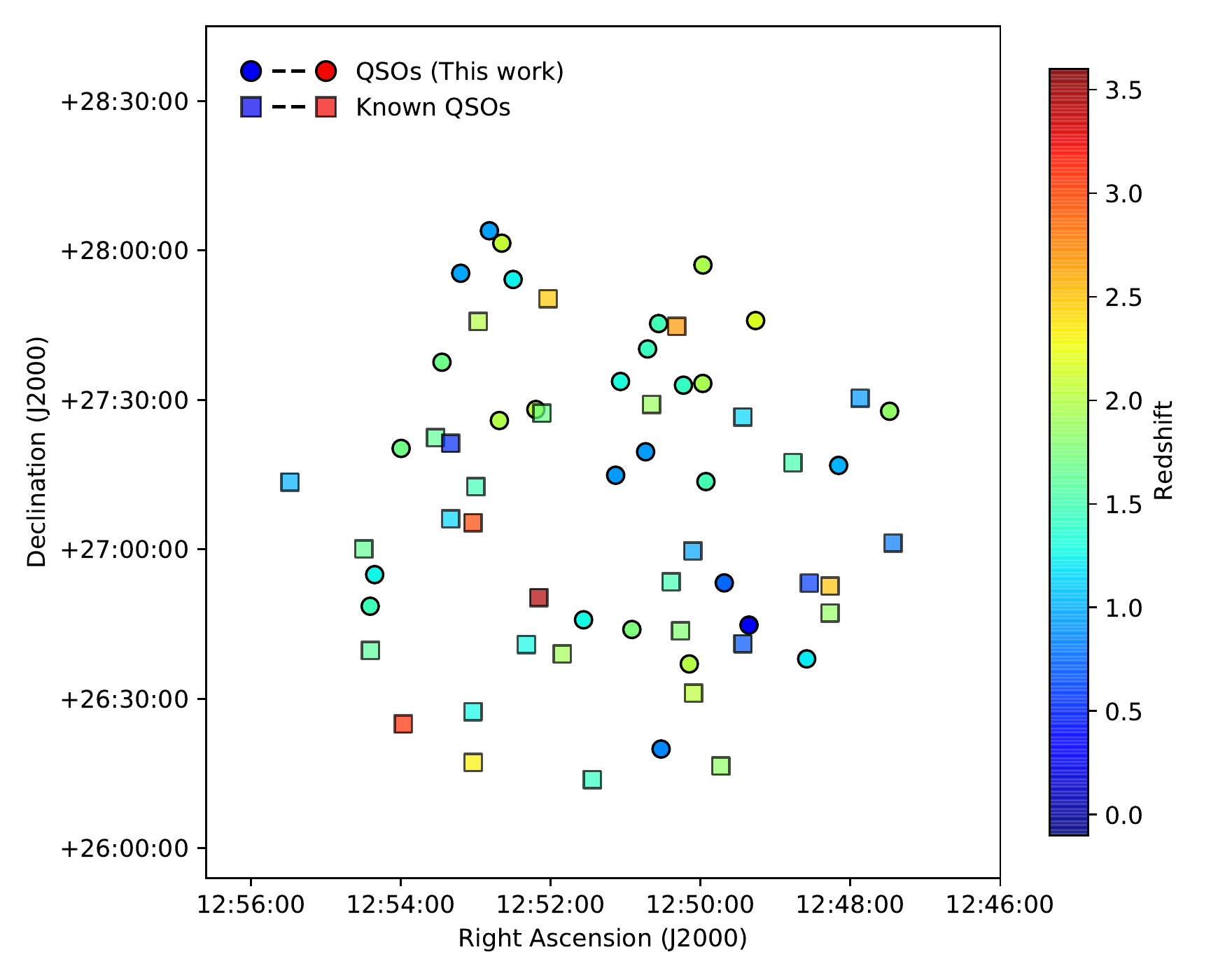}\includegraphics[height=8.0cm]{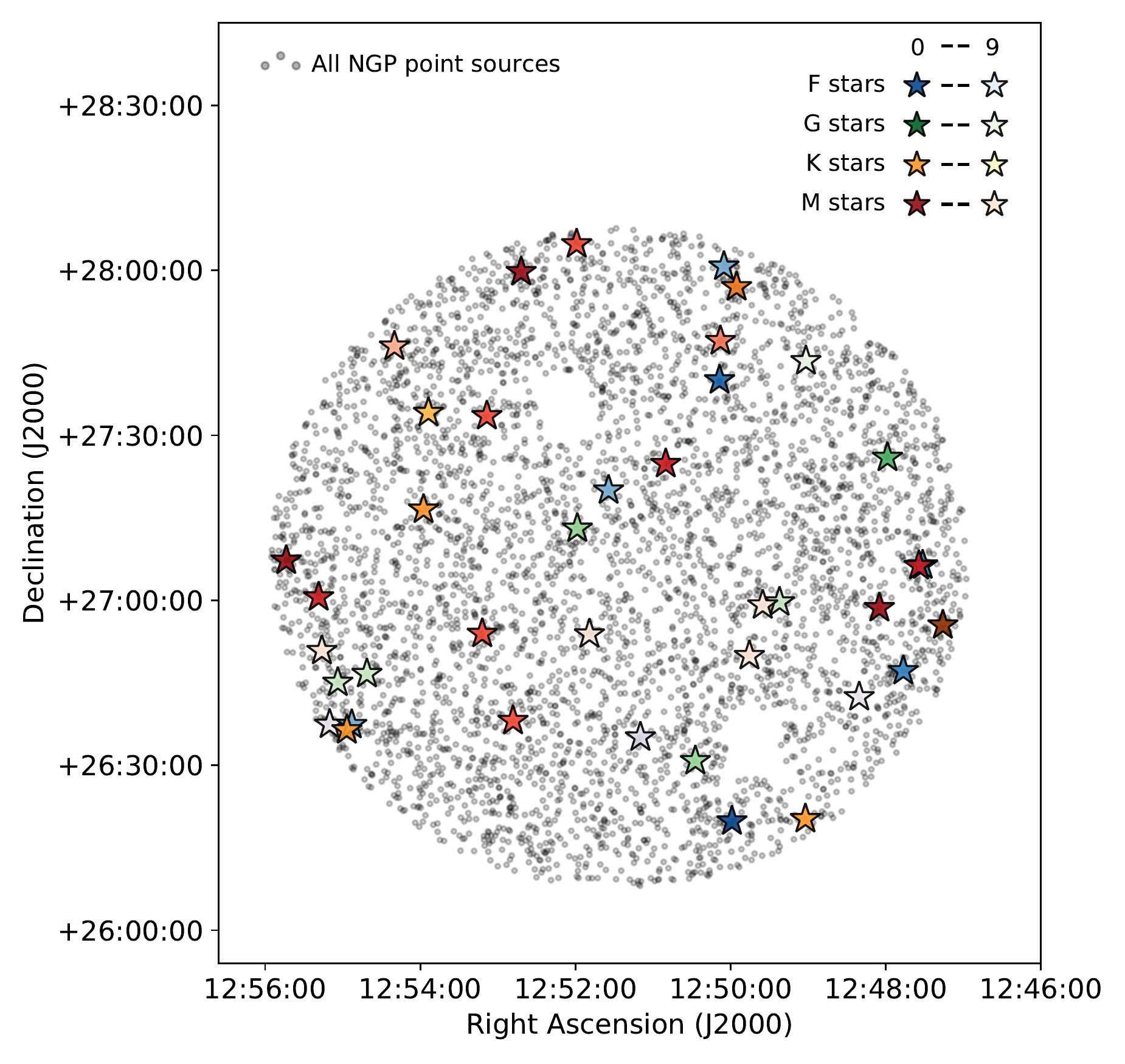}
    \caption{Skymaps of all point sources with measured proper motions from {\it Gaia} and $G<20$\,mag. The left panel shows the full sample of quasars, which are colour-coded by redshift; different symbols are used to illustrate quasars that have been found in the this work (filled circles) and previously known quasars (square symbols). The right panel shows all point sources (also with significant proper motions) as small black symbols, and the zero-proper-motion stars are marked with a colour-code reflecting their stellar type and sub-class.}
        \label{fig:skymap}
\end{figure*}

Tables~\ref{qsotab} and \ref{startab} list the full list of source classifications for quasars and stars, respectively. For each of the identified quasars, the redshift, the inferred visual extinction $A_V$, and a subset of their photometric colours, those that are commonly used for other quasar selection techniques, are provided. For the stars, the spectral type and sub-class are listed. Fig.~\ref{fig:skymap} shows a sky map of the classified targets, together with the full underlying distribution of point sources in the {\it Gaia}-DR2 sample, which are brighter than $G=20$\,mag. 

Out of the total 104 stationary point sources, 63 ($\approx 60\%$) are quasars. This yields a surface density of quasars that are brighter than $G < 20$\,mag of 20 deg$^{-2}$. To remove significant contamination from extended galaxies at low-$z$, our survey only targeted point sources, which, therefore, also excludes extended Type-1 AGN. We estimate this `missing' fraction to be  $\approx 30-45\%$ at most (i.e. 6 -- 9 extended quasars deg$^{-2}$), based on the complete quasar sample extracted from the VIMOS VLT Deep Survey (VVDS) by \citet{Gavignaud06,Gavignaud08}, considering only sources that are brighter than our survey limit. The true surface density of all quasars in this field which are brighter than $G < 20$\,mag, both extended and point-like, might thus be as high as $\lesssim 30$ deg$^{-2}$. We note that for unknown reasons, the region around the NGP is partly incomplete (see the right panel of Fig.~\ref{fig:skymap}) and we might therefore slightly underestimate the actual quasar surface density. We estimate that the two incomplete regions both roughly cover $0.03$\,deg$^2$, such that $\approx 2\%$ of the field is not included in our survey. In the following, we present the results for the newly identified quasars and the sources that have been identified as stars in turn.

\subsection{Newly identified quasars} \label{ssec:newqsos}

Quasars are readily identified based on the presence of strong, broad emission lines, typically with velocity widths of several 1000\,km\,s$^{-1}$. The redshifts for most quasars are easy to determine with a precision of a few percent from several distinct emission features. For a few of the sources, observations with additional grisms were obtained to confirm ambiguous redshift determinations from the presence of other emission lines (e.g. GQ124728+272742 and GQ125043+271934). 

Out of the 63 quasars, 27 are new spectroscopic quasar identifications from this work.
Individual spectra of all the quasars in the field can be inspected in Fig.~\ref{fig:spectra} (newly identified quasars), Fig.~\ref{fig:sdssspectra} \citep[quasars observed as part of the Sloan Digital Sky Survey, SDSS;][]{York00,Richards02,Paris18}, and Fig.~\ref{fig:otherspectra} (quasars from the CFHT survey of \citealt{Crampton1987}, but with no available spectra in the literature and hence re-observed in this work). 

\begin{figure}[!t]
\centering
    \includegraphics[width=8.7cm]{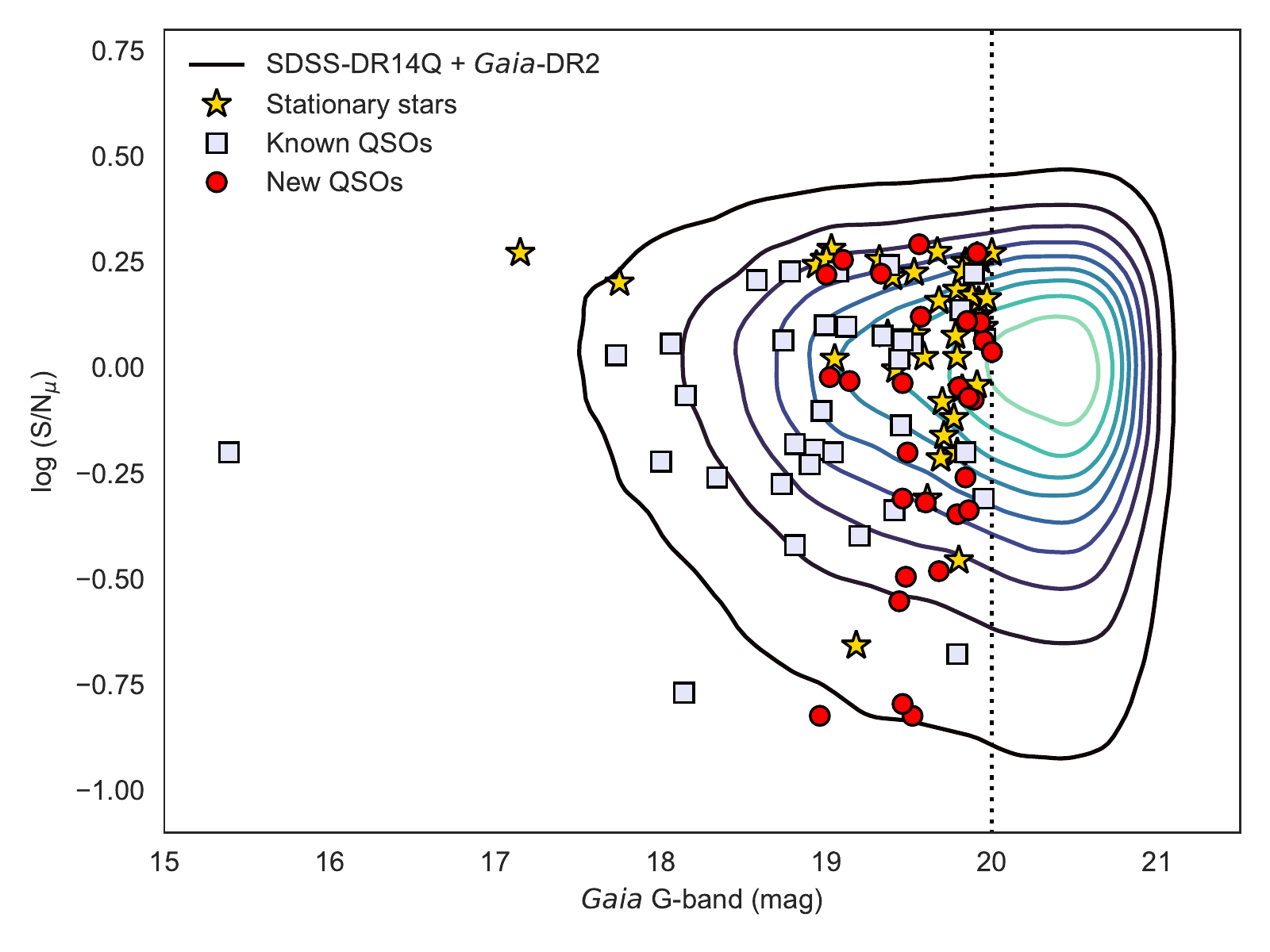}
    \caption{Signal-to-noise (S/N) measurements of the proper motion S/N$_{\mu} = \mu / \mu_{\rm err}$ vs. $G$-band magnitude of the zero proper motion sources. The symbols represent the following different classifications: quasars observed in this work (red filled circles), previously identified quasars (grey squares), and stellar sources (yellow star symbols). The contours represent the full SDSS-DR14 quasar sample cross-matched to the {\it Gaia}-DR2 catalogue. Our imposed limiting magnitude ($G<20$\,mag) is shown by the dotted line.
    }
        \label{fig:Gaiacomp}
\end{figure}

In Fig.~\ref{fig:Gaiacomp} we show the signal-to-noise (S/N) measurements of the proper motion S/N$_{\mu} = \mu / \mu_{\rm err}$ versus $G$-band magnitude of the zero proper motion sources. First, it is clear that the newly identified quasars are all part of the faintest zero proper motion sources, which is also occupied by the most significant stellar contamination. Then, we note that at brighter magnitudes ($G<19$\,mag), the SDSS is spectroscopically complete concerning the zero-proper-motion sources; whereas at $19<G<20$\,mag, only $\sim 38\%$ of the now confirmed quasars has been previously classified as such
spectroscopically by the SDSS.
For comparison, we show the underlying distribution of the full SDSS-DR14 quasar sample \citep[DR14Q;][]{Paris18}, cross-matched with the {\it Gaia}-DR2 catalogue as the background contours.

\begin{figure}[!t]
\centering
    \includegraphics[width=9cm]{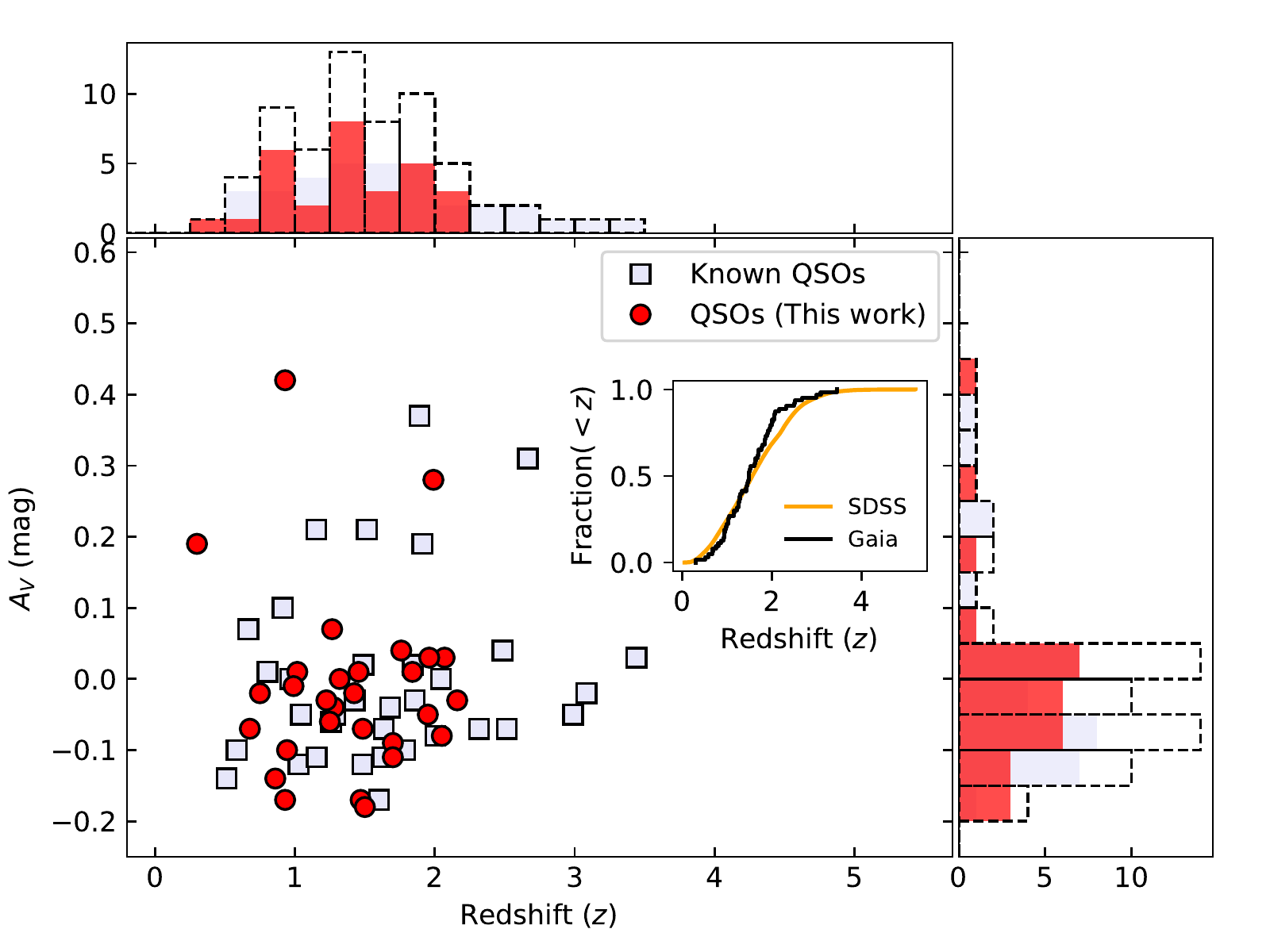}
    \caption{Scatter plot and histograms showing the measured redshift and $A_V$ distributions of quasars observed in this work (red filled circles) and those previously identified (grey squares). The black dashed lines in the two histograms show the combined redshift and $A_V$ distributions, respectively. In the inset, we show the CDF of the full {\it Gaia}-selected quasar sample (black) compared to the CDF of the SDSS-DR14Q sample (orange), considering only the quasars that are brighter than $G<20$\,mag.
    }
        \label{fig:zav}
\end{figure}

For each quasar, both observed as  part of this work and with existing archival spectra, the line-of-sight visual extinction (quantified by the extinction in the $V$-band, $A_V$) is determined in the following way. The quasar composite spectrum from \cite{Selsing2016} is fitted to the observed spectra and to available photometry from SDSS and UKIDSS. The Small Magellanic Cloud (SMC) extinction curve, as parametrized by \citet{Gordon2003}, is applied to determine the reddening in the rest-frame of the quasar. Since the observed spectral shapes may differ from the average spectral template, negative $A_V$ values are permitted in the fits to allow for intrinsically bluer spectral slopes than in the composite spectrum \citep[see methodology by][]{Krogager2015}\footnote{Another way to represent these intrinsic variations is by varying the power-law slope of the template before applying the reddening \citep[see][]{Krogager2016a}. However, this requires an extra free parameter which we are not able to constrain due to the lack of spectral data at longer wavelengths.}. The redshifted and reddened model is then compared to the data in order to obtain the best-fitting $A_V$. The formal uncertainty on the best-fitting $A_V$ is very small (of the order of 0.01~mag). However, the true uncertainty is dominated by the systematics related to the unknown intrinsic shape. This uncertainty is reflected statistically as the scatter around $A_V = 0$\,mag due to the inclusion of negative $A_V$ values. This systematic uncertainty amounts to 0.07~mag, which is a conservative estimate of the uncertainties on the best-fitting $A_V$ measurements.
Table~\ref{qsotab} provides a summary of the results and Fig.~\ref{fig:zav} shows a scatter plot and histograms of the redshift and $A_V$ distributions. The full sample spans a redshift range of $0.2<z<3.5$, and the majority of quasars show broad emission lines and negligible dust extinction ($A_V<0.1$\,mag). Only one quasar, GQ\,125150+263900, shows broad absorption line (BAL) features, which are otherwise found in abundance in targeted searches for reddened quasars \citep{Fynbo2013,Krogager2015}.

\begin{figure}[!t]
\centering
    \includegraphics[width=9cm]{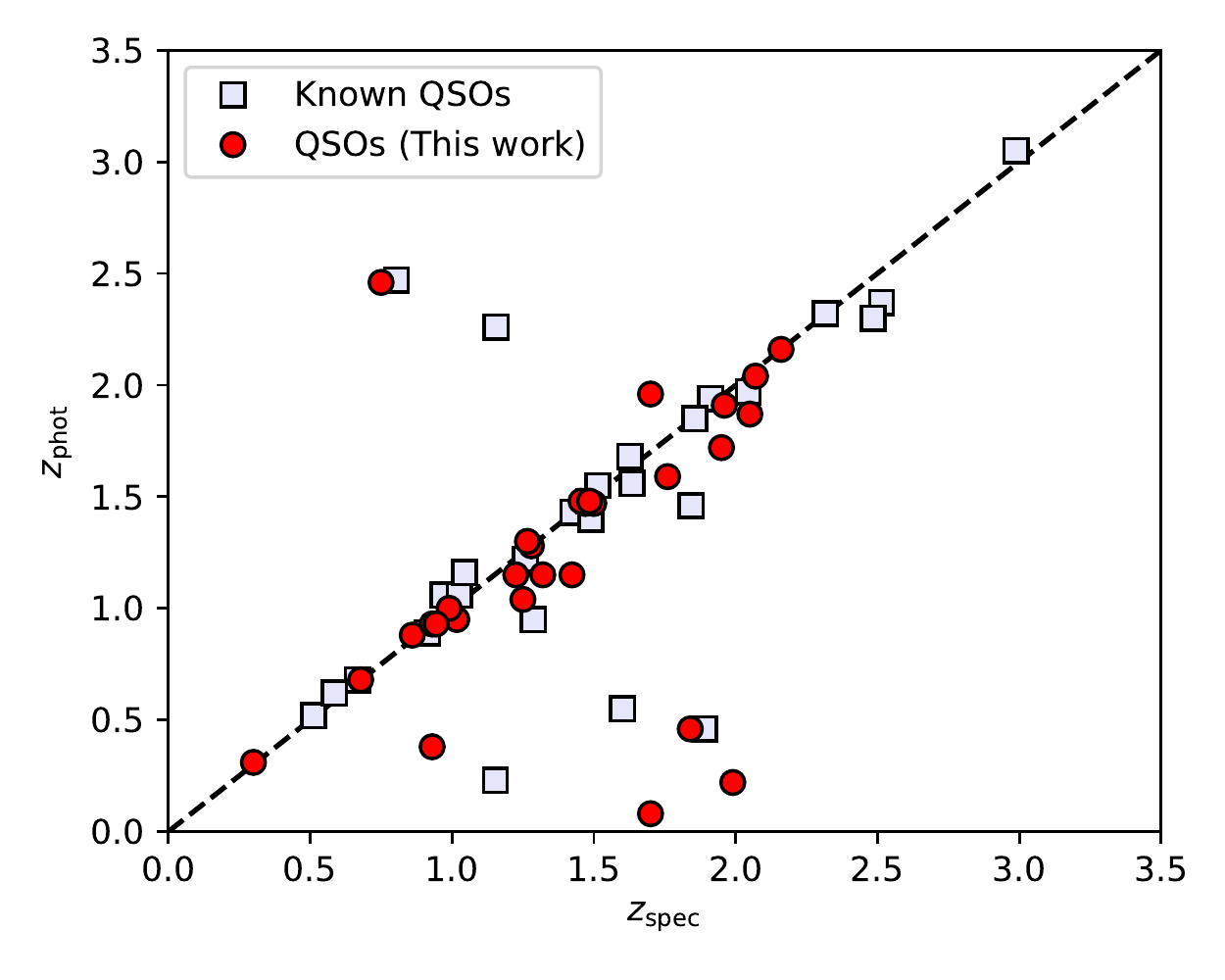}
    \caption{Comparison between the photometric redshifts estimated by \citet{Richards2009} and the spectroscopic redshifts determined in this work (red filled circles) or in previous surveys (grey squares).
    }
        \label{fig:zcomp}
\end{figure}

For all of the new targeted quasar candidates in this survey, \citet{Richards2009} had previously classified them as quasars based on their photometry and determined photometric redshifts. 
In Fig.~\ref{fig:zcomp} the photometric redshifts from \citet{Richards2009} are compared to our spectroscopic redshift measurements. The agreement is overall very good. The outliers are typically systems with significant amounts of reddening.

\subsection{GQ124728+272742}

\begin{figure*}[!t]
\centering
    \includegraphics[width=7.0cm]{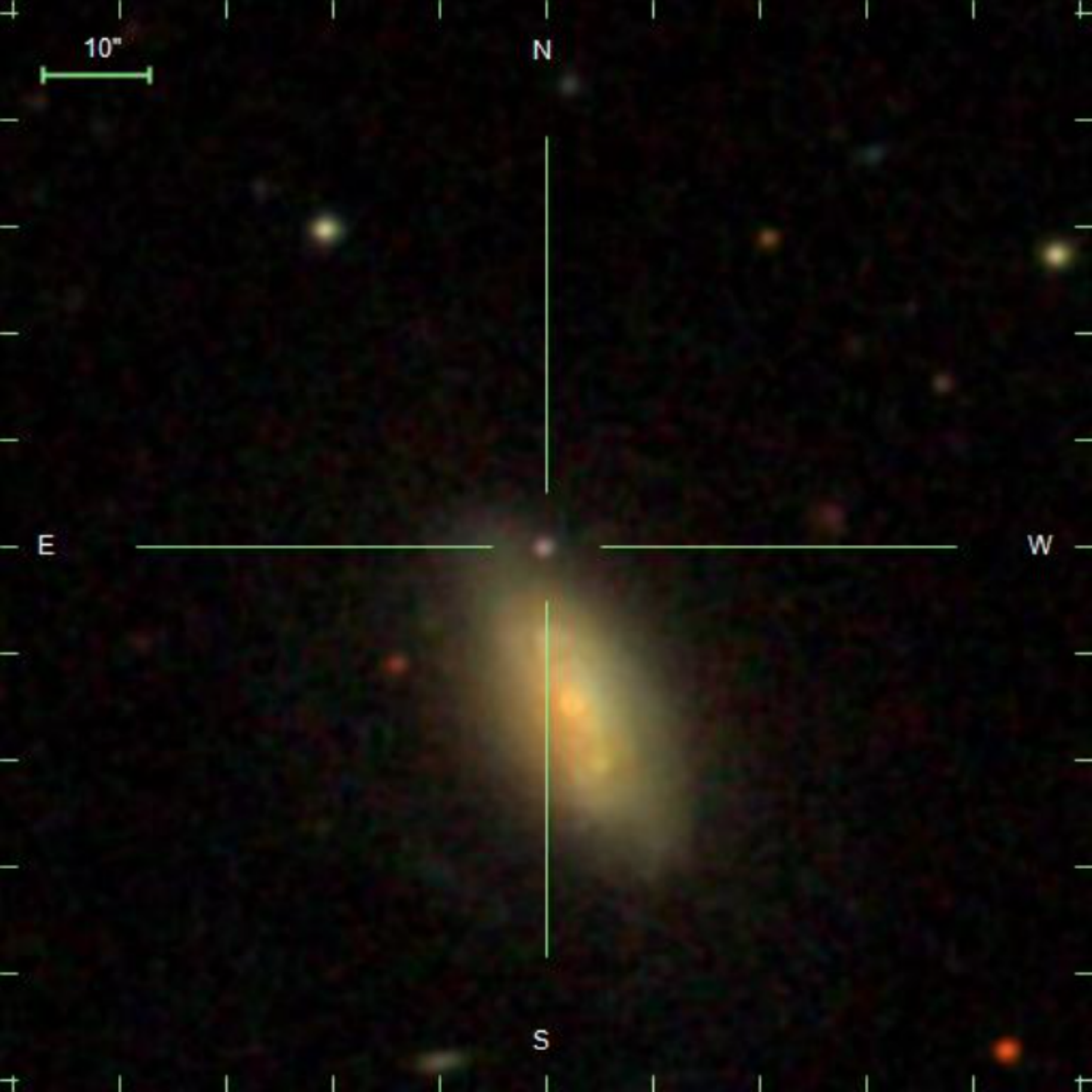}
    \includegraphics[width=7.2cm]{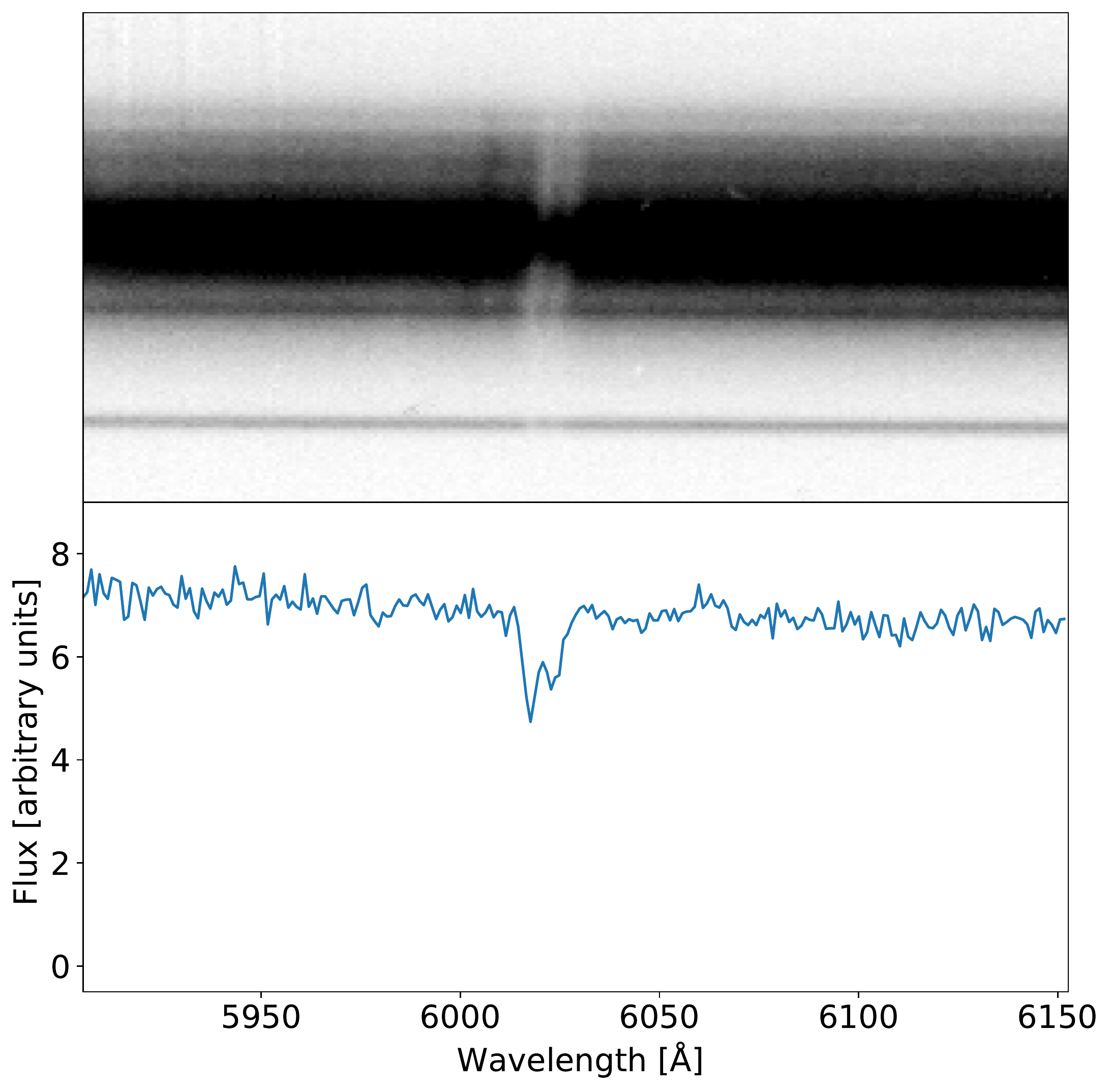}
    \includegraphics[width=13.0cm,height=8cm]{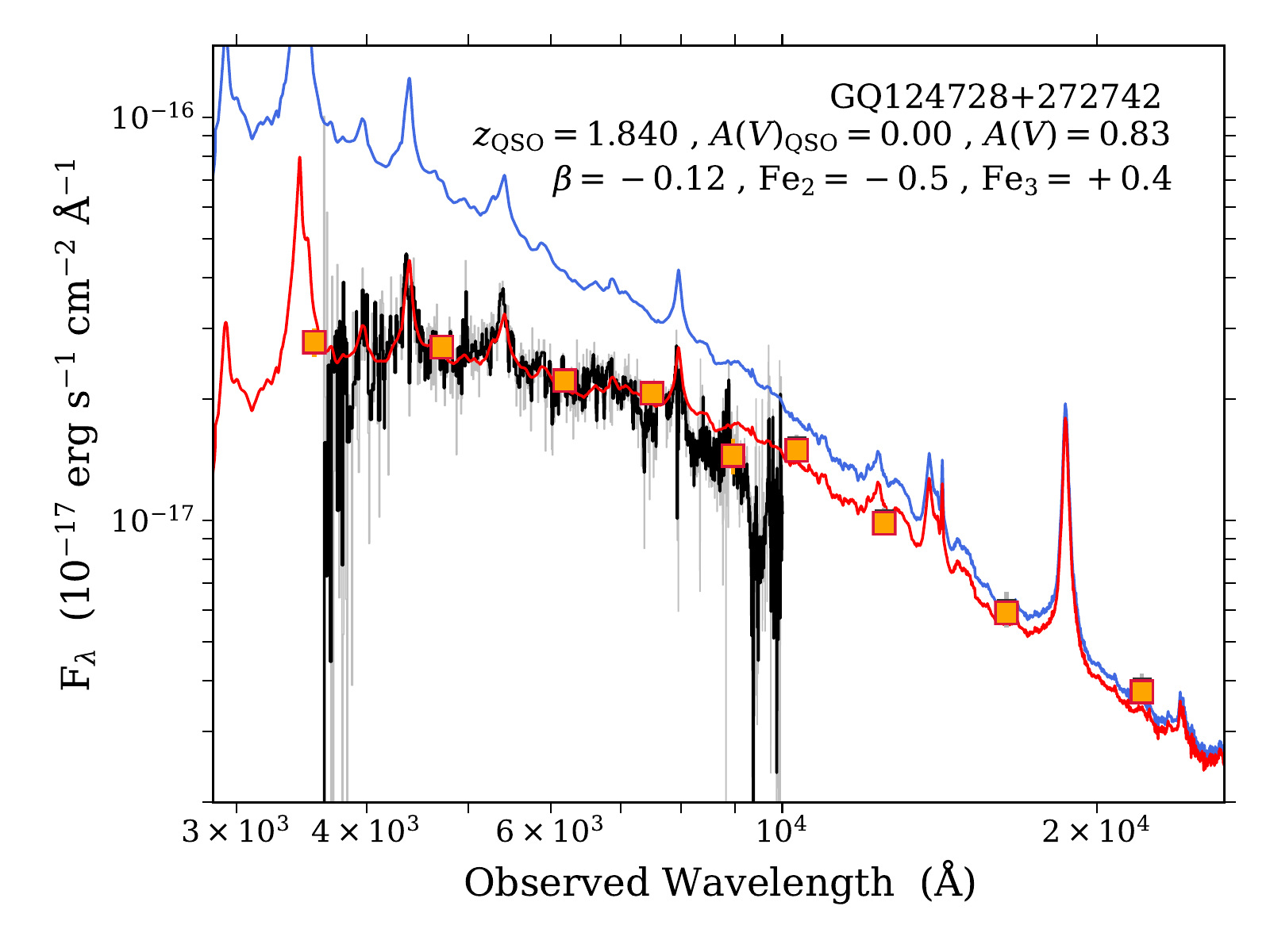}
    \caption{{\it Top Left:} Image of the field of GQ124728+272742 from SDSS. The quasar is marked with a line-of-sight passing through the outskirts of the $z=0.022$
    foreground galaxy SDSS J124728.37+272728.0.
    {\it Top Right:} Spectra covering the spectral region around \ion{Na}{D}. The top panel shows the two-dimensional 
    spectrum covering the foreground galaxy and the quasar. The NaD doublet can be seen in absorption 
    against the stellar light of the foreground galaxy and against the quasar continuum.
    {\it Bottom:} Analysis of the spectral energy distribution of the quasar. The orange squares mark the photometric measurements from $u$ to $z$ (from SDSS), and from $Y$ to $K$ (from UKIDSS). The black curve represents our observed spectrum. The blue and the red are the quasar composite spectrum from \citet{Selsing2016} without and with extinction, respectively. The parameters of the best fit are given in the upper right corner \citep[following][]{Krogager2016}.
    }
        \label{fig:GQ124728+272742}
\end{figure*}

One of the newly identified quasars, GQ124728+272742, warrants special attention as the line-of-sight to this quasar intersects the outer part of the dusty, star-forming foreground galaxy SDSS J124728.37+272728.0 at $z=0.022$ (See Fig.~\ref{fig:GQ124728+272742}). In the spectrum of 
the quasar, we measure strong \ion{Na}{D} doublet absorption (with an equivalent width of $2.2\pm0.2$ \AA).
Modelling of the quasar spectrum, assuming a foreground dust screen at $z=0.022$ and using the methodology of \citet[][their sect. 4.1]{Krogager2016} to model the unextinguished quasar spectrum, shows that dust in the foreground galaxy likely causes extinction at the level of $A_\mathrm{V}=0.8$\,mag. The strength of the \ion{Na}{D} line appears consistent with the relation between extinction and \ion{Na}{D} found in other studies \citep[e.g.][]{Murga2015}, but we note that this study probes significantly weaker \ion{Na}{D} lines compared to what we see in the case of GQ124728+272742.

\subsection{Stellar sources}

We find that 41 sources ($\approx 40\%$) from the total zero-proper-motion sample are stars. These are readily identified due to the lack of broad emission line features and from the presence of stellar absorption lines. We classify the stellar types and sub-classes using {\tt PyHammer} \citep{Kesseli2017}. This code classifies stellar
spectra by cross-matching input spectra with a library of empirical stellar spectra, which was created using
spectra from the SDSS Baryon Oscillation Spectroscopic Survey (BOSS). The 41 stars in our sample cover most of the sub-classes within the spectral types F, G, K, and M (see Table~\ref{startab}). A fraction of these ($\approx 25\%$) show significant parallaxes consistent with their stellar nature. We discuss the use of parallax measurements as an additional selection criteria in Sect.~\ref{ssec:plx}.

\section{Discussion}
\label{sec:disc}

With our spectroscopic campaign, we have built a complete sample of point sources with zero proper motions (within 2$\sigma$ in {\it Gaia}-DR2), one degree from the NGP (see paper I for further details on the definition of the sample). The magnitude limit is set by the {\it Gaia} $G$-band measurement at $G<20$\,mag and this in turn is set by the requirement of having a precise measurement of the proper motion.

In the field that we have surveyed, that is 3.14 square degrees centred on the NGP, there are more than 4500 point sources at $G < 20$\,mag in the broad {\it Gaia} $G$-band. Of these, only the 104 sources ($2.3\%$) targeted here have proper motions consistent with zero to within 2$\sigma$. The simple criterion requiring the sources to be stationary on the sky is thus already a powerful discriminator, which allows one to remove the vast majority of stars from the large population of point sources. The astrometric selection and our complete spectroscopic follow-up allow us to state with certainty that we now have a complete quasar sample down to the imposed $G$-band magnitude of 20 mag. In the following, we first discuss the properties of this sample and then discuss the perspectives for astrometric quasar selection.

\subsection{Quasar properties}

Firstly, we can now investigate the basic properties of the underlying quasar population within the imposed $G$-band limit. Quantities, such as the redshift distribution and the fraction of dust-reddened quasars, provide important constraints on the formation and evolution of galaxies and the interplay between their central supermassive black hole.

\subsubsection{Redshift distribution}

The redshift distribution of the {\it Gaia}-NGP quasar sample is shown in Fig.~\ref{fig:zav}. By including both the new quasars observed as part of this work and those extracted from the literature, we find that the distribution peaks at $z\sim 1.5$. In the inset of Fig.~\ref{fig:zav}, we compare the redshift cumulative distribution function (CDF) of the {\it Gaia}-selected quasars, with the redshift CDF of the cross-matched SDSS-DR14Q and {\it Gaia}-DR2 catalogue, considering only the quasars that are brighter than $G<20$\,mag. We find a median redshift for the {\it Gaia}-selected quasar sample of $z = 1.49$, compared to the median $z = 1.56$ of the SDSS-DR14Q sample. Performing a two-sample Kolmogorov-Smirnov (KS) test for the two CDFs to be drawn from the same distribution yields $P=0.07$. We therefore cannot reject the null hypothesis that the two populations sample the same redshift distribution. This suggests that any biases in the SDSS-DR14Q selection has a small impact on the redshift distribution.

\subsubsection{The fraction of red quasars}

\begin{figure}[!t]
\centering
    \includegraphics[width=9.2cm]{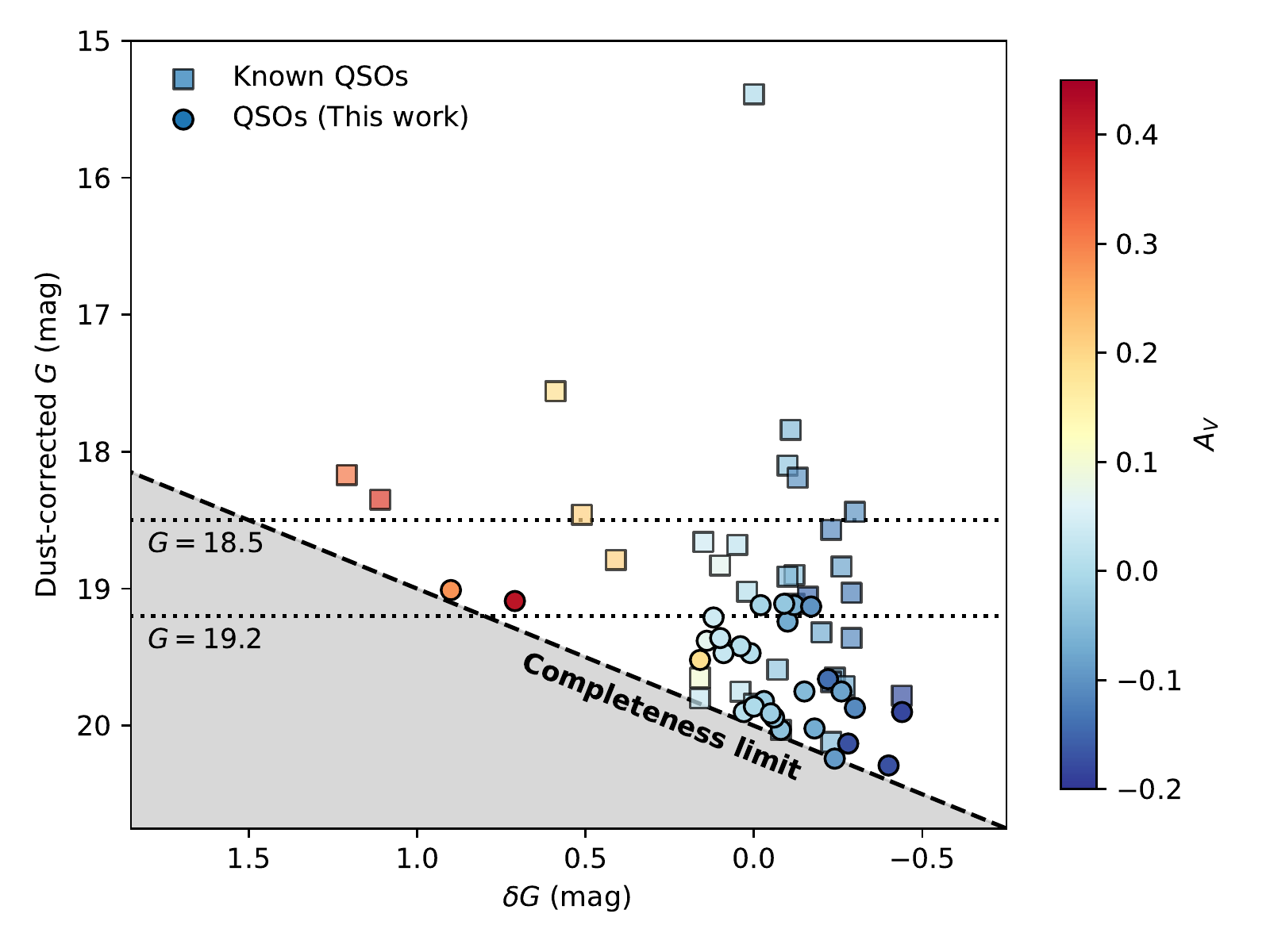}
    \caption{Dust-corrected $G$-band magnitudes vs. the correction factor $\delta G$ of the full quasar sample, colour-coded as a function of $A_V$. Our completeness limit of $G_{\rm obs} < 20$\,mag is shown by the grey-shaded area. 
    }
        \label{fig:gmagcorr}
\end{figure}

Out of the total 63 spectroscopically confirmed quasars in our NGP survey, only eight systems ($13^{+5}_{-3}\%$) show significant dust reddening with $A_V > 0.1$\,mag. It is important to recall here that a dust reddened quasar also becomes fainter in the $G$-band and therefore could drop out of the sample definition. That is, while the observed fraction of dust-reddened quasars is 13\% in the sample presented here, the 'intrinsic reddened fraction' could be much higher. We illustrate this in Fig~\ref{fig:gmagcorr} where, for each object in our sample, we have corrected the observed $G$-band magnitude to the intrinsic (dust-corrected) $G$ magnitude and plotted it against the 'dust-correction at $G$' ($\delta$G). Visualising our sample in this way means that the intrinsic completeness limit is now a function of the absorption in the $G$ band. The symbols in Fig.~\ref{fig:gmagcorr} are colour-coded following the amount of dust absorption. Dust absorption moves an object towards the left, and increasingly so with larger dust absorption. However, the effect is also a function of redshift, which is why the reddest points are not necessarily the ones furthest to the left.

\begin{figure*}[!t]
\centering
    \includegraphics[width=6cm]{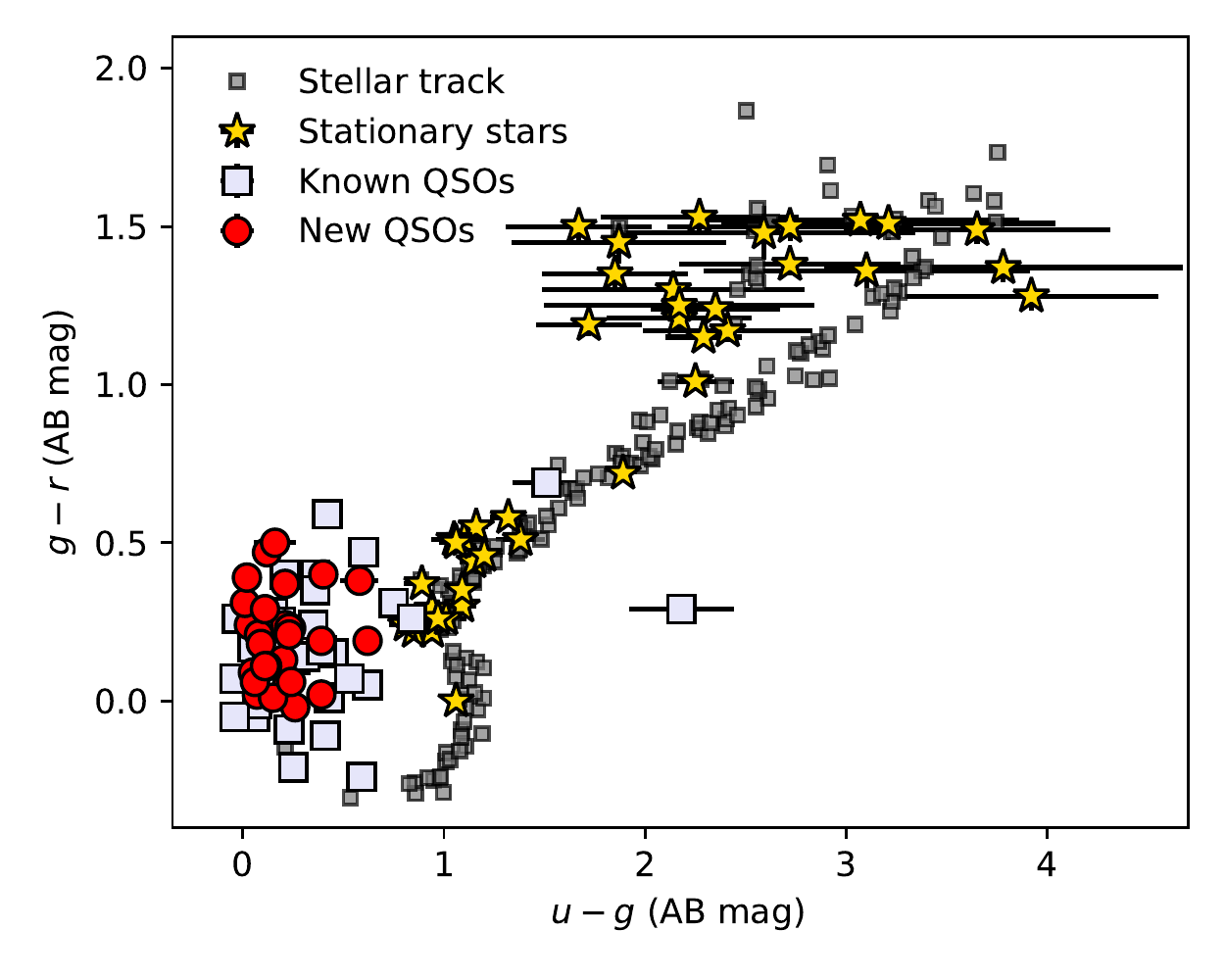}
    \includegraphics[width=6cm]{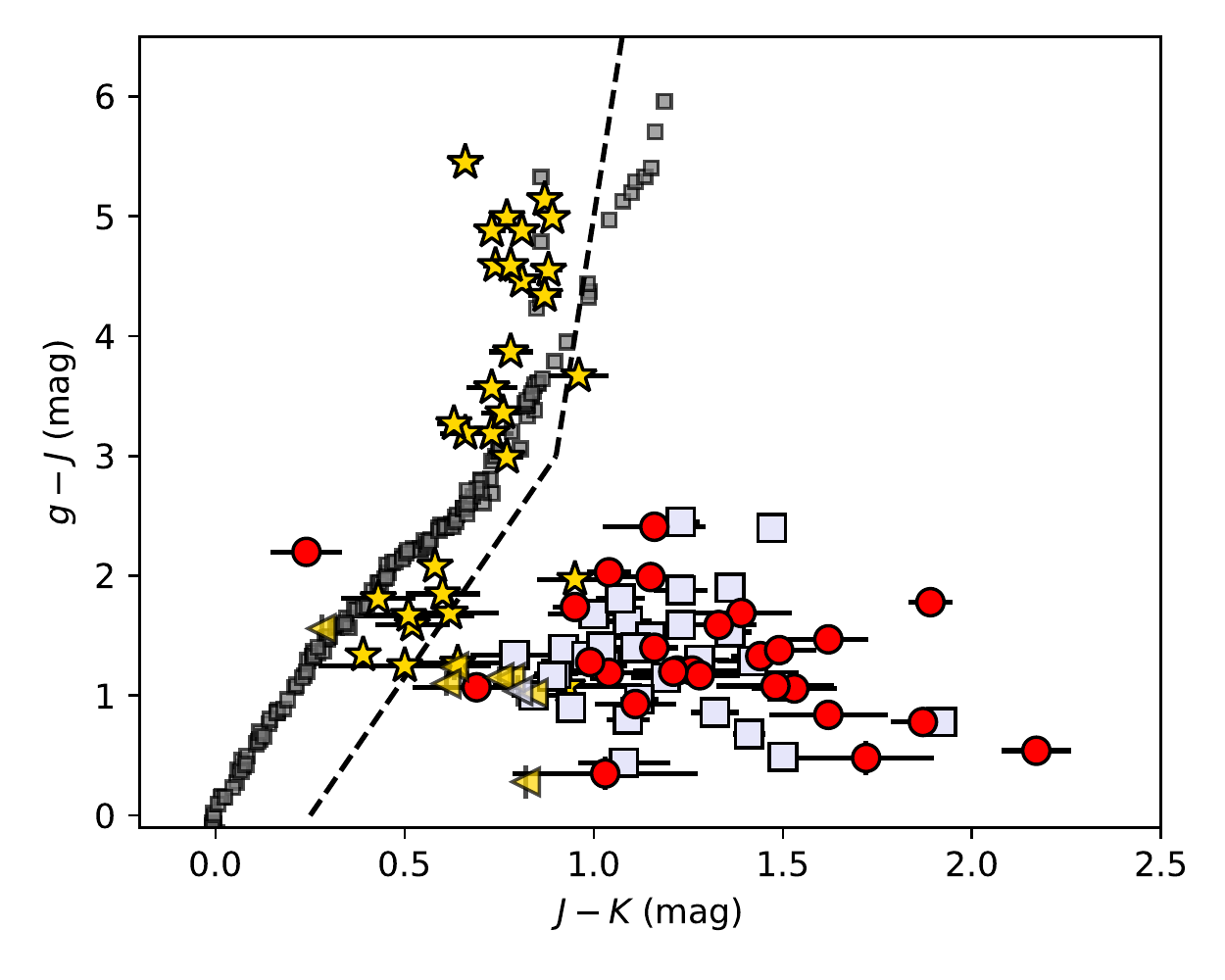}
    \includegraphics[width=6cm]{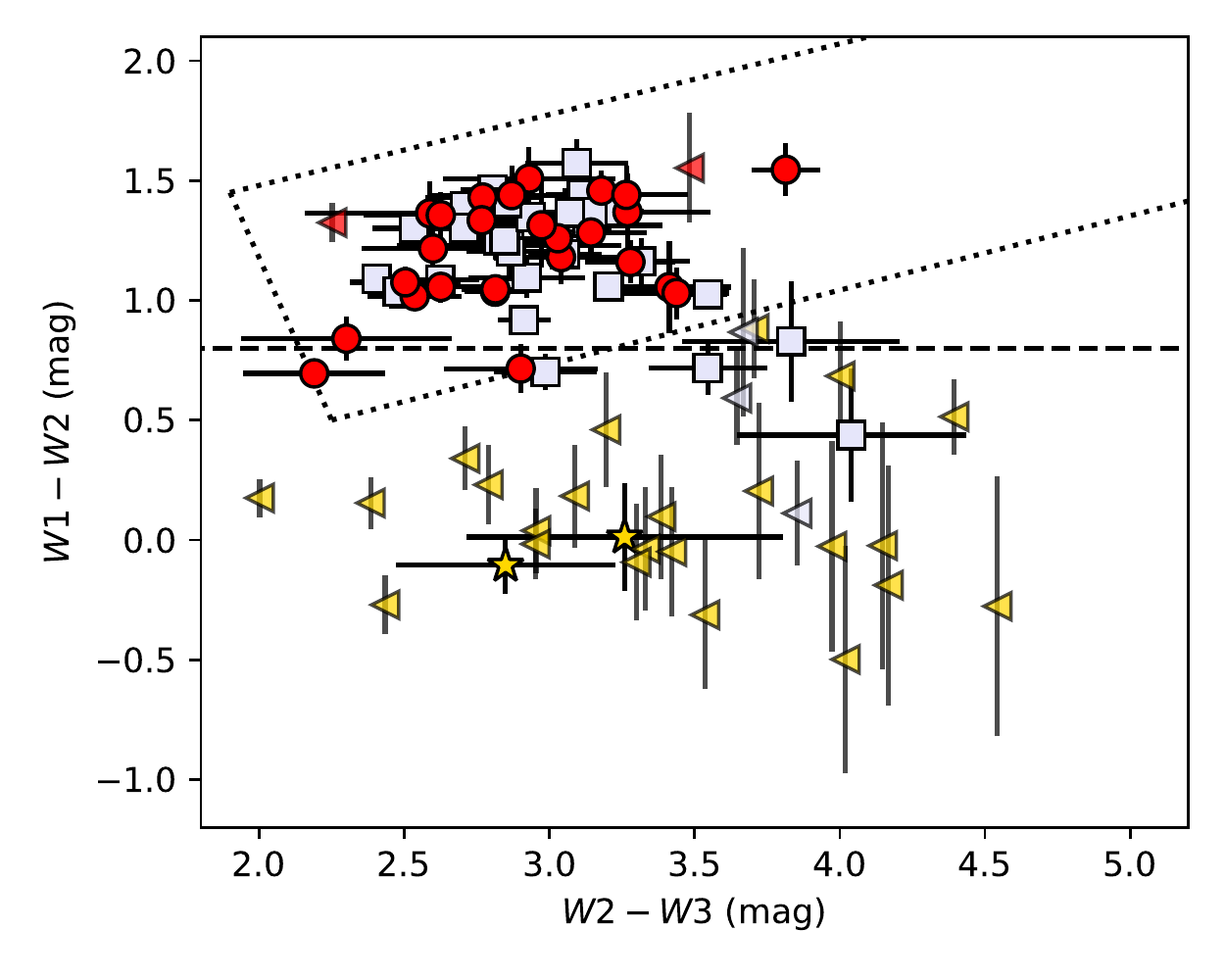}
    \caption{Colour-colour diagrams of the zero proper motion sources. The quasars observed in this work are shown by the red filled circles, the previously known quasars by the grey squares, and the stellar sources by the yellow star symbols. Limits are denoted by triangles, which are colour-coded in the same way. The left panel shows the $u-g$ vs. $g-r$ colours of the targets, which represent the UVX selection, the middle panel illustrates the $g-J$ vs. $J-K$ colours (KX selection), and the right panel shows the WISE colours $W1-W2$ vs. $W2-W3$. For comparison, typical stellar colours from the catalogue of \citet{Hewett06} are shown as the dark grey squares in the left and middle panel. In the middle panel, we also show the selection line from \citet{Maddox08,Maddox12}, selecting targets with a $K$-band excess relative to typical stellar colours. In the right panel, we show the selection criterion of $W1-W2 > 0.8$ presented by \citet{Stern12} as the dashed line and the selection wedge defined by \citet{Mateos12} as the dotted line.
    }
        \label{fig:multicol}
\end{figure*}

From Fig.~\ref{fig:gmagcorr} it is first seen that the blue (non-absorbed) points are clustered to the right and mostly down at the faint end. This distribution simply reflects the quasar luminosity function. The reddened quasars are seen to hover in the middle left side of the figure. The lack of intrinsically faint reddened quasars is evident but also logical as even a small amount of reddening of a faint quasar pushes it below our completeness limit. Simply computing the dust-reddened quasar fraction using the few intrinsically bright reddened quasars versus all of the non-reddened ones is clearly strongly under-estimating the intrinsic value. To get a better estimate, we instead make two horizontal cuts in Fig.~\ref{fig:gmagcorr} at intrinsic $G$ magnitudes 19.2 and 18.5 (shown by the dotted lines), capturing the regions where significantly dust-reddened quasars can still enter our sample. At those cuts, we find fractions of $26^{+10}_{-7}\%$ and $44^{+16}_{-14}\%$, respectively, for red quasars relative to the total number of quasars above these intrinsic thresholds. In conclusion, the intrinsic red fraction of this sample is much higher than the overall observed fraction; however, based on these data alone, we cannot quantify by exactly how much. 

In an earlier complete quasar sample study, \cite{Heintz2016} found 
that $39^{+9}_{-8}\%$ of all quasars had
$A_V > 0.1$\,mag. In that study, the quasars were selected through 
a wide range of methods to ensure the completeness of the sample
(radio, X-rays, optical, near-infrared, and mid-infrared colours)
and the completeness limit was defined in the $J$-band where the
effect of reddening is less severe. The consistency between the
result from the highest intrinsic cut ($G=18.5$\,mag) in Fig.~\ref{fig:gmagcorr} and that of the $J$-band limited sample suggests that the true intrinsic red fraction is at least a factor of $3-4$ higher than the observed fraction of $\approx 13\%$ in the $G$ magnitude limited sample presented here. In a future work, we will discuss how one may determine the true intrinsic red quasar fraction from a larger {\em Gaia}-selected sample in detail.

\subsection{Efficiency of photometric selection of quasars} \label{ssec:photsel}

Based on this {\it Gaia}-selected sample of quasars, the completeness of complementary searches for quasars using other, commonly adopted colour-selection criteria can be quantified. First, we note that we found two additional quasars in the NGP field in Paper I, both with $S/N_{\mu} > 2$. This sample is therefore only complete within the statistics from the $2\sigma$ cut on the proper motion, but it is unbiased in terms of colours. It is also only possible to assess the completeness of these other surveys within the same limiting magnitude $G<20$\,mag. For this analysis, we thus cross-match the full {\it Gaia}-NGP sample with the optical SDSS-DR14 survey \citep{Paris18}, the near-infrared UKIRT Infrared Deep Sky Survey (UKIDSS) Large Area Survey (LAS) \citep[DR10;][]{Lawrence2007}, and the mid-infrared Wide-field Infrared Survey Explorer ({\it WISE}) auxiliary data \citep[from the AllWISE catalogue;][]{Cutri13}. 

In Fig.~\ref{fig:multicol}, the optical, near-, and mid-infrared photometry in colour-colour diagrams of all of the zero proper motion sources examined as part of this work are shown. The first `radio-quiet' low-$z$, unobscured quasars were identified by their ultraviolet (UV) excess (UVX) relative to that of main-sequence stars \citep{Sandage1965,Schmidt83}. This is represented as the typical bluer $u-g$ colours in the left panel of Fig.~\ref{fig:multicol}. Only five quasars ($\sim 8\%$ of the full quasar sample) show optical $u-g$ colours, which is inconsistent with the majority of the quasar population. A typical UVX selection is thus complete to about $92\%$ down to the limiting magnitude of $G<20$\,mag. Although only about half of the {\it Gaia}-selected quasars had existing spectra from the SDSS, this is likely mainly due to the issue of fibre-collisions in the SDSS spectroscopic follow-up \citep[e.g.][]{Strauss2002}. For example, \citet{Richards2009} classified 58 out of the 63 ($\approx 92\%$) confirmed quasars as such based on their photometric algorithm specifically targeting sources with a high UV excess. Moreover, none of the sources classified spectroscopically as stars in our work were flagged as quasars in their work. This specific UVX selection thus also has a completeness of $92\%$, with a selection efficiency of $\approx 100\%$ within the survey limit.

A complementary approach for identifying quasars, which is less sensitive to dust or Lyman-$\alpha$ forest blanketing, is based on the observed $K$-band excess \citep[KX;][]{Warren2000} of quasars. To examine this selection, the middle panel of Fig.~\ref{fig:multicol} shows the demarcation line separating quasars and stars as defined by \citet{Maddox08,Maddox12} in $g-J$ versus $J-K$ colour-colour space together with all of the zero proper motion sources for which a near-infrared counterpart was detected. Only two of the spectroscopically confirmed quasars ($3\%$) have colours that are inconsistent with the bulk of the quasar population (i.e. are located to the left of the demarcation line). There were five additional {\it Gaia}-identified sources with no near-infrared counterpart, which is likely due to their intrinsic faintness. Including these, we thus estimate the completeness of the KX method to be about $90\%$ for sources that are brighter than $G<20$\,mag.

Another approach to identify quasars, which is also less biased against extremely dust-obscured sources compared to optical surveys, is via a selection based on mid-infrared colours. This has the added advantage that quasar candidates can be selected from the all-sky {\it WISE} survey. In the right panel of Fig.~\ref{fig:multicol}, the {\it Gaia}-selected sample is compared to the {\it WISE} selection criteria of \citet{Stern12} and \citet{Secrest15}. \citet{Stern12} used a simple criterion of $W1-W2 > 0.8$ to identify quasars (shown as the dashed line in the figure), whereas \citet{Secrest15} adopted the $W1-W2$ versus $W2-W3$ selection wedge defined by \citet{Mateos12} (dotted line). We find that 55 of the spectroscopically confirmed quasars ($86\%$) meet the criteria that $W1-W2 > 0.8$ of \citet{Stern12} and 56 ($88\%$) would be selected following the same $W1-W2$ versus $W2-W3$ colour criteria as \citet{Secrest15}. In addition, one confirmed quasar does not have a {\it WISE} detection in any of the $W1$, $W2,$ or $W3$-bands (GQ\,124752+273018). Including this source, we thus estimate a survey completeness of $\approx 85\%$ using typical mid-infrared {\it WISE} selection criteria for quasars that are brighter than $G<20$\,mag.

\subsection{Improving the astrometric selection} \label{ssec:plx}

\begin{figure*}[!t]
\centering
    \includegraphics[width=15cm]{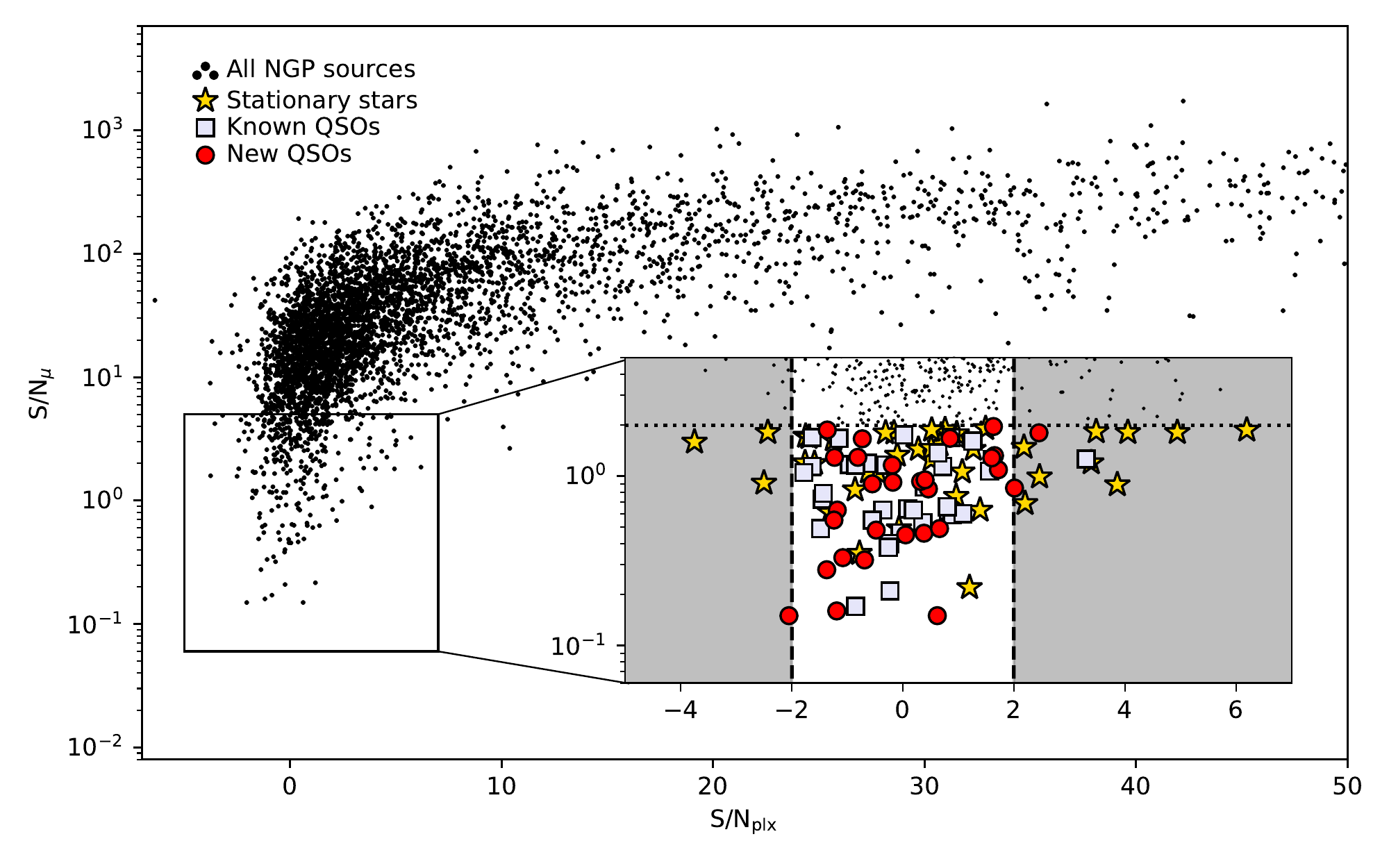}
    \caption{Significance on the proper motions vs. the parallaxes for the full sample of {\it Gaia} sources within 1 deg of the NGP, all brighter than $G < 20$\,mag. In the inset we show the classifications of the full sample of zero proper motion sources (defined as ${\rm S/N}_{\mu} < 2$; dotted line), where red dots and grey squares denote quasars and star symbols represent the stellar contaminants. The grey-shaded regions show the additional proposed criterion (see Sect.~\ref{ssec:plx}), requiring parallaxes to be consistent with zero (within $2\sigma$). It is clear that imposing this additional criteria would exclude a large fraction ($12/41 \sim 30\%$) of the confirmed stellar sources.
    }
        \label{fig:Gaia_astromet}
\end{figure*}

Paper I found that few of the sources with apparent proper motions on the sky had significant parallax measurements, indicating that they were likely stars. To quantify this, Fig.~\ref{fig:Gaia_astromet} shows the significance on the proper motions versus the parallaxes of all of the sources within the NGP field. The inset zooms in on the sources we followed up on spectroscopically in this work. From the figure, it is clear that a large fraction of the classified stellar sources have significant parallaxes. Requiring that the parallaxes be consistent with zero within $2\sigma$ (similar to the proper motion cut) yields a total sample size of 89 sources, of which 61 are classified as quasars and 28 are classified as stars. Imposing this additional zero-parallax criterion, without introducing any biases, would therefore increase the quasar selection efficiency from 61\% to 69\%. We caution that this extra criterion would still make the sample less complete, though only in a statistical sense (i.e. not related to any specific intrinsic quasar properties). 

\begin{figure}[!t]
\centering
    \includegraphics[width=9cm]{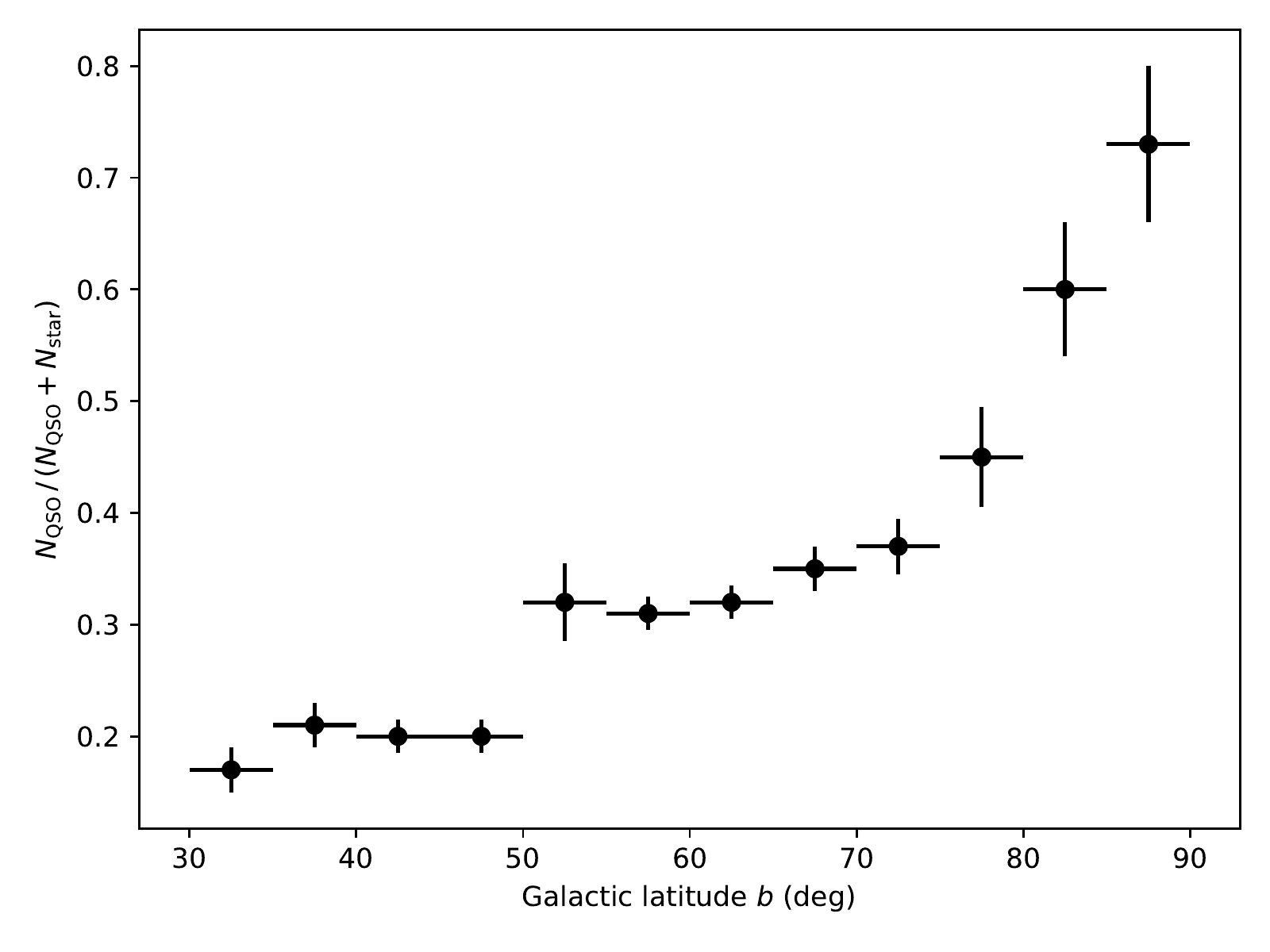}
    \caption{Fraction of likely quasars to the total number of sources with zero proper motions and parallaxes as a function of Galactic latitude $b$. The quasars are identified via their $K$-band excess (see Sect.~\ref{ssec:plx}). The fractions are shown in bins of $\Delta b = 5$\,deg. The horizontal errorbars show the widths of the bins and the vertical errorbars indicate the $1\sigma$ Poisson errors for each bin. A selection efficiency of $N_{\rm QSO} / N_{\rm tot} > 50\%$, using a purely astrometric identification for quasars, can only be reached at $b\gtrsim 80$\,deg.
    }
        \label{fig:QSOeffb}
\end{figure}

Motivated by the preliminary study of \citet{Heintz2015}, we have built the sample of zero proper motions sources around the NGP to limit the contamination of stellar sources. Here, we now assess the actual selection efficiency as a function of Galactic latitude $b$ based on the {\it Gaia}-DR2 astrometric data. To quantify the selection efficiency, we extract a representative set of sources from the {\it Gaia}-DR2 catalogue with proper motions and parallaxes consistent with zero (again within $2\sigma$) and cross-match it to the UKIDSS DR10 LAS catalogue. We then apply the KX selection (see Sect.~\ref{ssec:photsel}) to separate the candidates that are most likely stars or quasars. In Fig.~\ref{fig:QSOeffb} we show the fraction of quasars to the total number of sources with zero proper motions and parallaxes as a function of $b$, ranging from $b = 30-90$\,deg in steps of $\Delta b = 5$\,deg. At $b>85$\,deg, we find that the selection efficiency of a purely astrometric selection is $73\pm14\%$, which is consistent with the confirmed fraction of quasars within the NGP 3.14 square degree sample. However, already at $b \lesssim 80$\,deg, the selection efficiency drops to $\sim 50\%$. Closer to the Galactic plane at $b = 30-50$\,deg, only $\approx 20\%$ of the sources with zero proper motions and parallaxes are likely quasars. We expect a similar trend between the selection efficiency and Galactic latitude going from the south Galactic pole to the Galactic disc \citep[see also][]{Heintz2015}. A purely astrometric selection of quasars based on the measurements from {\it Gaia}-DR2 is thus mainly efficient at high Galactic latitudes, $|b|\gtrsim 80$\,deg. We note, however, that with the release of future {\it Gaia} data with more sensitive measurements of the proper motions, the fraction of recovered quasars to stars will increase as the measurement error decreases.

\section{Summary and outlook} \label{sec:conc}

In this work, we have presented a spectroscopically complete sample of quasar candidates, which have been identified purely from their apparent zero proper motion on the sky. The targets were extracted from the {\it Gaia}-DR2 catalogue, and we find a total of 104 sources to be brighter than $G=20$\,mag within a 3.14 square degree survey area around the NGP. Based on spectroscopic data of the full set of candidates, we classified 63 of the sources as quasars and 41 as stars. This yields a surface density of quasars with $G<20$\,mag of 20 deg$^{-2}$. Based solely on the zero proper motion criterion, we thus estimated a quasar selection efficiency of $\approx 60\%$ based on the astrometric measurements from {\it Gaia}-DR2 close to the NGP. Imposing an additional astrometric cut, requiring parallaxes consistent with zero (within a $2\sigma$ confidence level), increases the selection efficiency to $\approx 70\%$. We found that the fraction of quasars to the total number of sources with zero proper motions and parallaxes, however, is strongly correlated with the Galactic latitude of the field. At $b \lesssim 80$\,deg, the selection efficiency drops to $\sim 50\%$; furthermore, even closer to the Galactic plane ($b = 30-50$\,deg), only $\approx 20\%$ of the sources with zero proper motions and parallaxes are likely quasars. A purely astrometric selection of quasars based on the {\it Gaia}-DR2 data is thus mainly feasible at $|b| \gtrsim 80$\,deg.

For each of the identified quasars, we determined the redshift $z$ and visual extinction $A_V$. Based on this sample, we inferred the distributions of the underlying quasar properties. This is motivated by the simple sample criteria adopted here, which is unbiased in terms of colours or intrinsic energy outputs of the quasar. These therefore provide a complete census of the quasar population within the imposed magnitude limit of $G<20$\,mag. We found that the observed redshift distribution spans from $z = 0.2$ to $z=3.5$ and peaks at $z\sim 1.5$. The observed fraction of quasars that are significantly reddened by dust was found to be $13^{+5}_{-3}\%$ in this sample, but the intrinsic red fraction is at least a factor of $3-4$ higher.

Based on the complete {\it Gaia}-selected quasar sample, we then investigated the relative efficiency of other, more common photometric selection techniques. Particularly, we inferred the completeness of selecting quasars based on their UV excess, $K$-band excess, and typical red mid-infrared colours. We found that the UVX and KX methods would identify $\approx 90\%$ of the confirmed quasars, whereas the mid-infrared selection was complete at up to $\approx 85\%$. Inferring the basic properties of the underlying sample of quasars based on a photometrically-selected quasar sample would therefore only yield a biased census, typically missing either quasars at high-$z$ or with significant amounts of dust.

For now, the main advantage of a {\it Gaia}-based astrometric search for quasars seems to be the ability to reject stellar contamination from candidate quasar lists. As we have demonstrated here, the use of astrometric information can reduce the number of stellar sources by several orders of magnitude. While the efficiency of a purely astrometric selection is still not feasible for a large fraction of the sky, the current release of the {\it Gaia} survey is still very valuable for other objectives, such as identifying heavily obscured BAL quasars, quasars with dusty and metal-rich foreground galaxies, or rare lensed systems \citep{Pindor2003,Lemon2017,Lemon2019,Geier2019,Delchambre2019,Fynbo2020}.

The present paper is based on the {\it Gaia} DR2 data from 22 months of observations. The next data release, DR3, will be based on 34 months of data and is expected to be released in late 2020 with smaller errors on proper motions by a factor of two. This will translate into a reduced selection radius around zero proper motion by the same factor, resulting in approximately a factor of four decrease in the number of contaminating stars, assuming a homogeneous distribution of proper motions at these scales (see also fig. 1 of \citealt{Heintz2015}). This will reduce the number of contaminating stars from the 41 (29 including the zero-parallax cut) observed here to $\approx 10$ (7) within the same surveyed sky region. Since the number of quasars with zero proper motions (63 in this work) is expected to stay constant, this will result in a quasar selection efficiency, that is the fraction of quasars relative to the total number of stationary point sources, of $\approx 85-90\%$ (at high Galactic latitudes) for a purely astrometric selection based on {\it Gaia} DR3 data.
The ideal astrometric selection would be a more precise (at least by a factor of ten), preferably deeper all sky mission observing in the infrared \citep{Heintz2015}. Such a 'GaiaNIR' mission is currently under consideration \citep{McArthur2019,Hobbs2019}.

\begin{acknowledgements}
KEH and PJ acknowledge support by a Project Grant (162948--051) from The Icelandic Research Fund. JPUF thanks the Carlsberg Foundation for support. The Cosmic DAWN center is funded by the DNRF. JK acknowledges support from the French {\sl Agence Nationale de la Recherche} under grant no ANR-17-CE31-0011-01 (project ``HIH2'' -- PI Noterdaeme). BMJ is supported in part by Independent Research Fund Denmark grant DFF - 7014-00017.
This work is based on observations made with the Gran Telescopio Canarias (GTC), the William Herschel Telescope (WHT), and with the Nordic Optical Telescope (NOT), installed in the Spanish Observatorio del Roque de los Muchachos of the Instituto de Astrofísica de Canarias, on the island of La Palma. 
Funding for the Sloan Digital Sky Survey IV has been provided by the Alfred P. Sloan Foundation, the U.S. Department of Energy Office of Science, and the Participating Institutions. SDSS-IV acknowledges
support and resources from the Center for High-Performance Computing at
the University of Utah. The SDSS web site is www.sdss.org.
SDSS-IV is managed by the Astrophysical Research Consortium for the 
Participating Institutions of the SDSS Collaboration including the 
Brazilian Participation Group, the Carnegie Institution for Science, 
Carnegie Mellon University, the Chilean Participation Group, the French Participation Group, Harvard-Smithsonian Center for Astrophysics, 
Instituto de Astrof\'isica de Canarias, The Johns Hopkins University, Kavli Institute for the Physics and Mathematics of the Universe (IPMU) / 
University of Tokyo, the Korean Participation Group, Lawrence Berkeley National Laboratory, 
Leibniz Institut f\"ur Astrophysik Potsdam (AIP),  
Max-Planck-Institut f\"ur Astronomie (MPIA Heidelberg), 
Max-Planck-Institut f\"ur Astrophysik (MPA Garching), 
Max-Planck-Institut f\"ur Extraterrestrische Physik (MPE), 
National Astronomical Observatories of China, New Mexico State University, New York University, University of Notre Dame, 
Observat\'ario Nacional / MCTI, The Ohio State University, 
Pennsylvania State University, Shanghai Astronomical Observatory, 
United Kingdom Participation Group,
Universidad Nacional Aut\'onoma de M\'exico, University of Arizona, 
University of Colorado Boulder, University of Oxford, University of Portsmouth, 
University of Utah, University of Virginia, University of Washington, University of Wisconsin, 
Vanderbilt University, and Yale University.
\end{acknowledgements}

\bibliographystyle{aa}
\bibliography{ref}

\begin{appendix}

\section{Spectroscopic data}

In this section we provide figures for the full set of spectroscopically observed quasars and two examples of the stellar spectra. For each quasar, we overplot the best-fit dust-reddened quasar template. 

\begin{figure*} [!b]
\centering
\epsfig{file=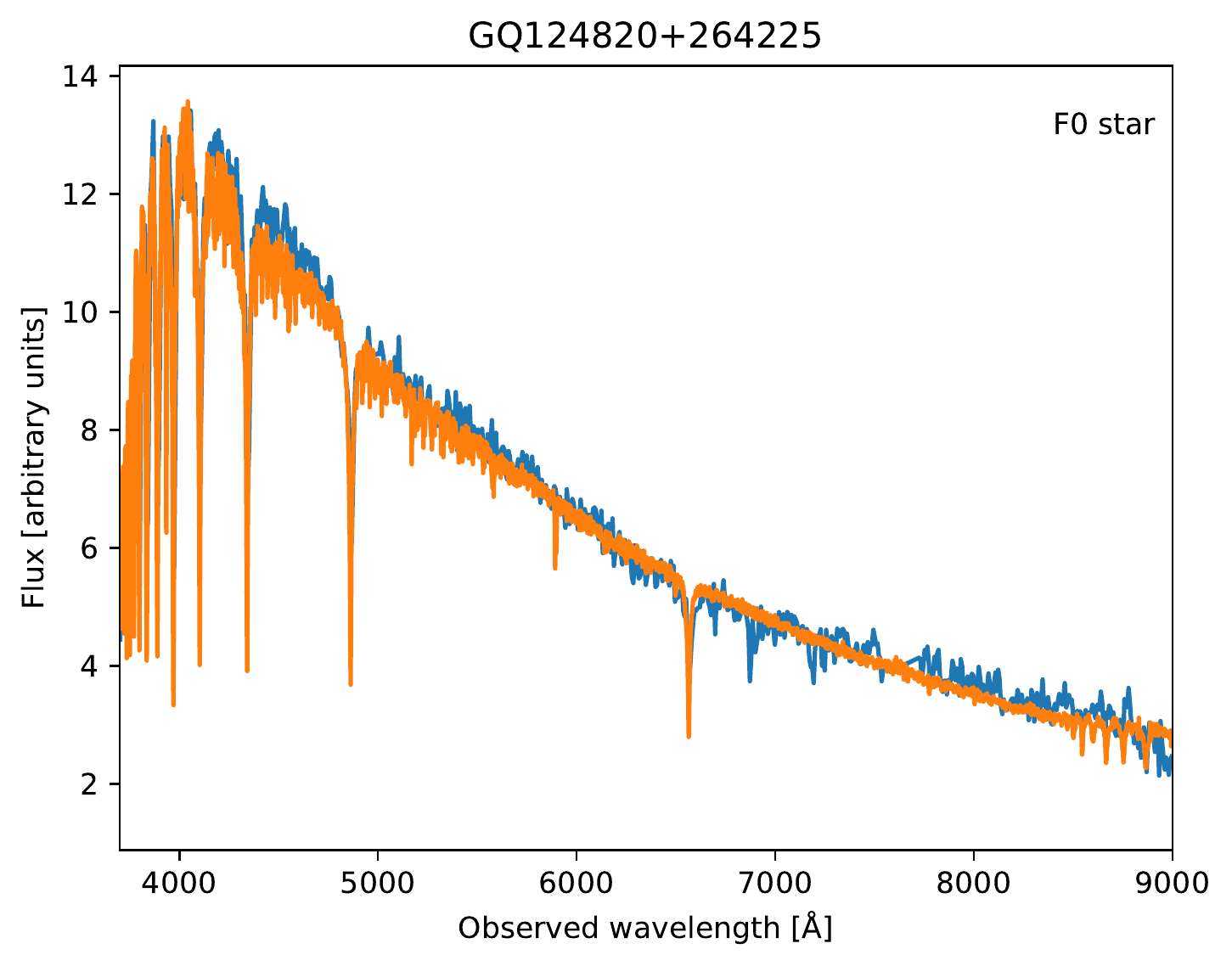,width=7.5cm}
\epsfig{file=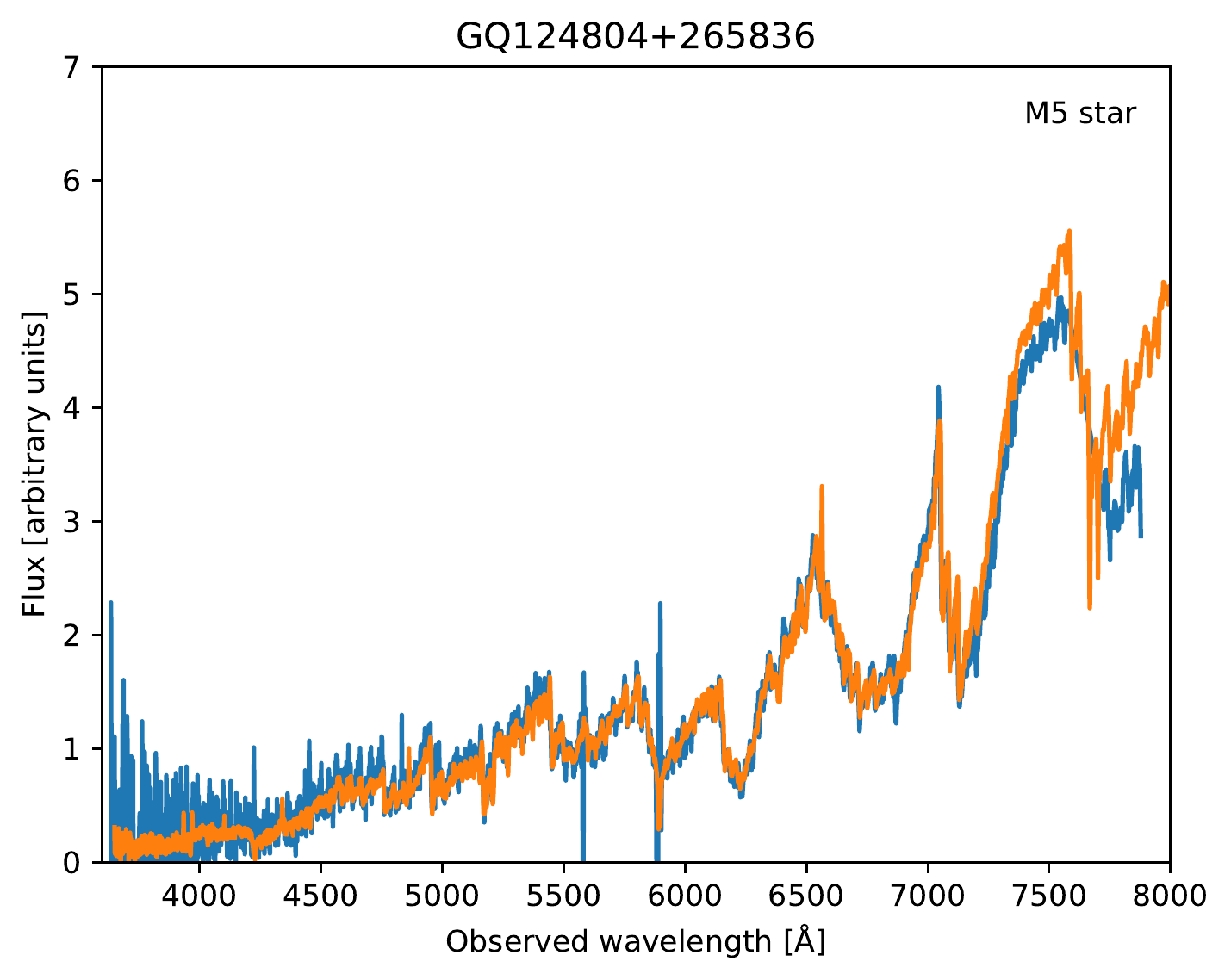,width=7.5cm}
\caption{Two examples of stellar spectra from our survey. Here, we have chosen the hottest and the coolest star in the sample (F0 and M5, respectively). The plots show the observed spectra in blue and the template spectra from PyHammer in orange.}
\label{fig:stellarspectra}
\end{figure*}

\begin{figure*} [!b]
\centering
\epsfig{file=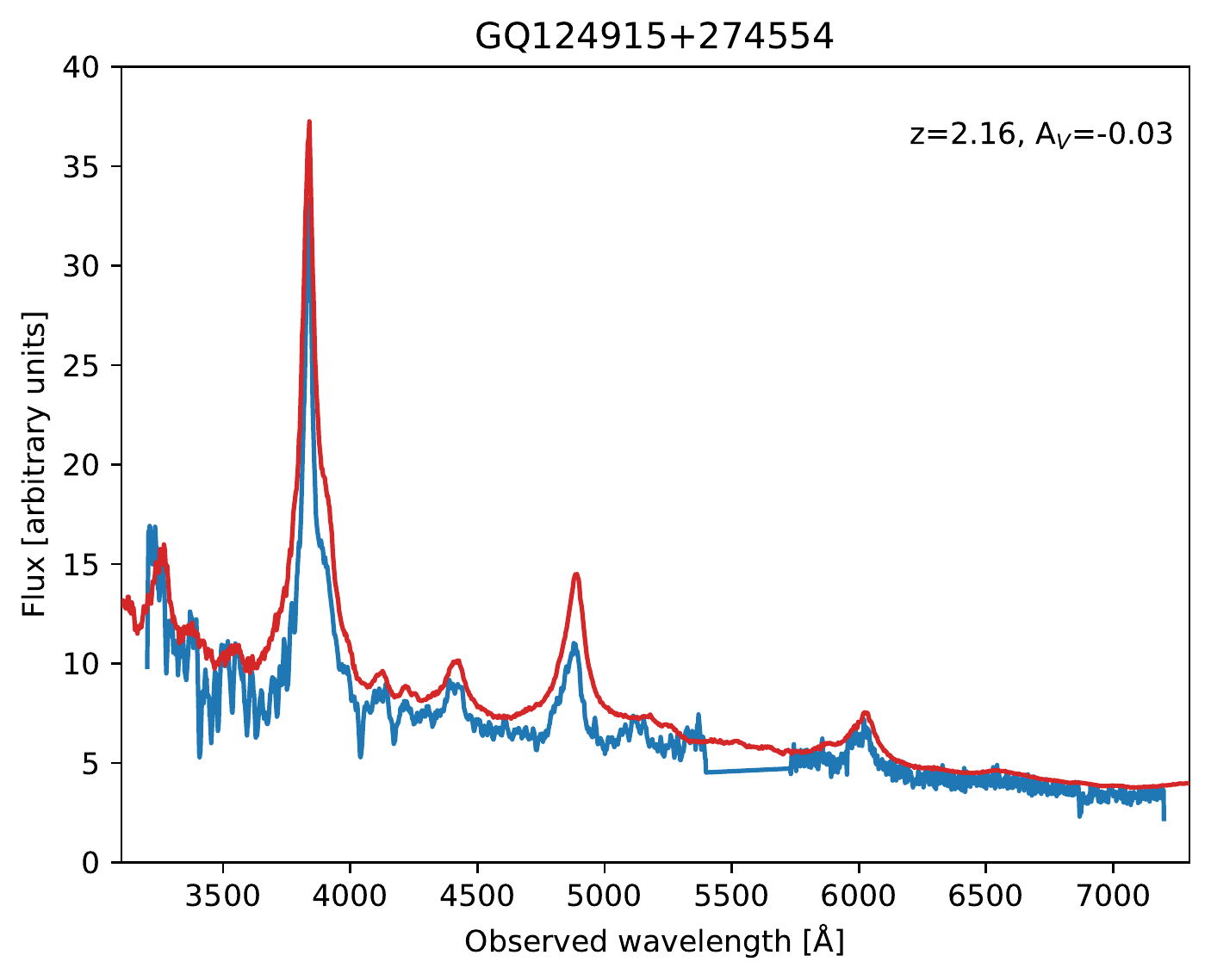,width=7.5cm}
\epsfig{file=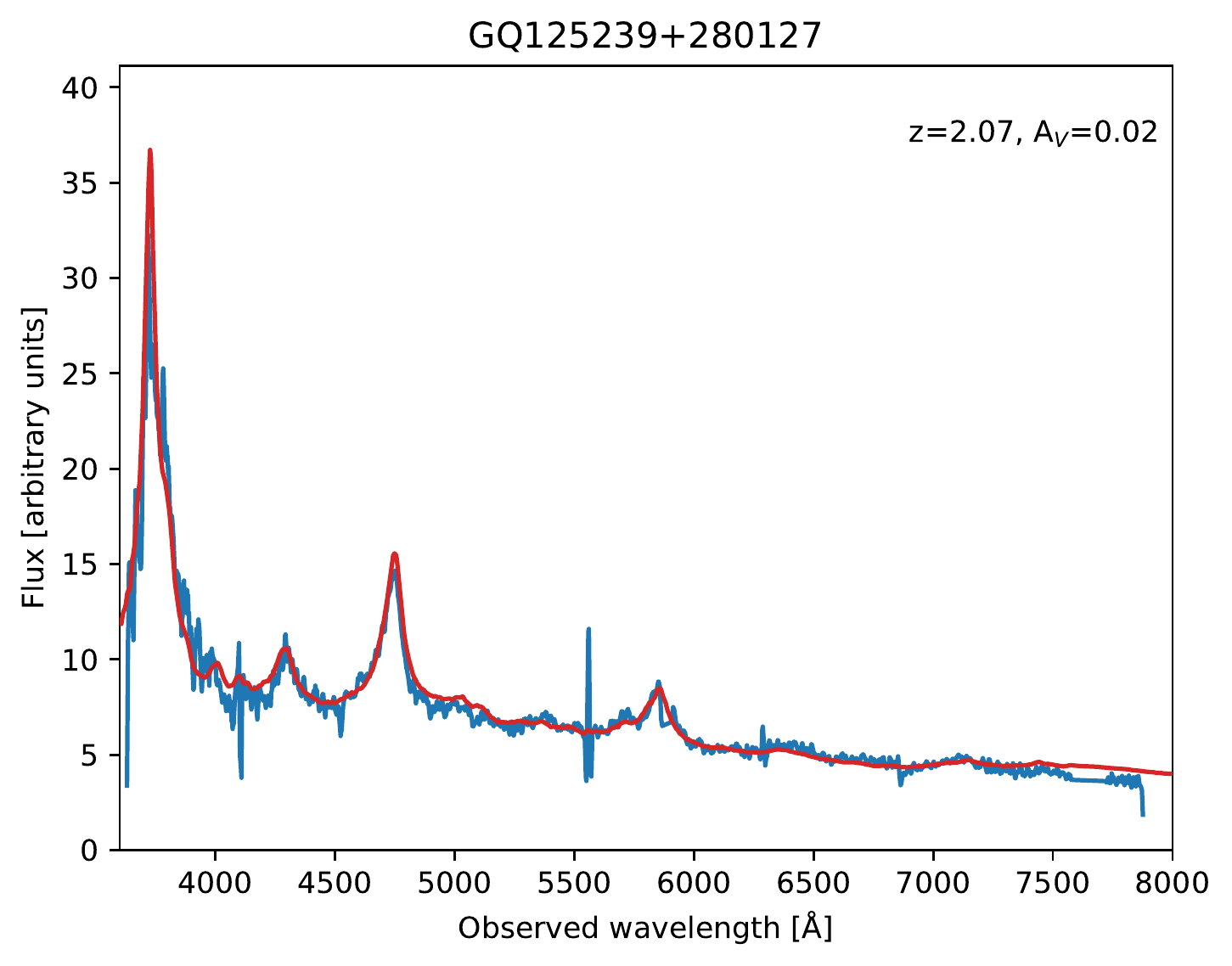,width=7.5cm}
\epsfig{file=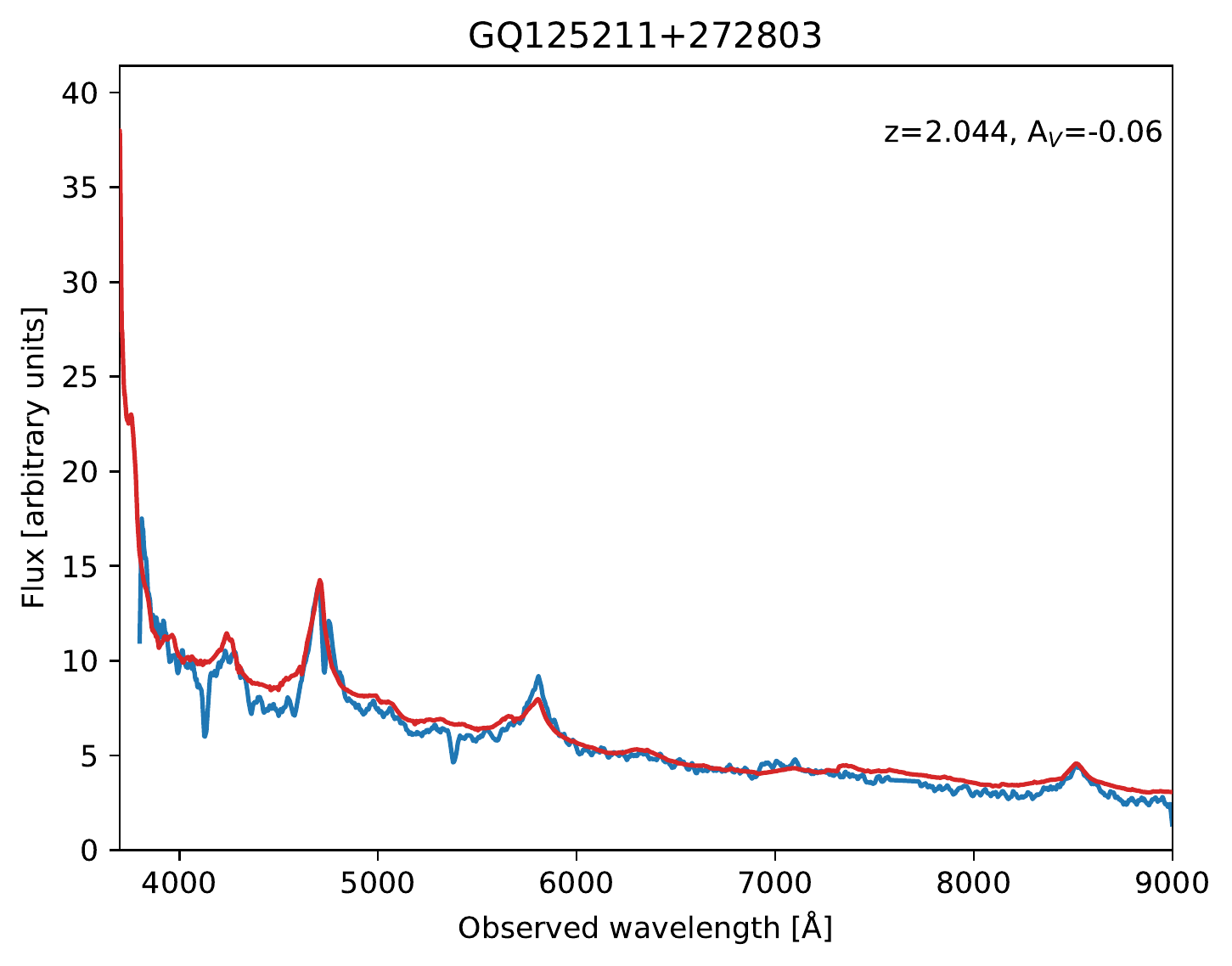,width=7.5cm}
\epsfig{file=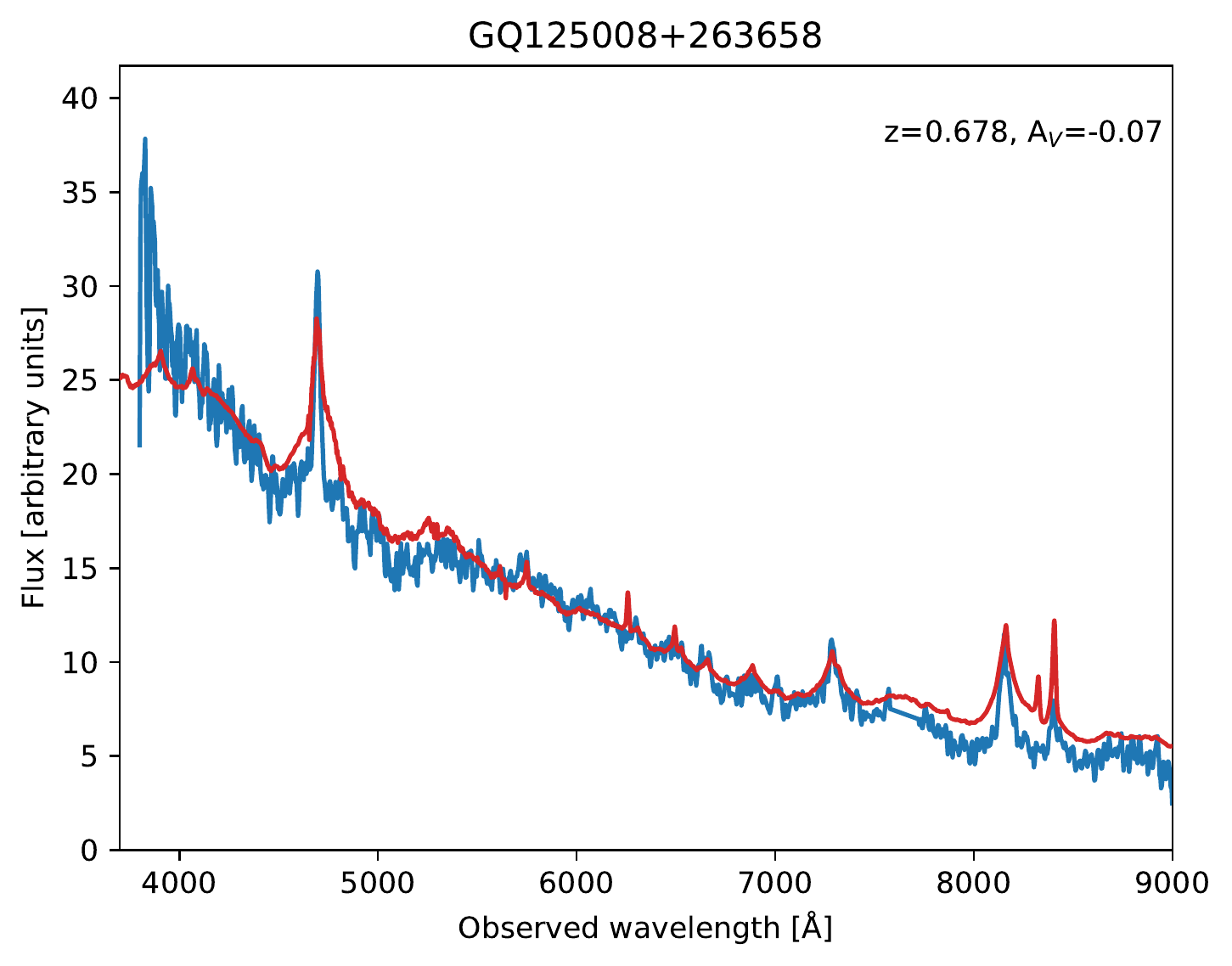,width=7.5cm}
\epsfig{file=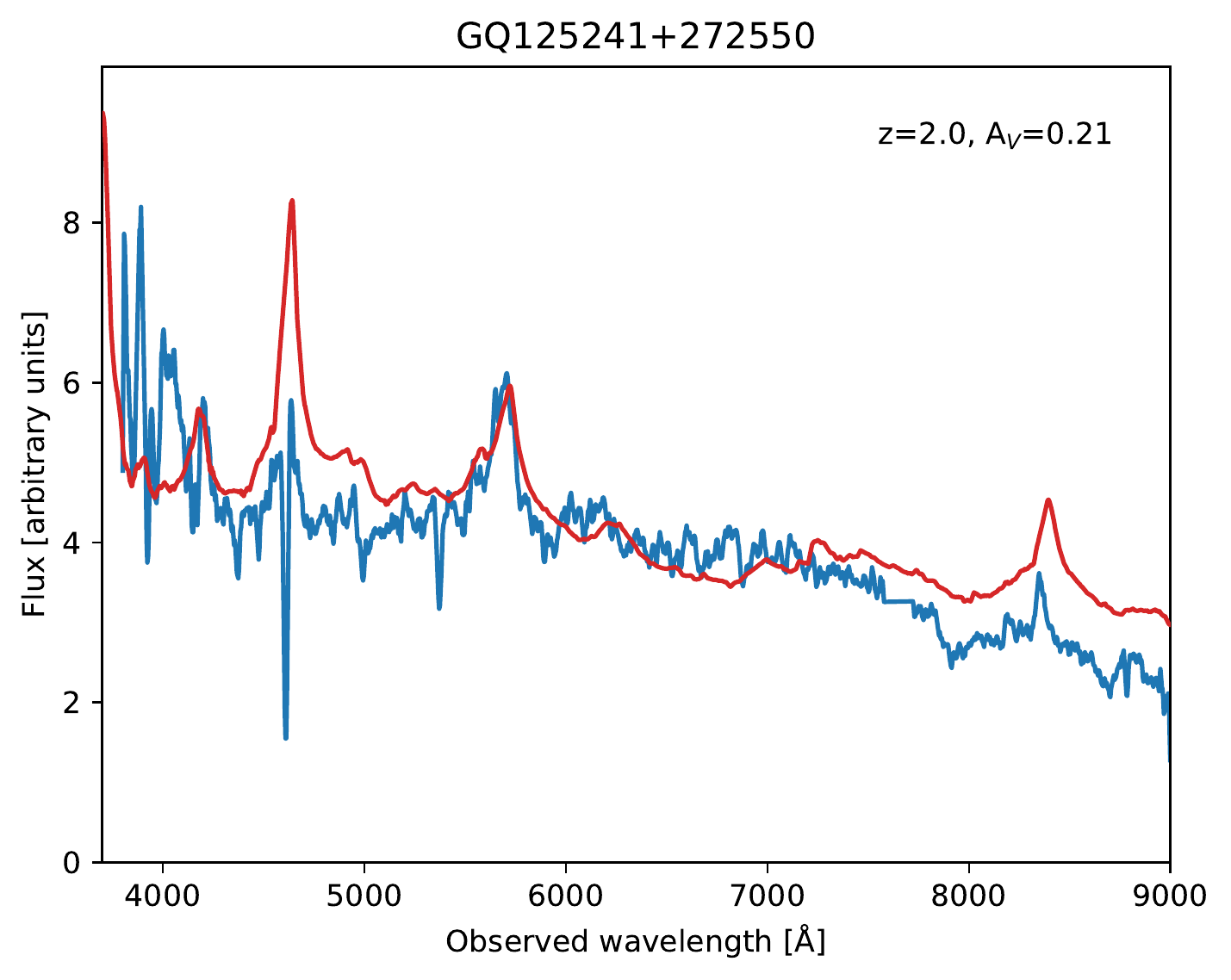,width=7.5cm}
\epsfig{file=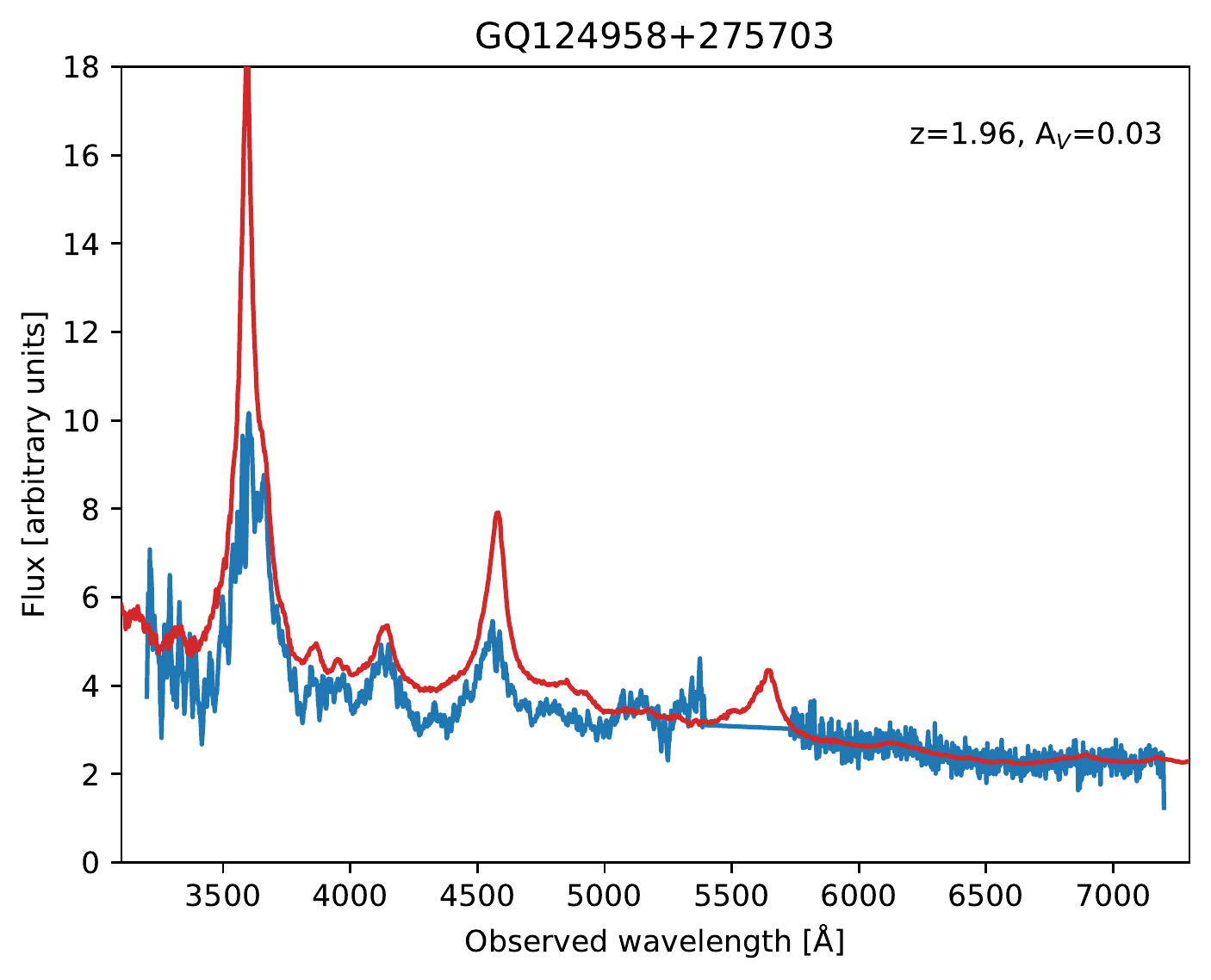,width=7.5cm}

\caption{Spectra of all the newly identified quasars in this study. The observed spectra are shown in blue and the redshifted composite model from \citet{Selsing2016} is shown in red. We have given the derived redshift and extinction in the upper right corner of each sub-figure. For the spectra secured with ISIS,
we have interpolated across the gap between the two arms of the spectrograph. We have also interpolated
across the atmospheric absorption band at 7600 \AA.}
\label{fig:spectra}
\end{figure*}

\begin{figure*} [!b]
\centering
\epsfig{file=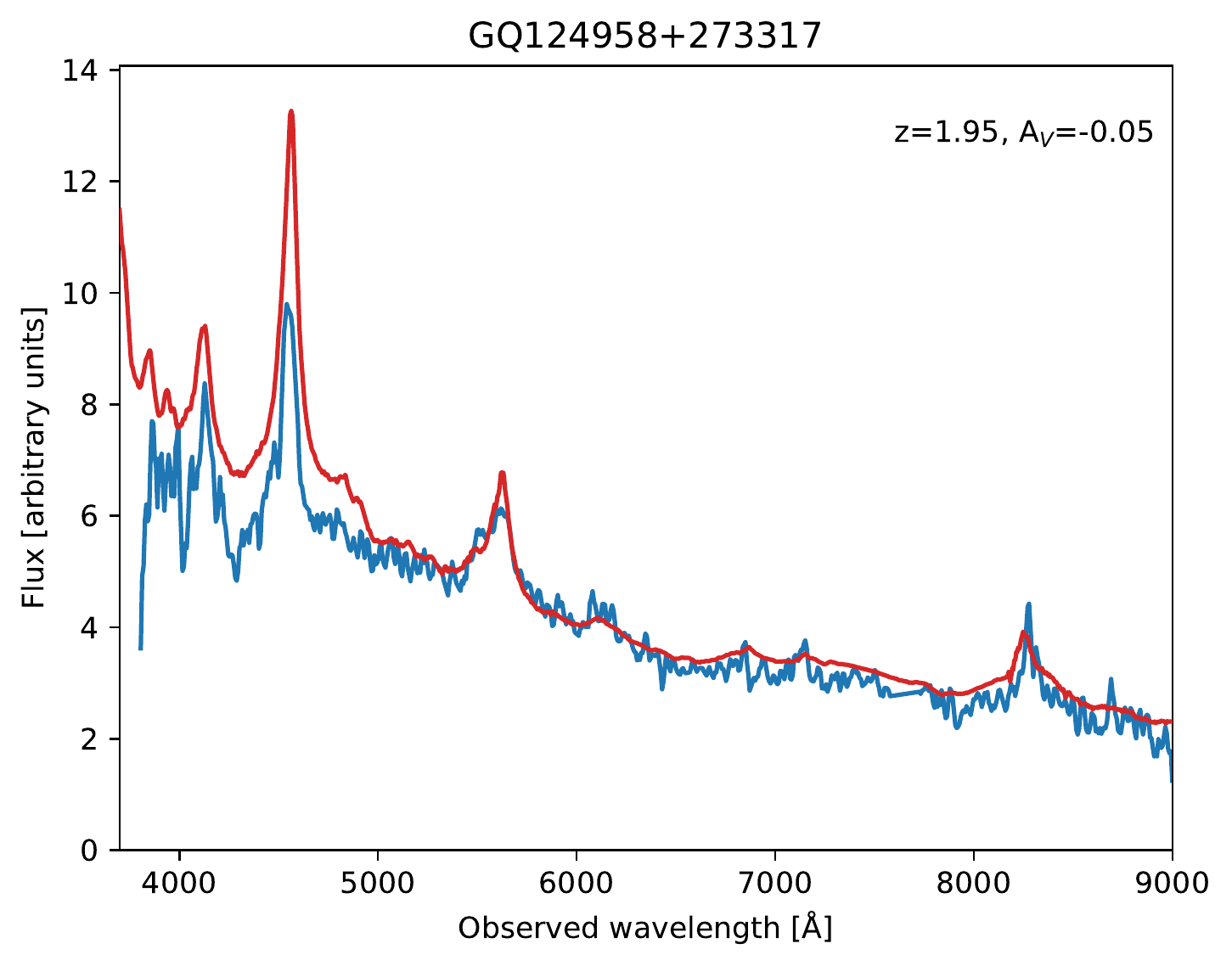,width=7.5cm}
\epsfig{file=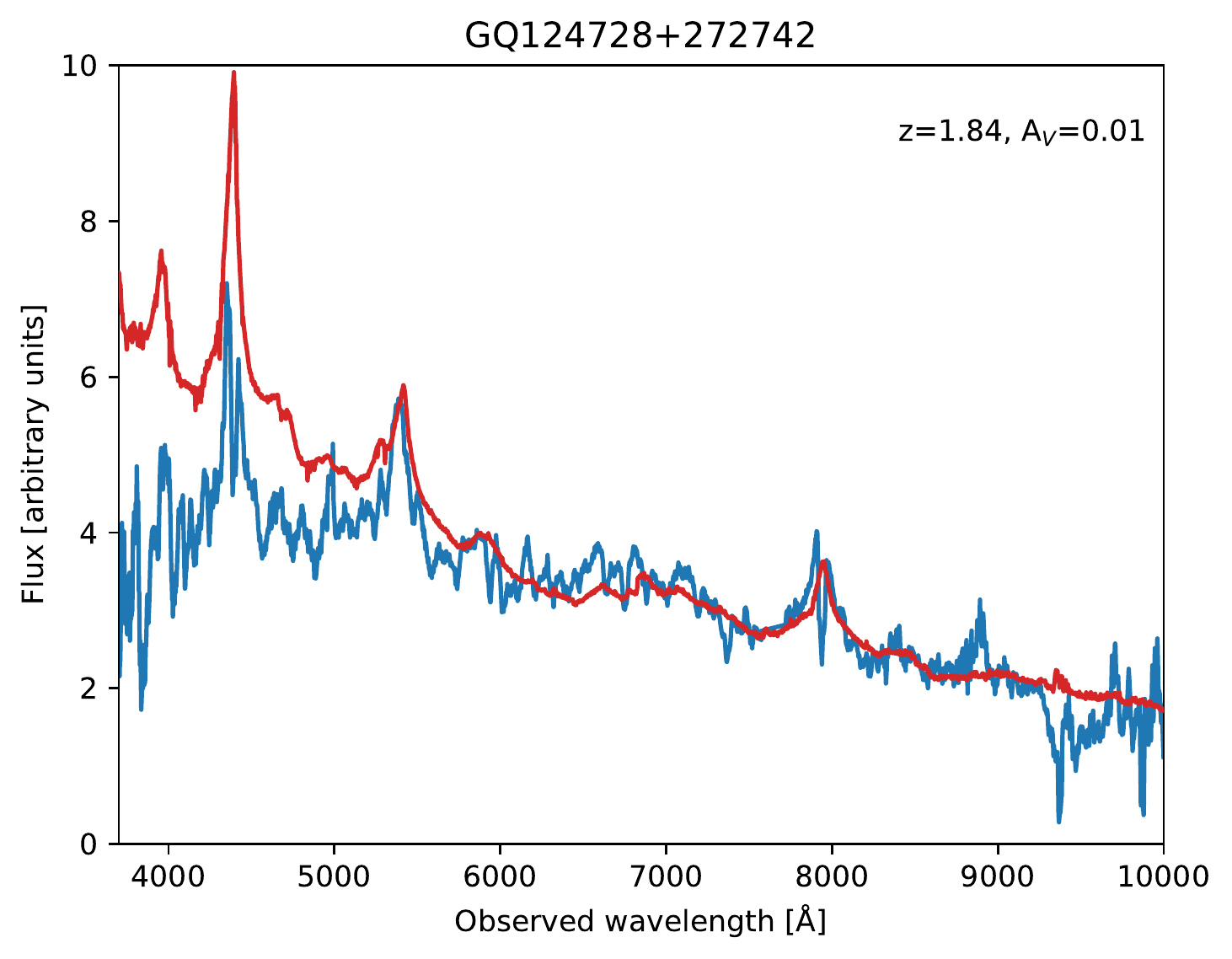,width=7.5cm}
\epsfig{file=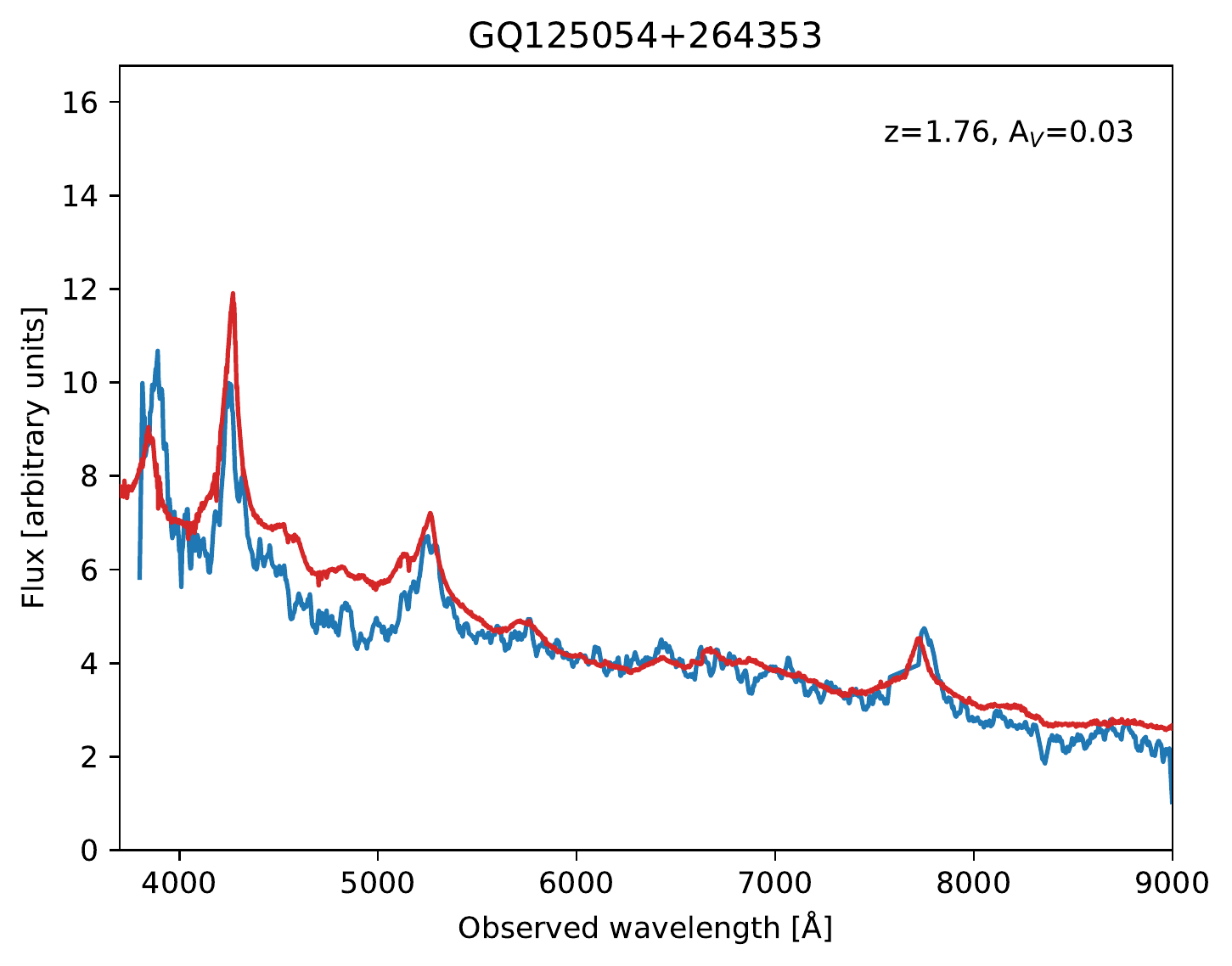,width=7.5cm}
\epsfig{file=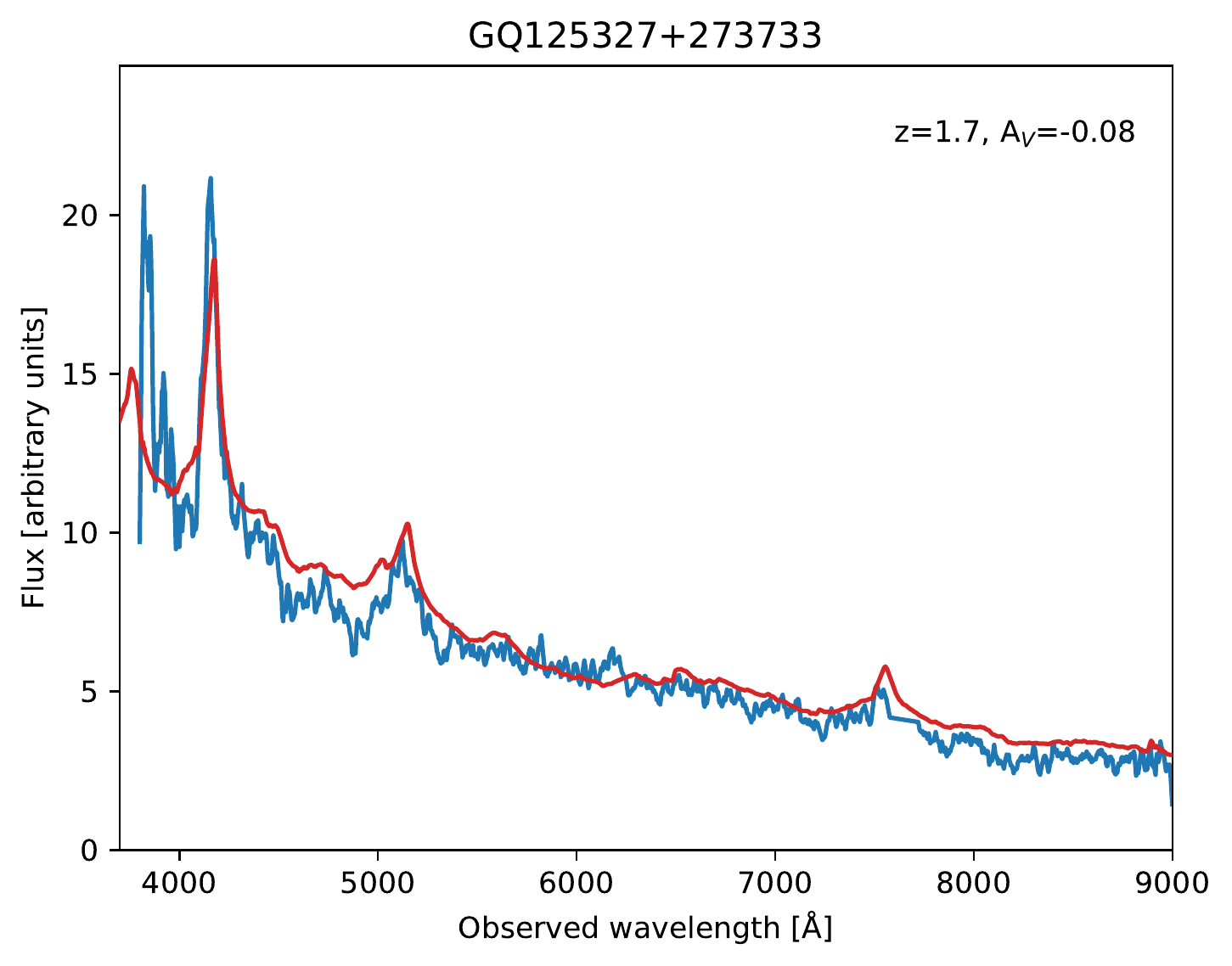,width=7.5cm}
\epsfig{file=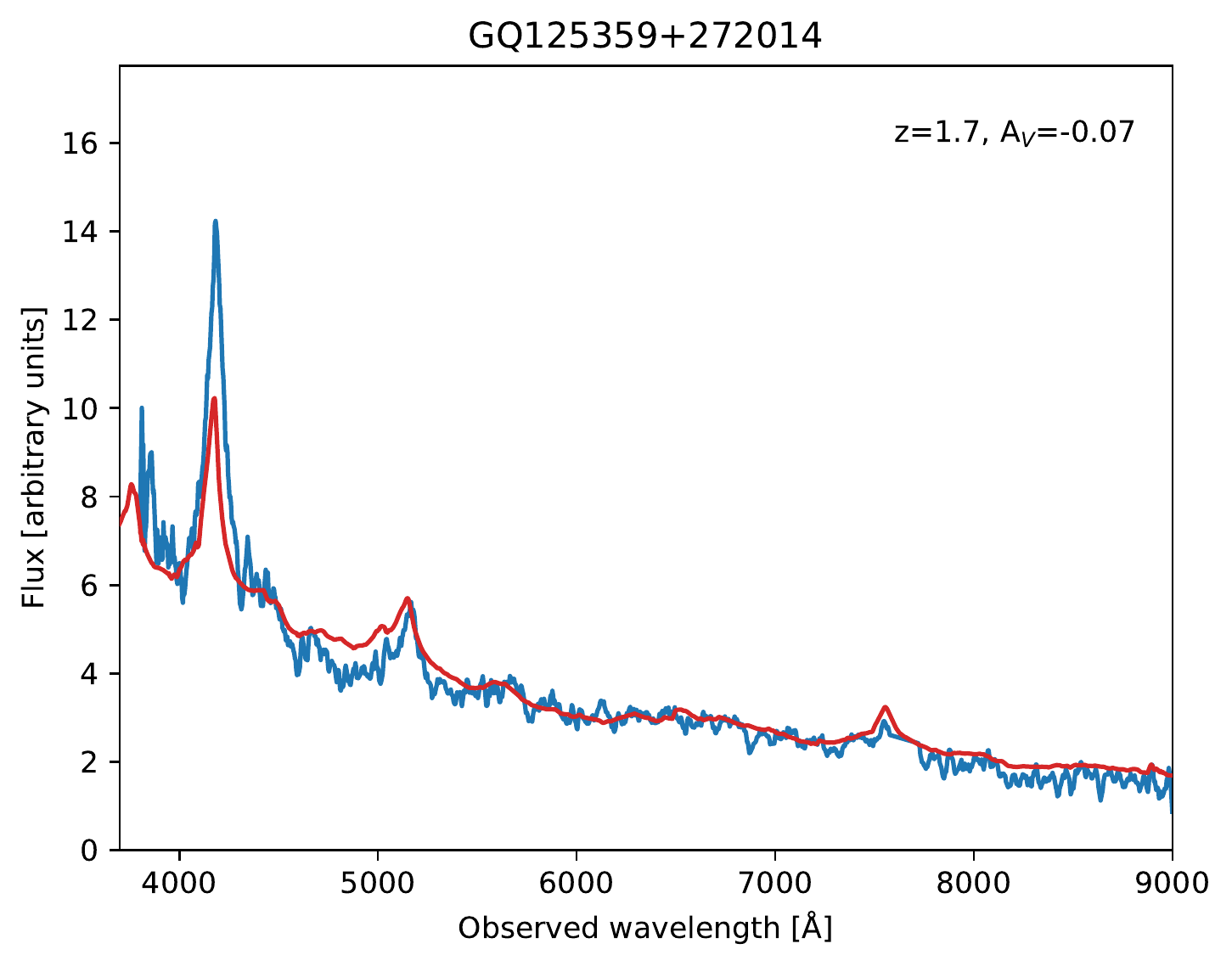,width=7.5cm}
\epsfig{file=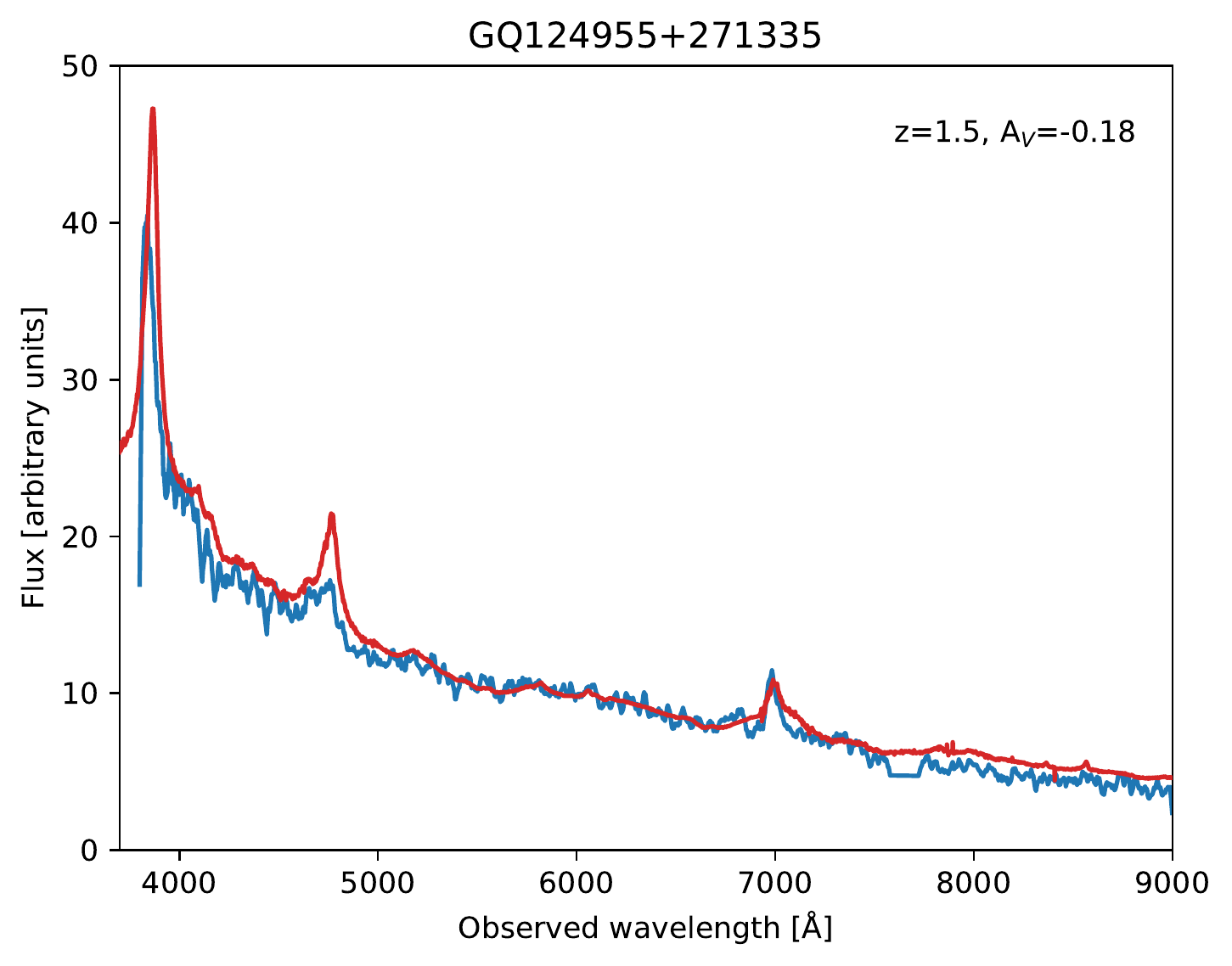,width=7.5cm}
\epsfig{file=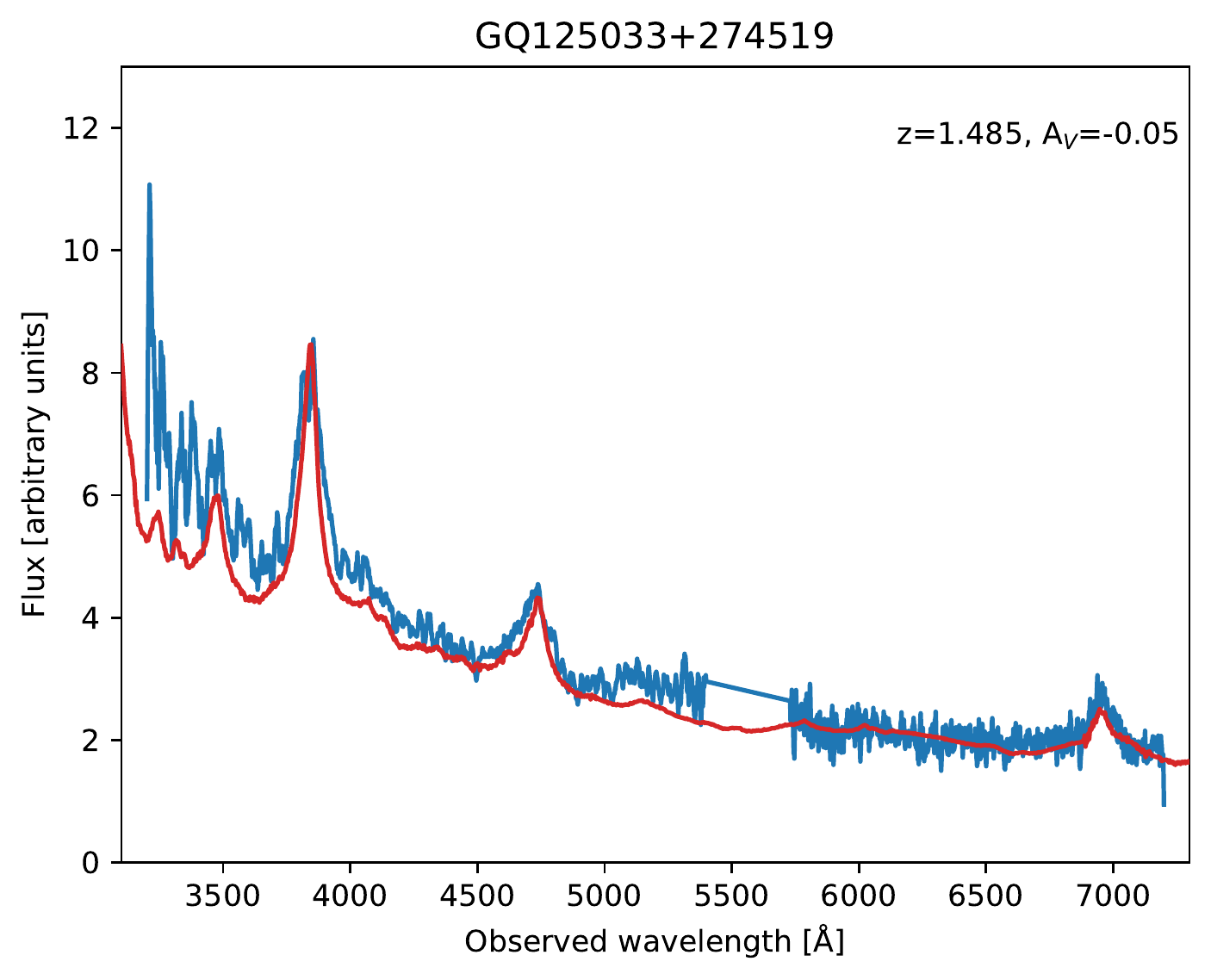,width=7.5cm}
\epsfig{file=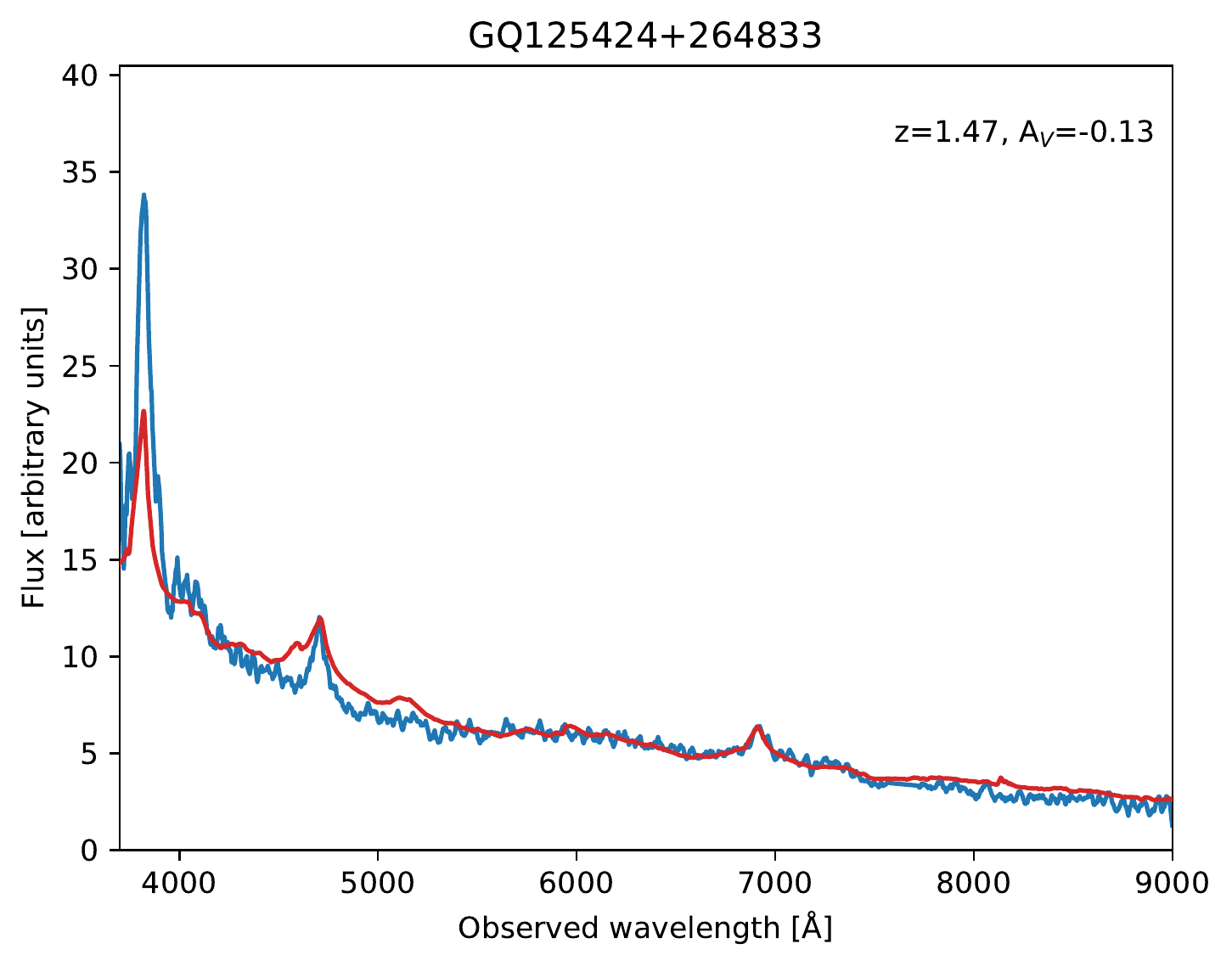,width=7.5cm}

\end{figure*}

\begin{figure*} [!b]
\centering
\epsfig{file=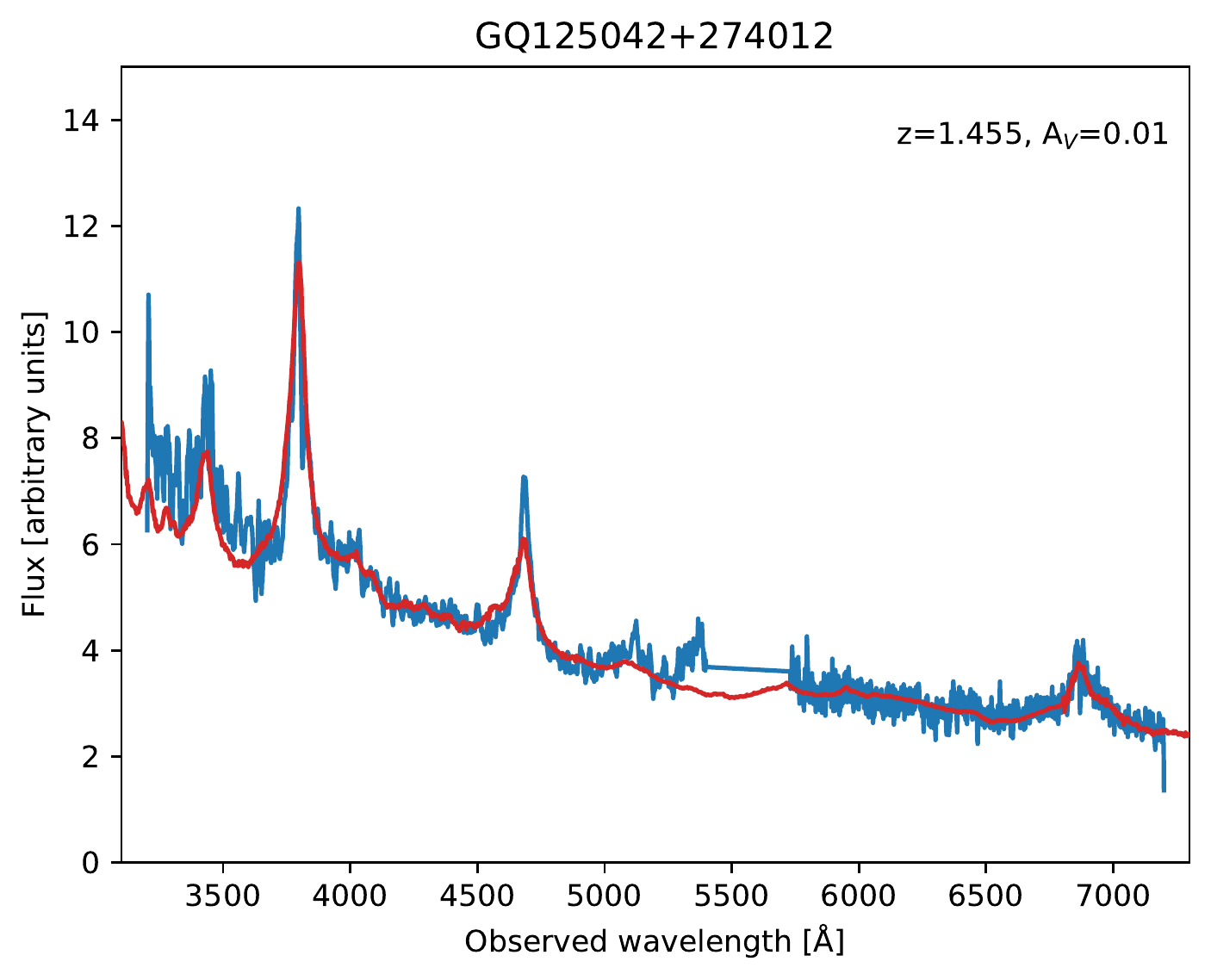,width=7.5cm}
\epsfig{file=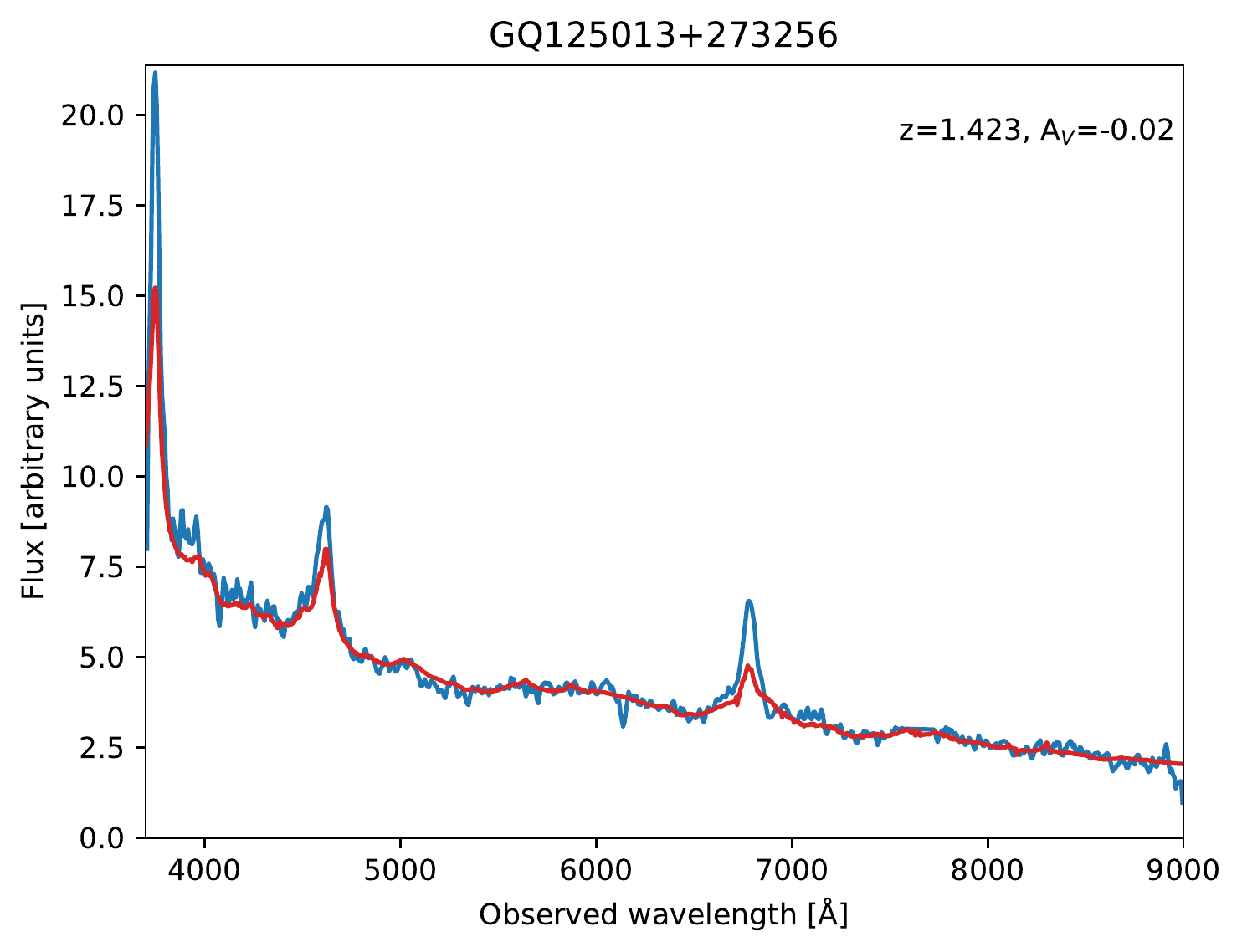,width=7.5cm}
\epsfig{file=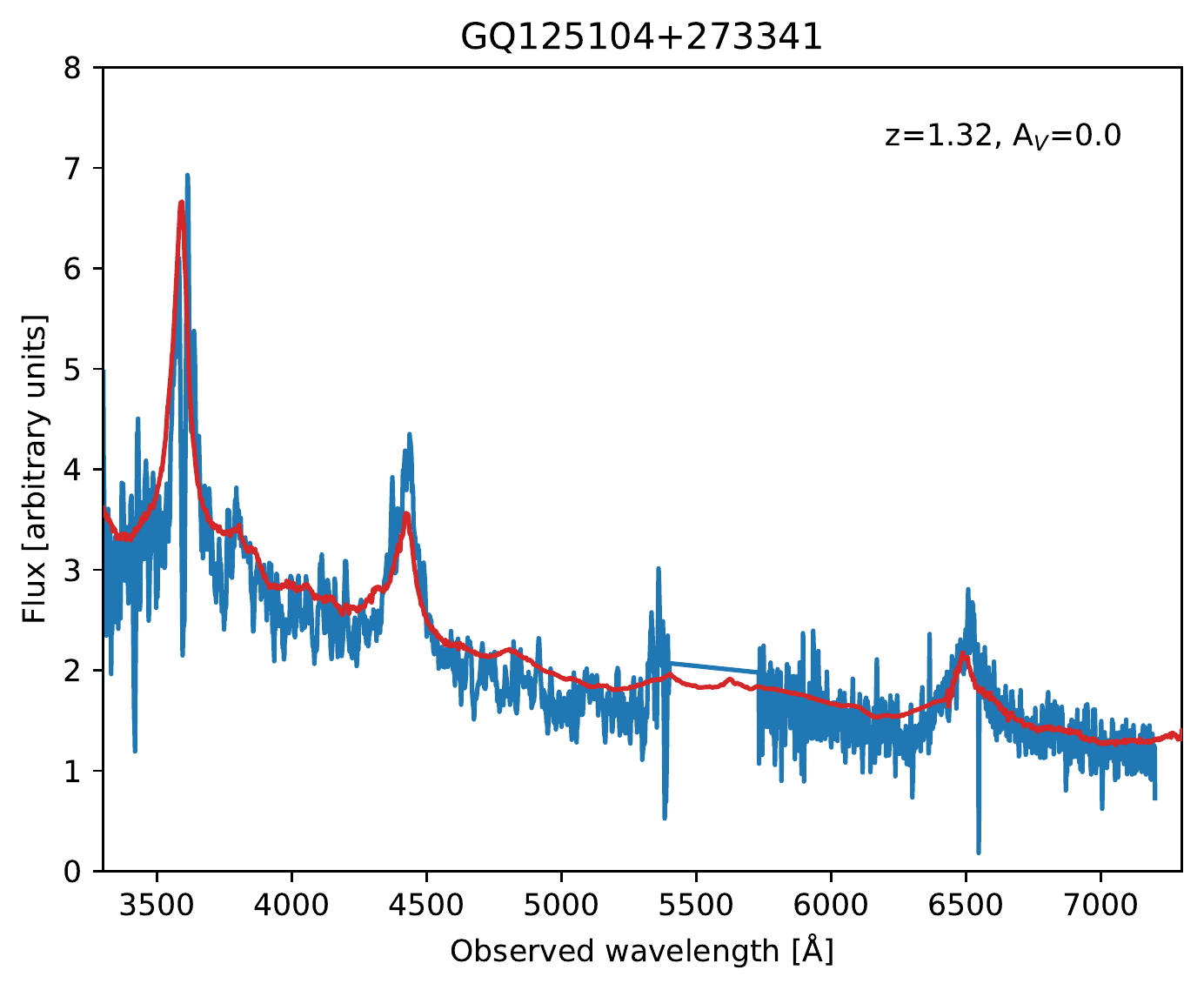,width=7.5cm}
\epsfig{file=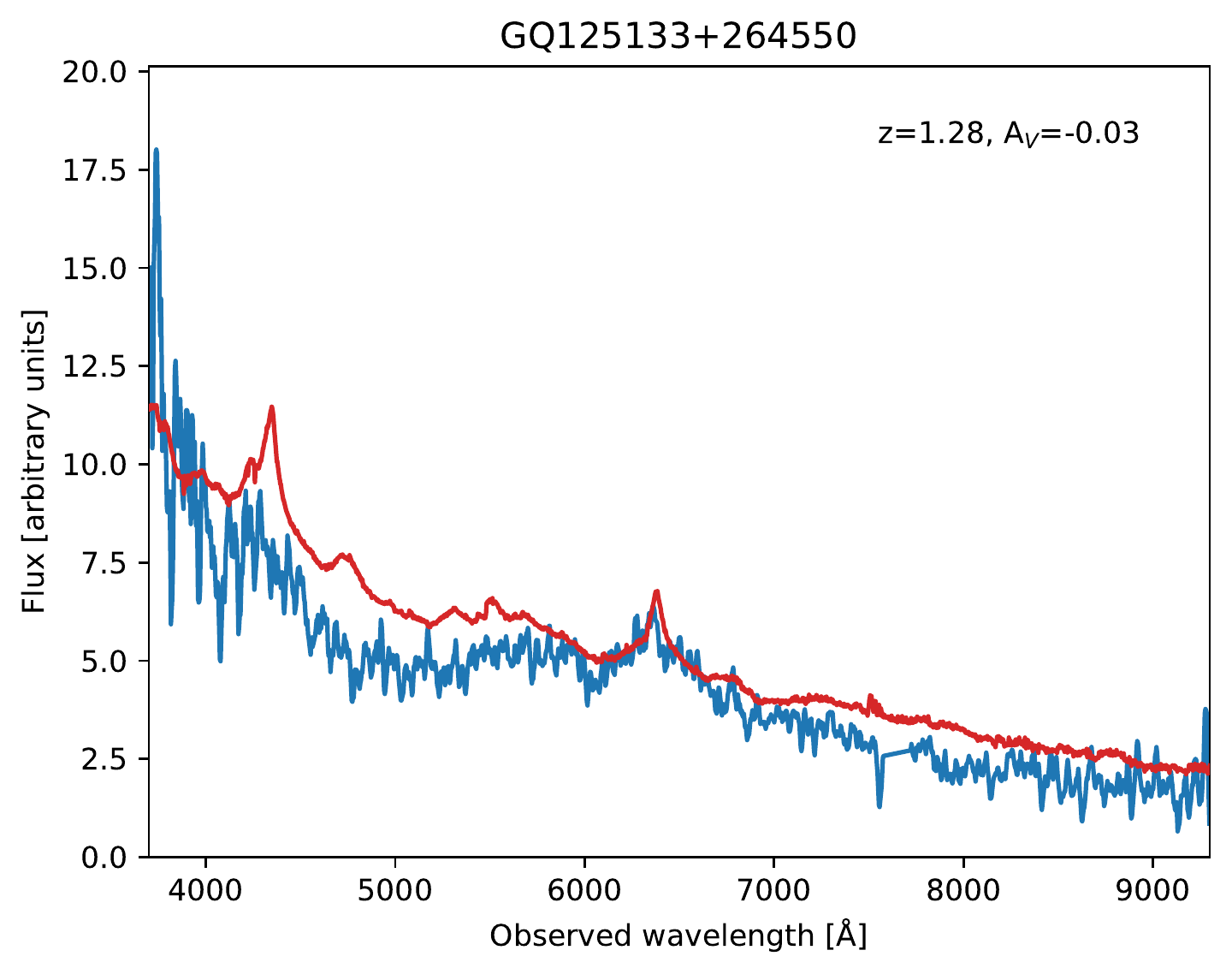,width=7.5cm}
\epsfig{file=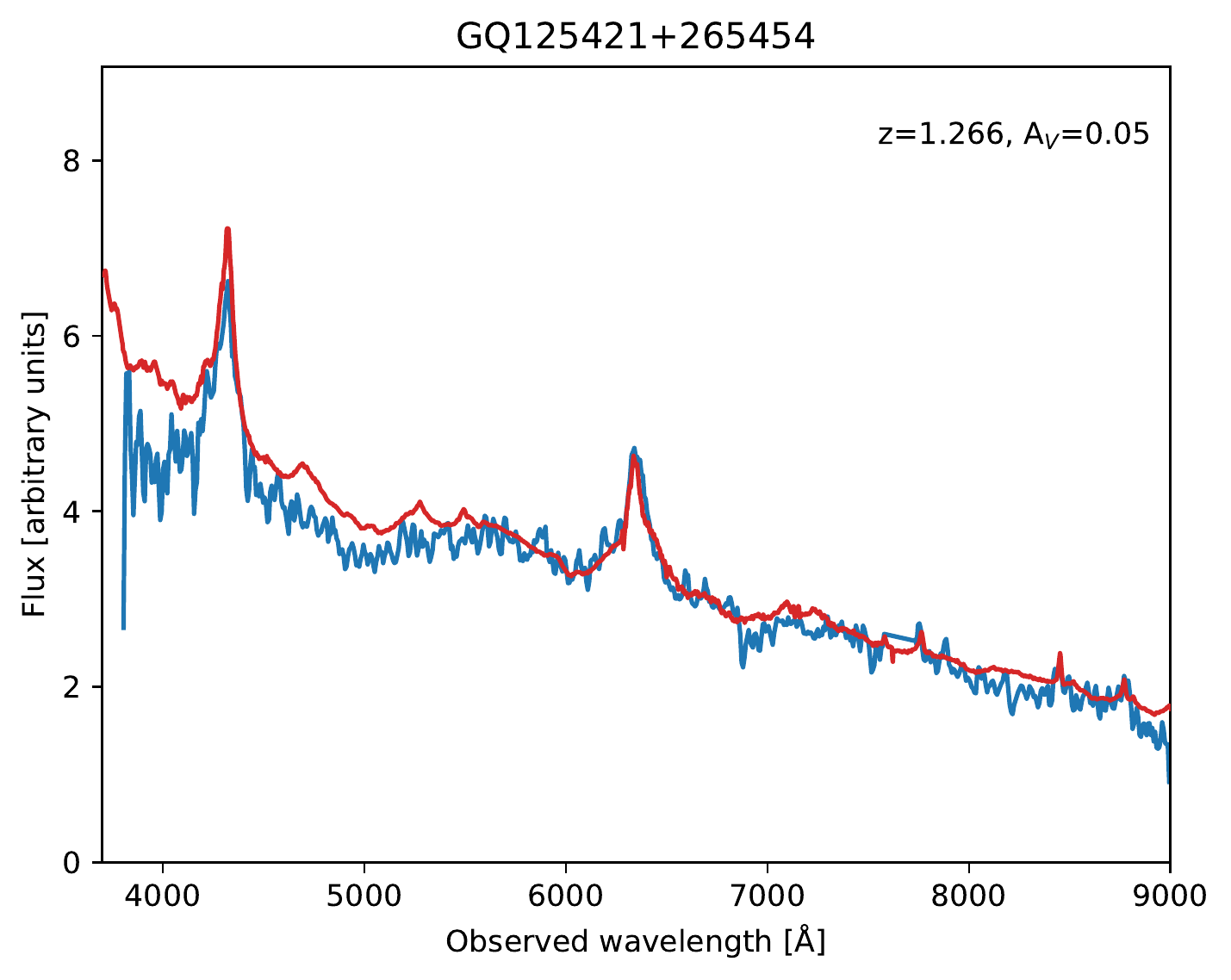,width=7.5cm}
\epsfig{file=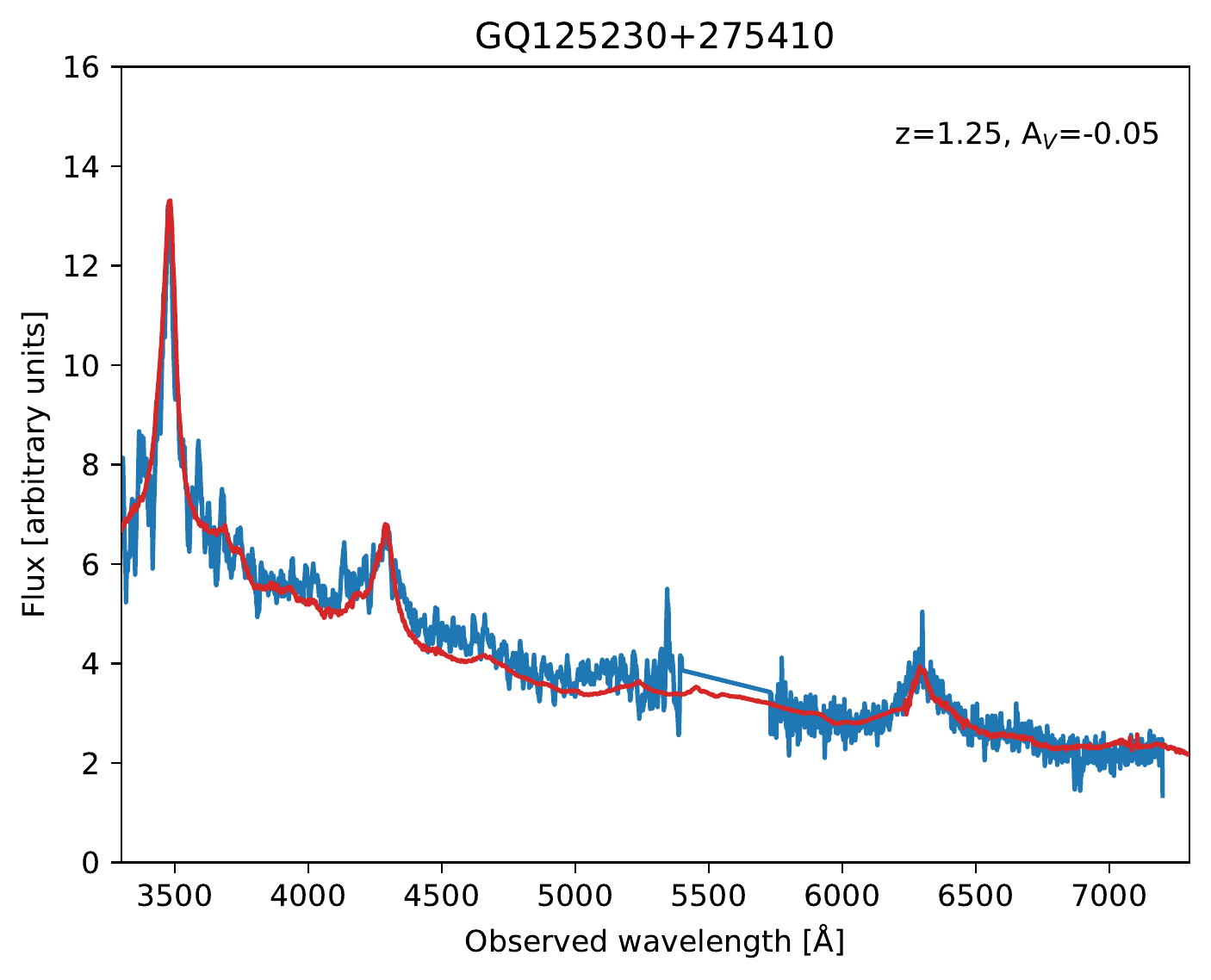,width=7.5cm}
\epsfig{file=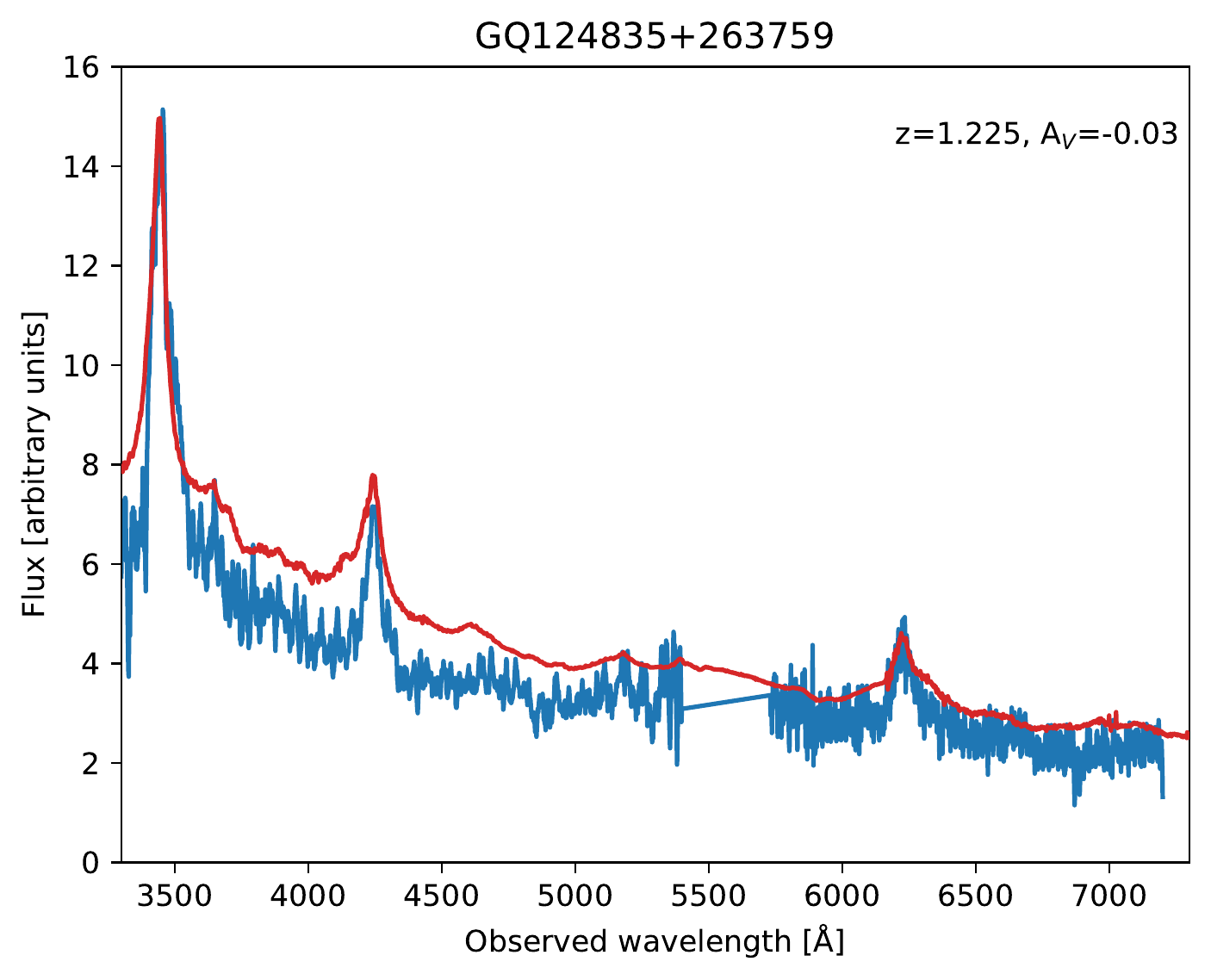,width=7.5cm}
\epsfig{file=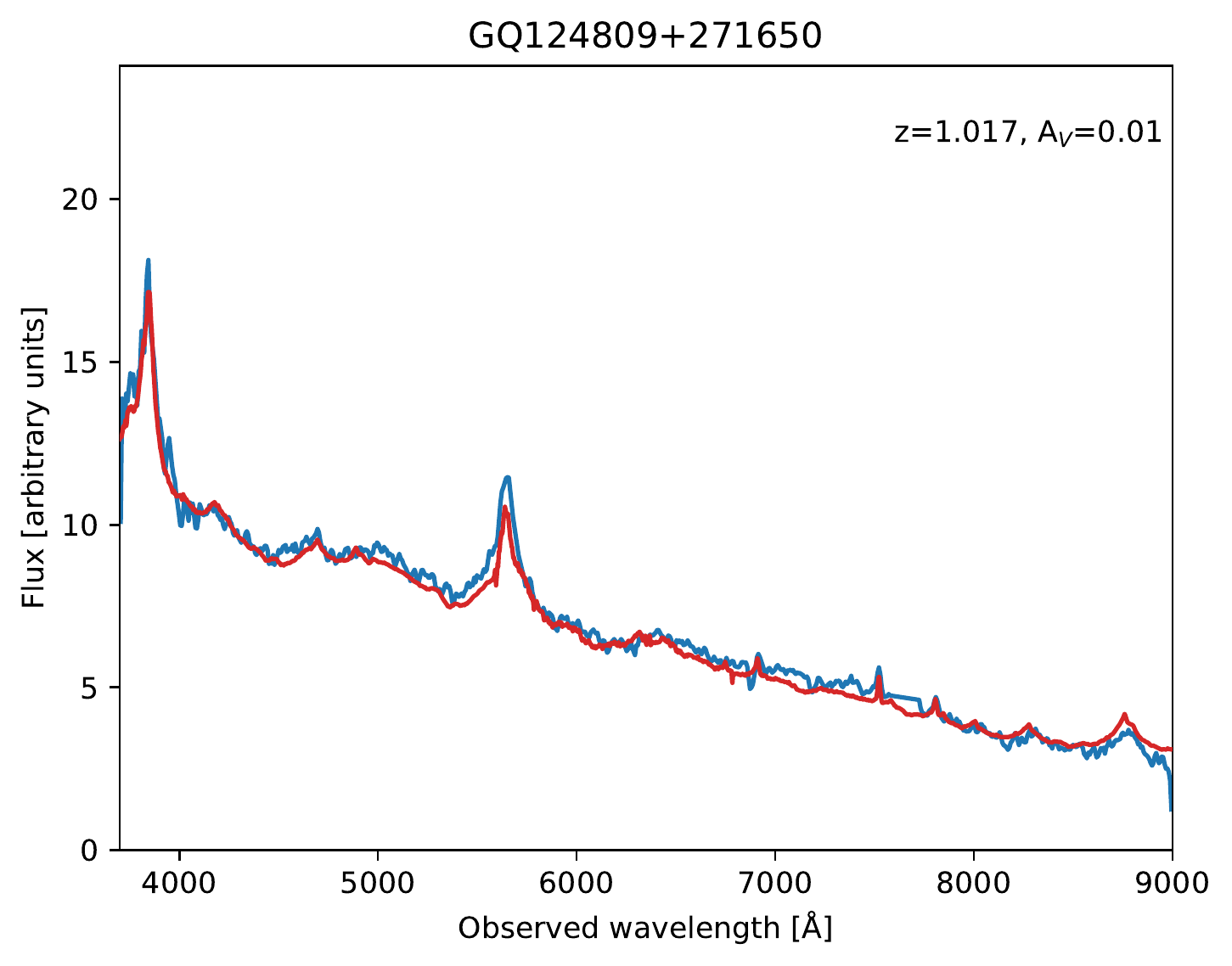,width=7.5cm}

\end{figure*}

\begin{figure*} [!b]
\centering
\epsfig{file=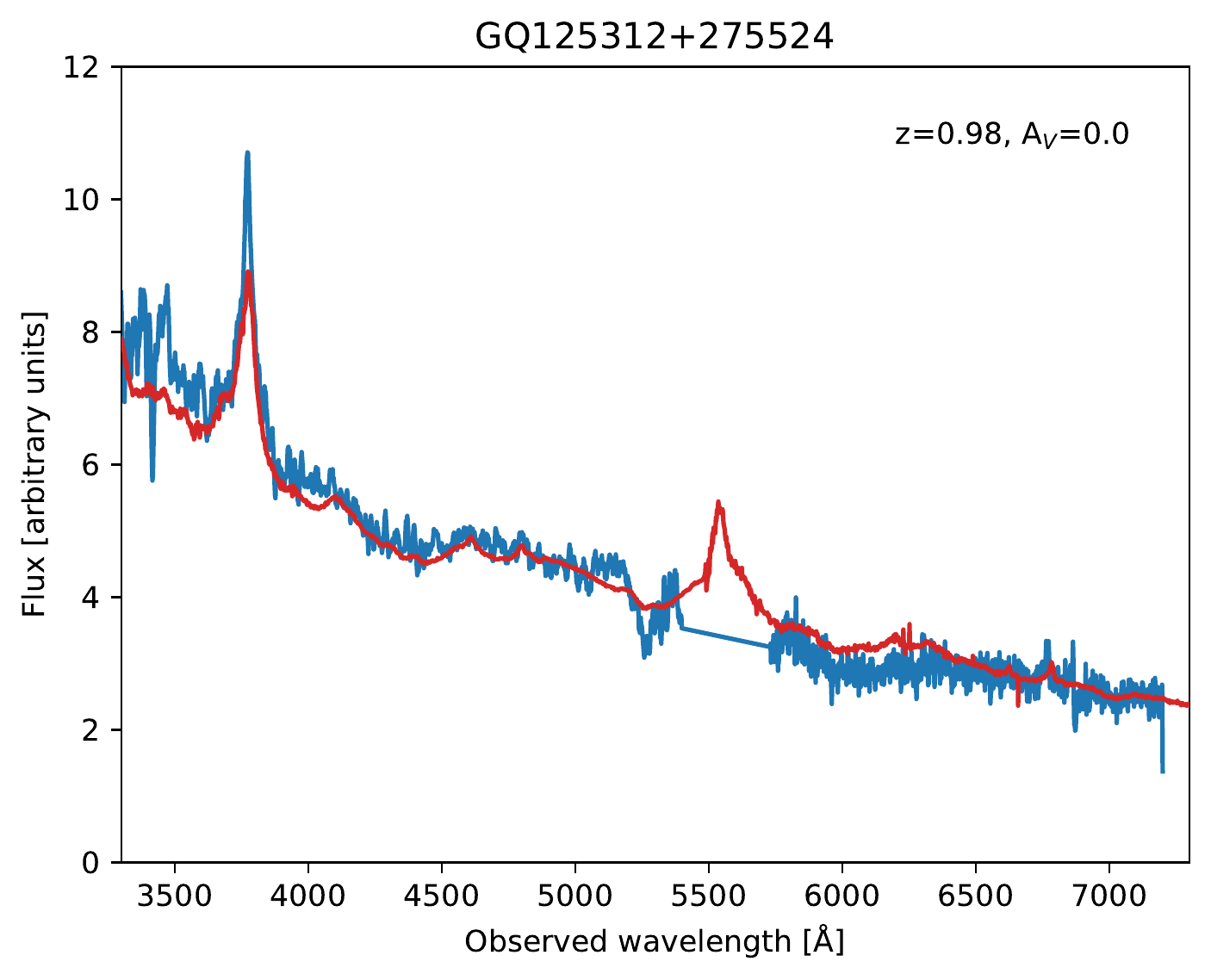,width=7.5cm}
\epsfig{file=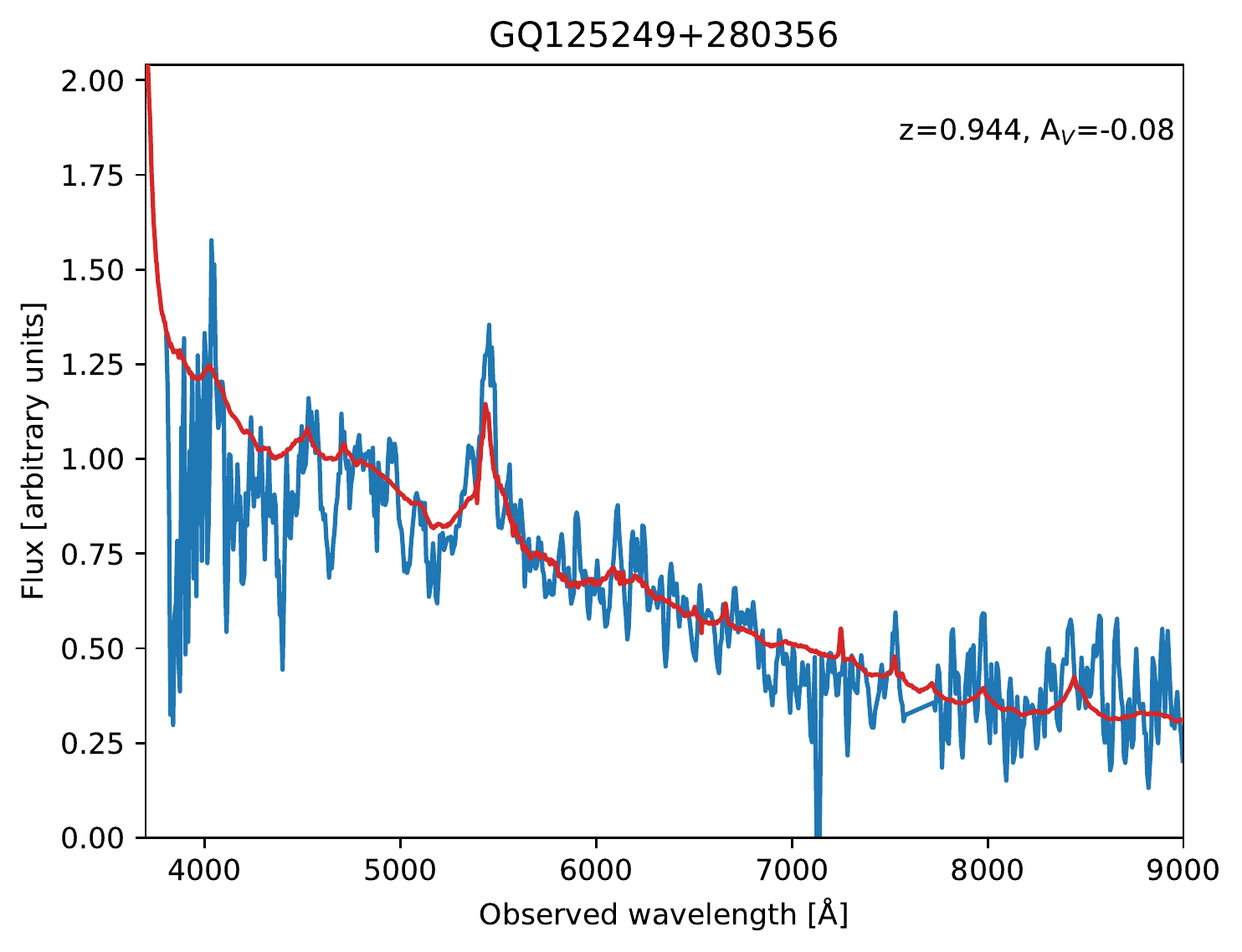,width=7.5cm}
\epsfig{file=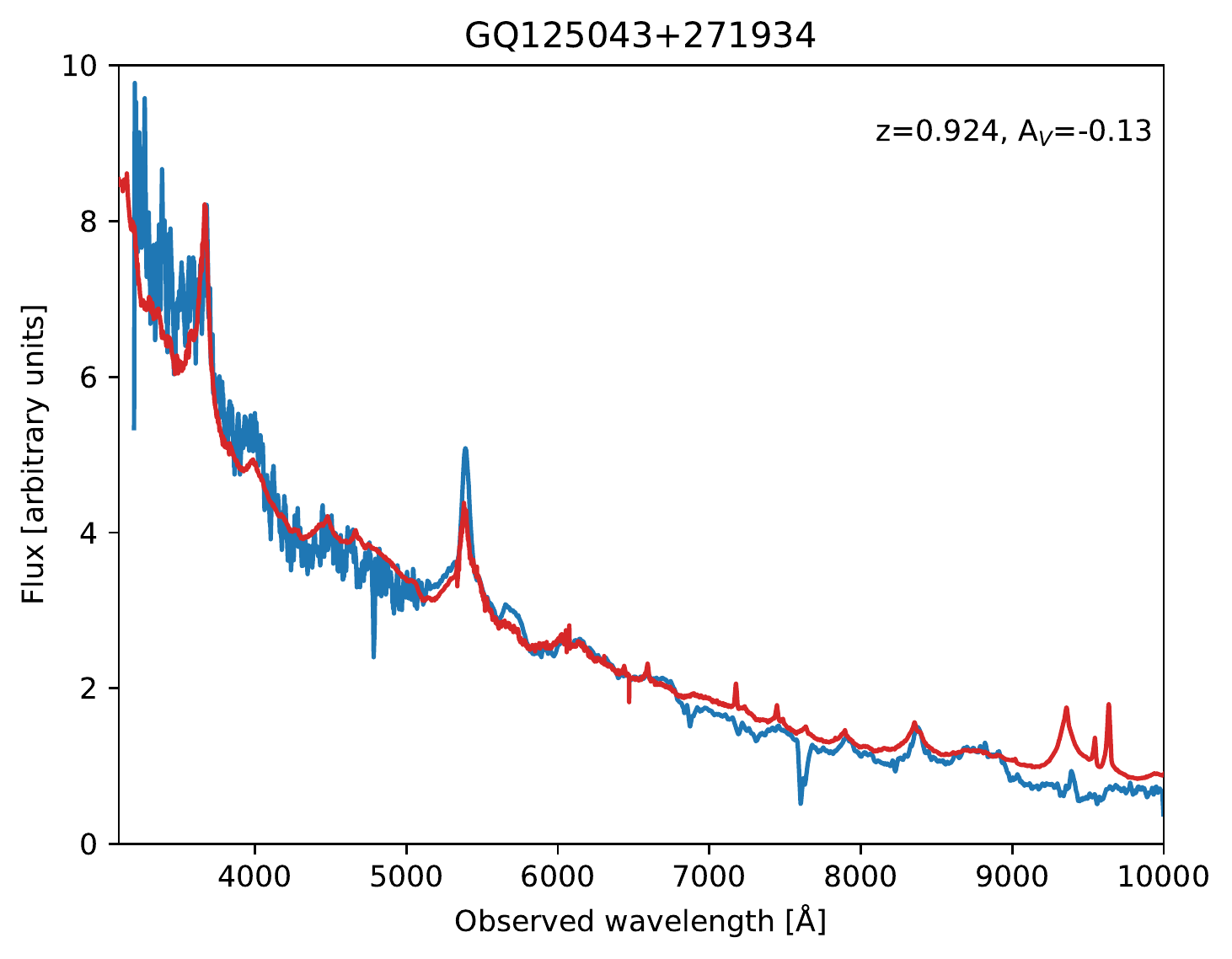,width=7.5cm}
\epsfig{file=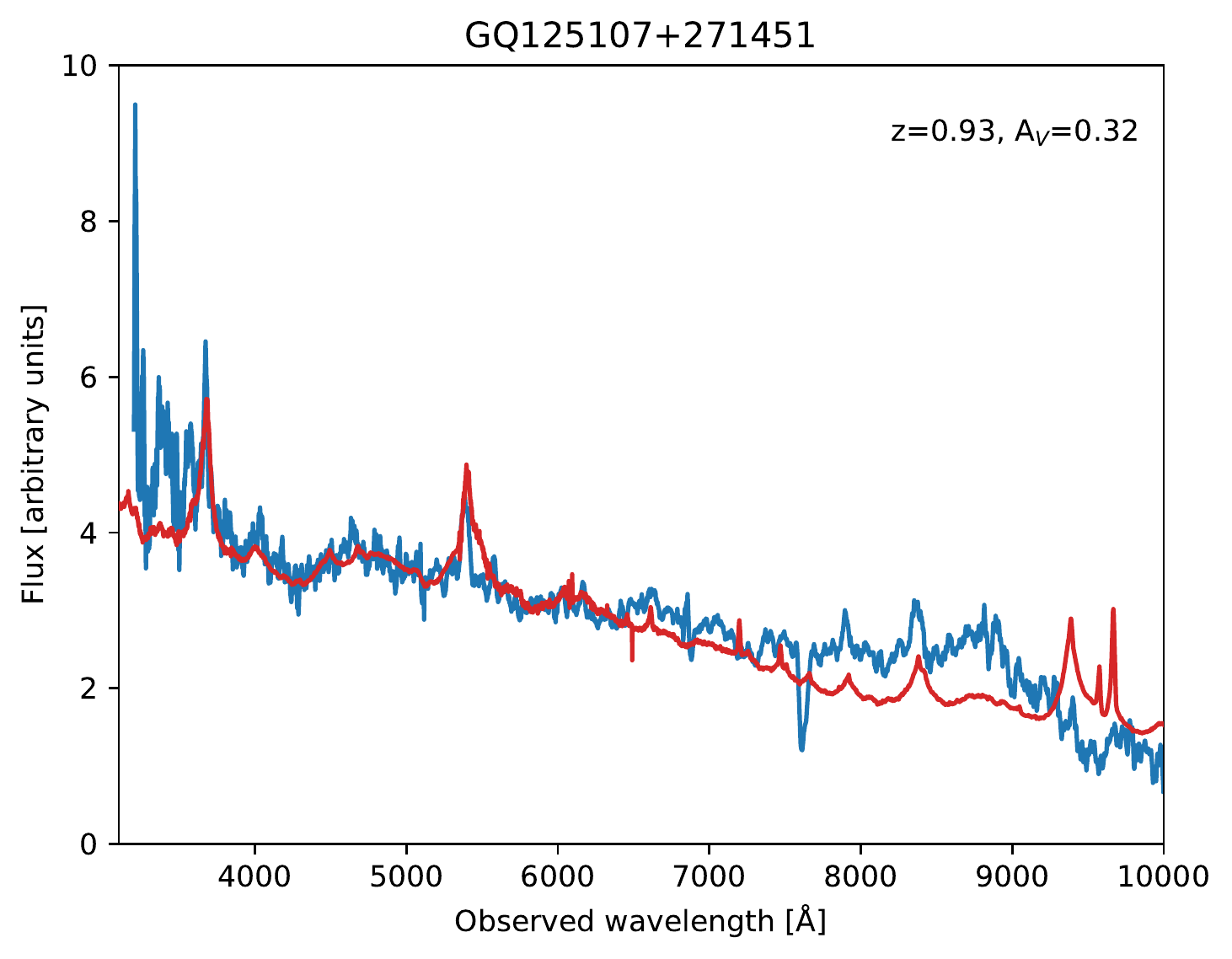,width=7.5cm}
\epsfig{file=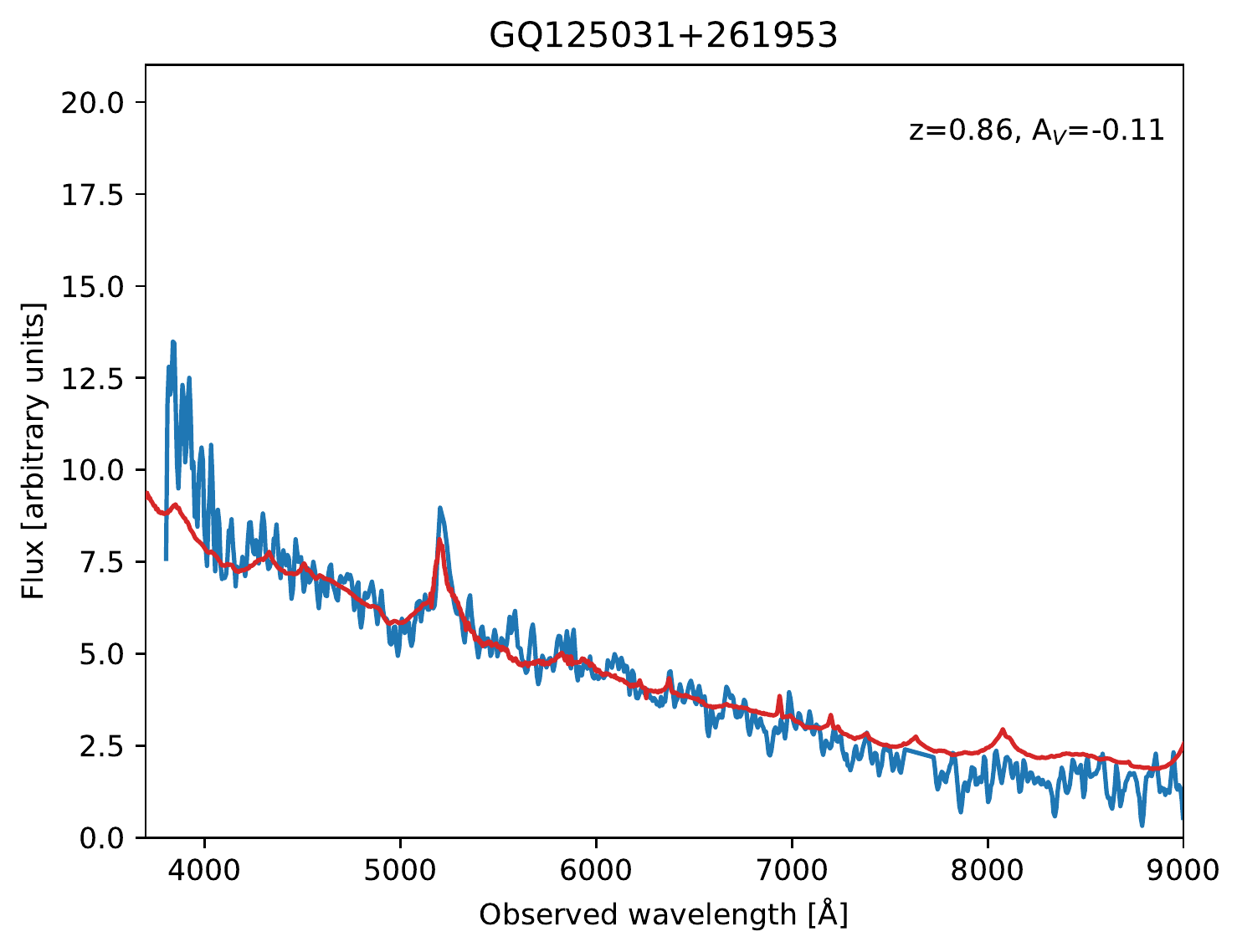,width=7.5cm}
\epsfig{file=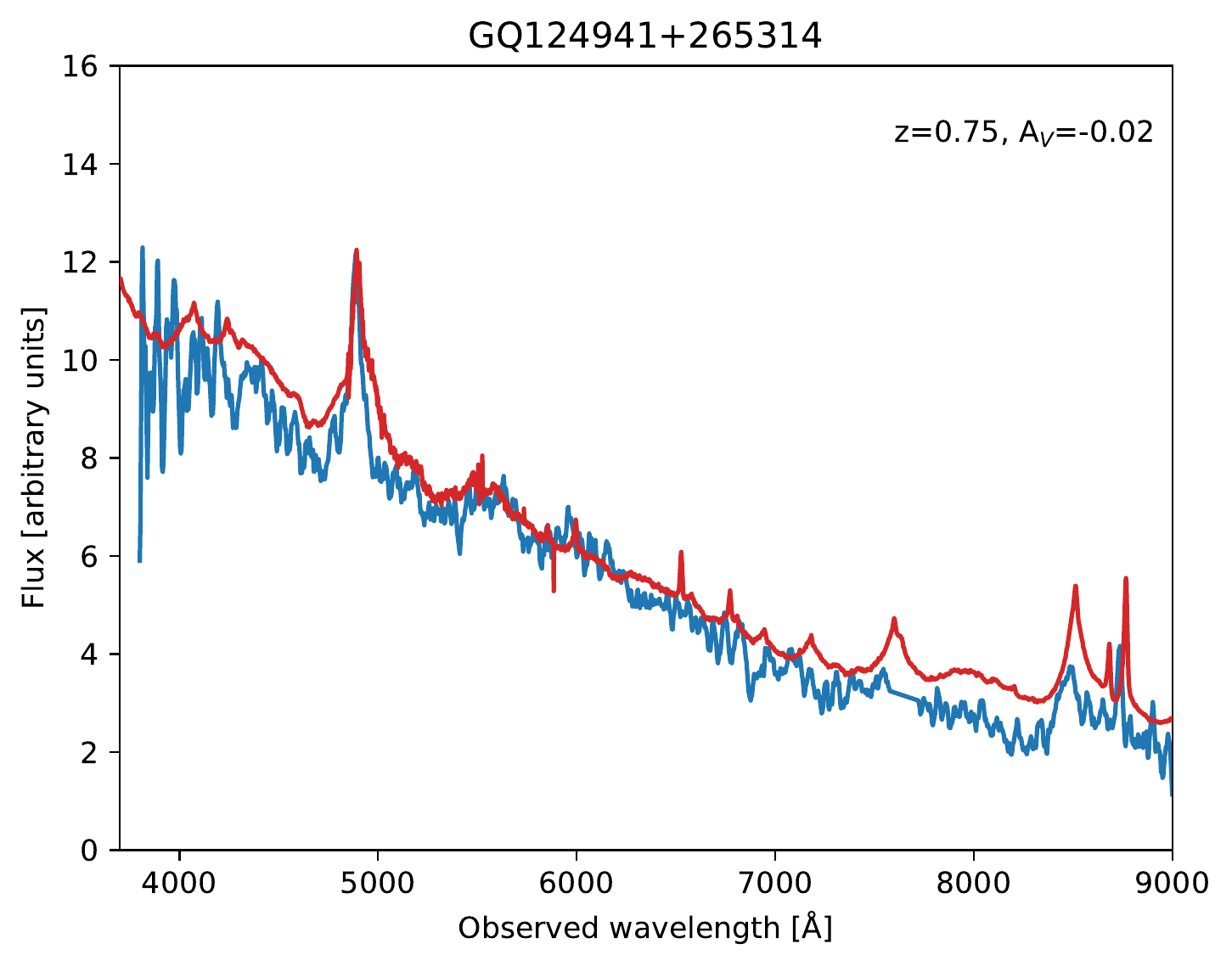,width=7.5cm}
\epsfig{file=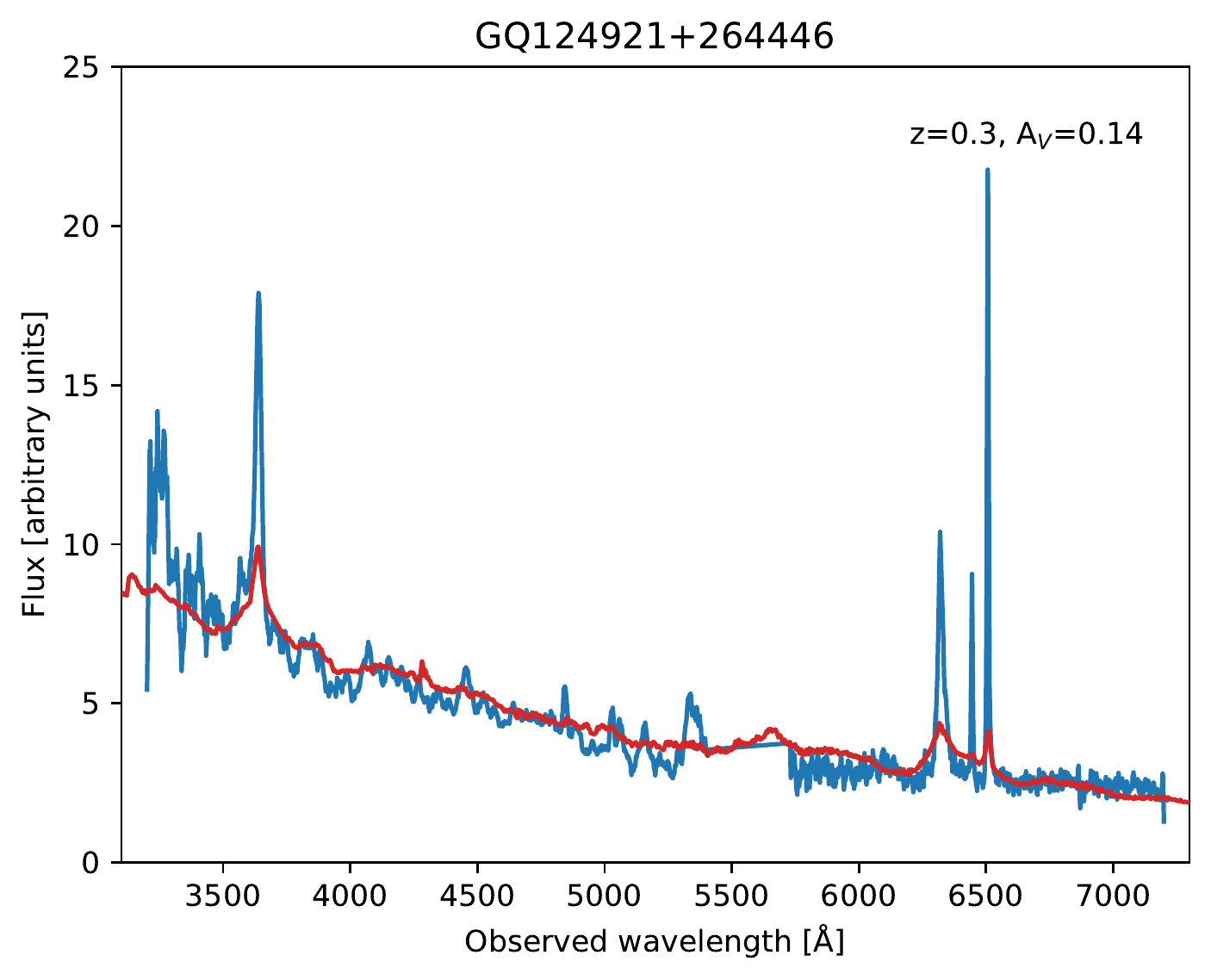,width=7.5cm}
\end{figure*}

\begin{figure*} [!b]
\centering
\epsfig{file=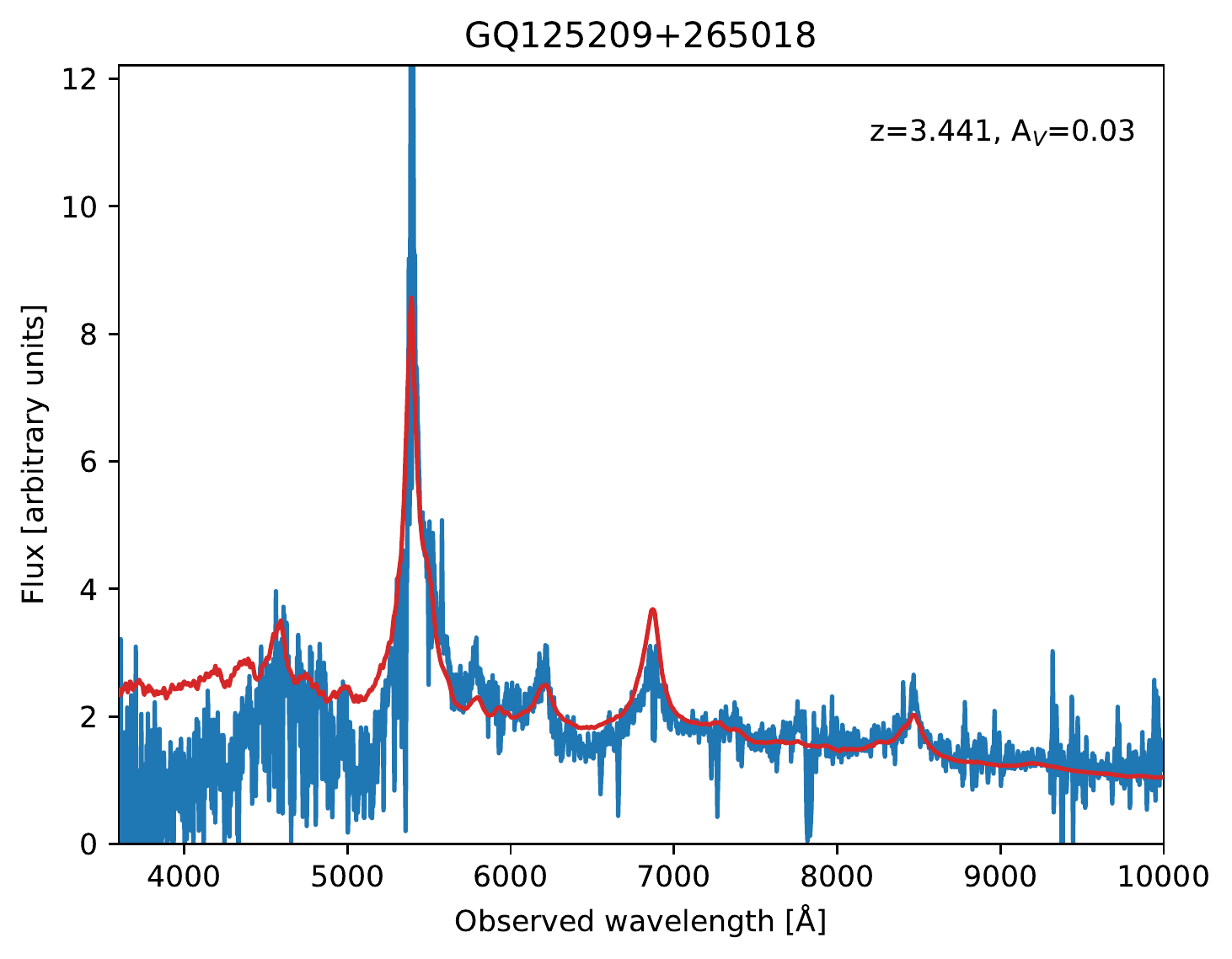,width=7.5cm}
\epsfig{file=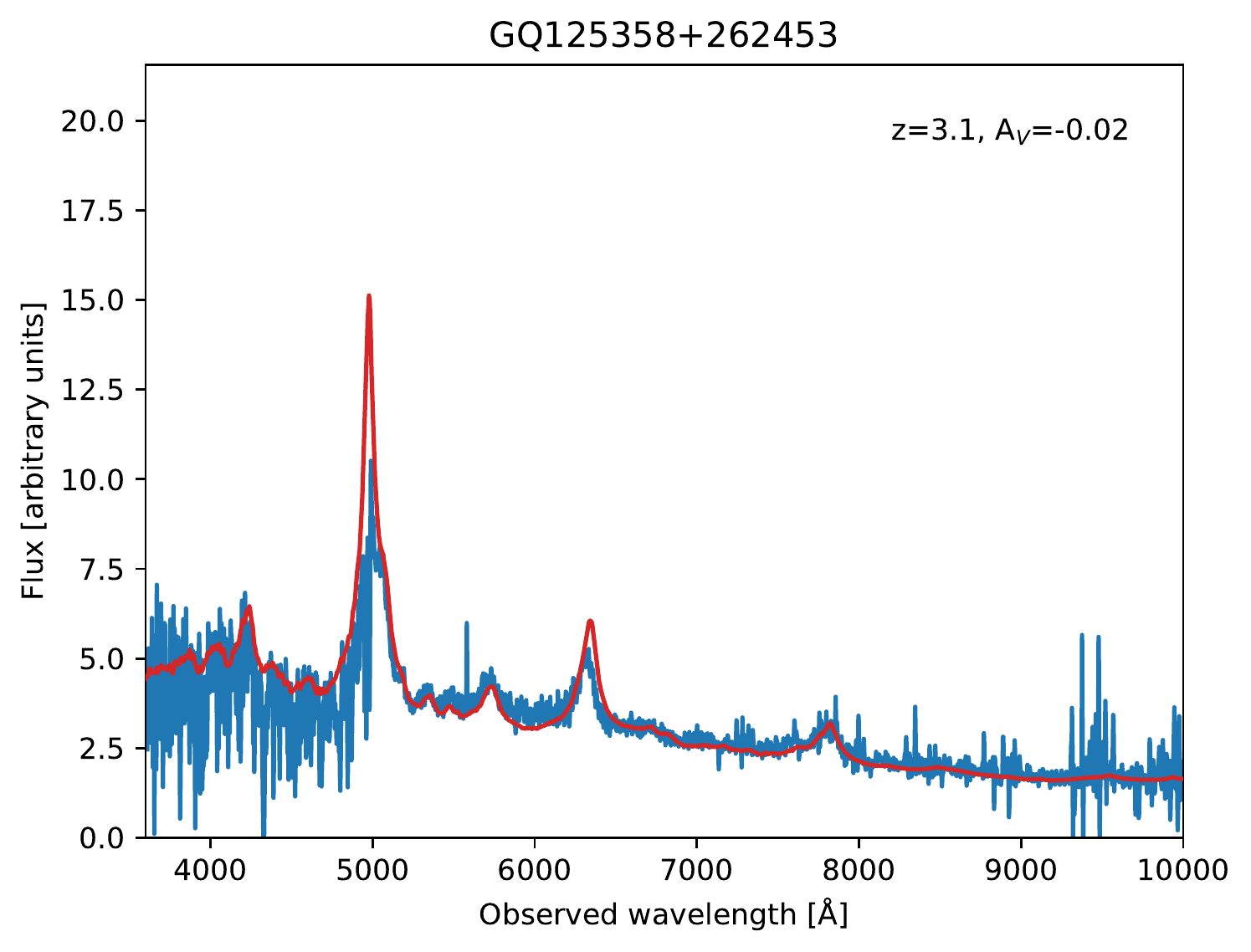,width=7.5cm}
\epsfig{file=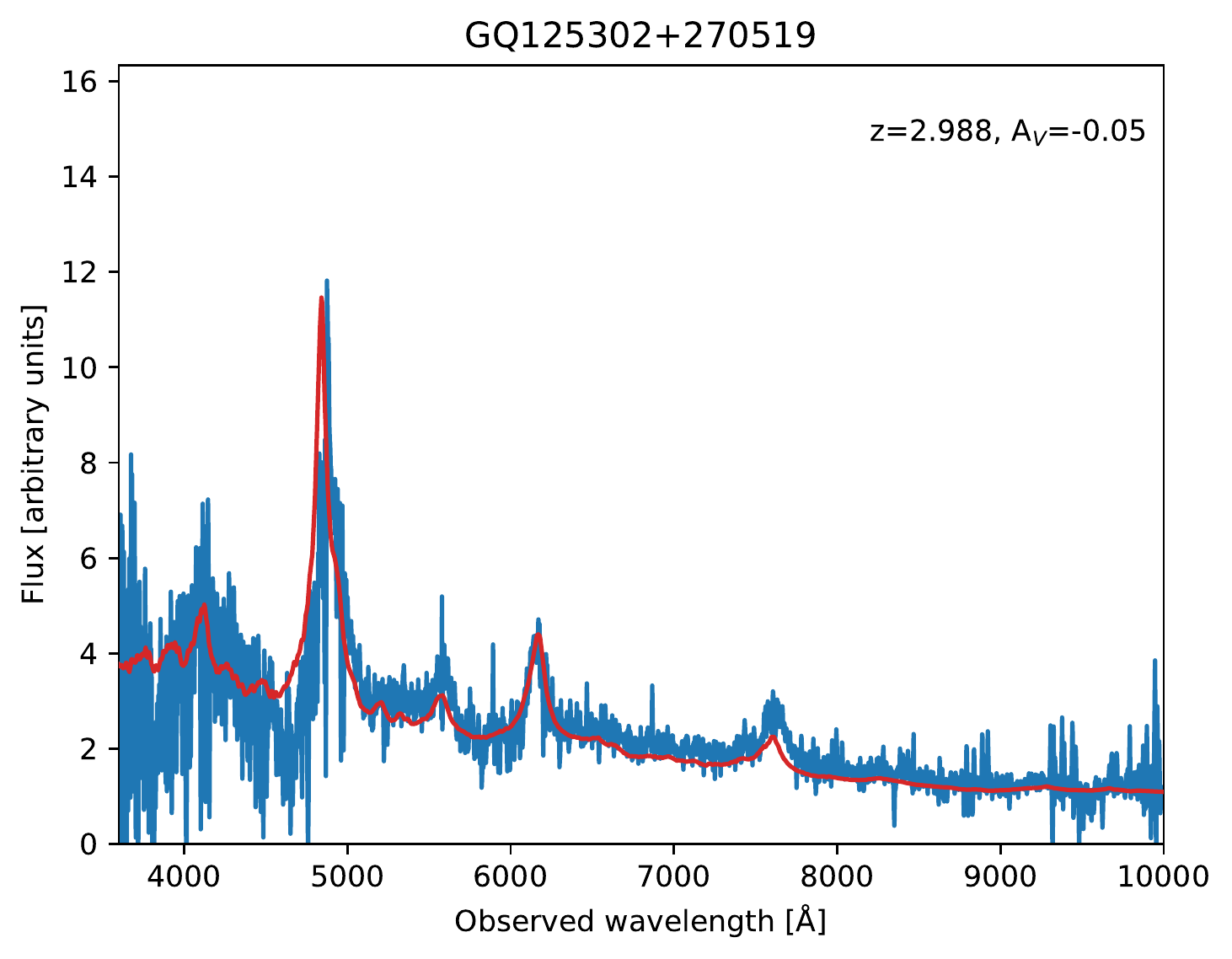,width=7.5cm}
\epsfig{file=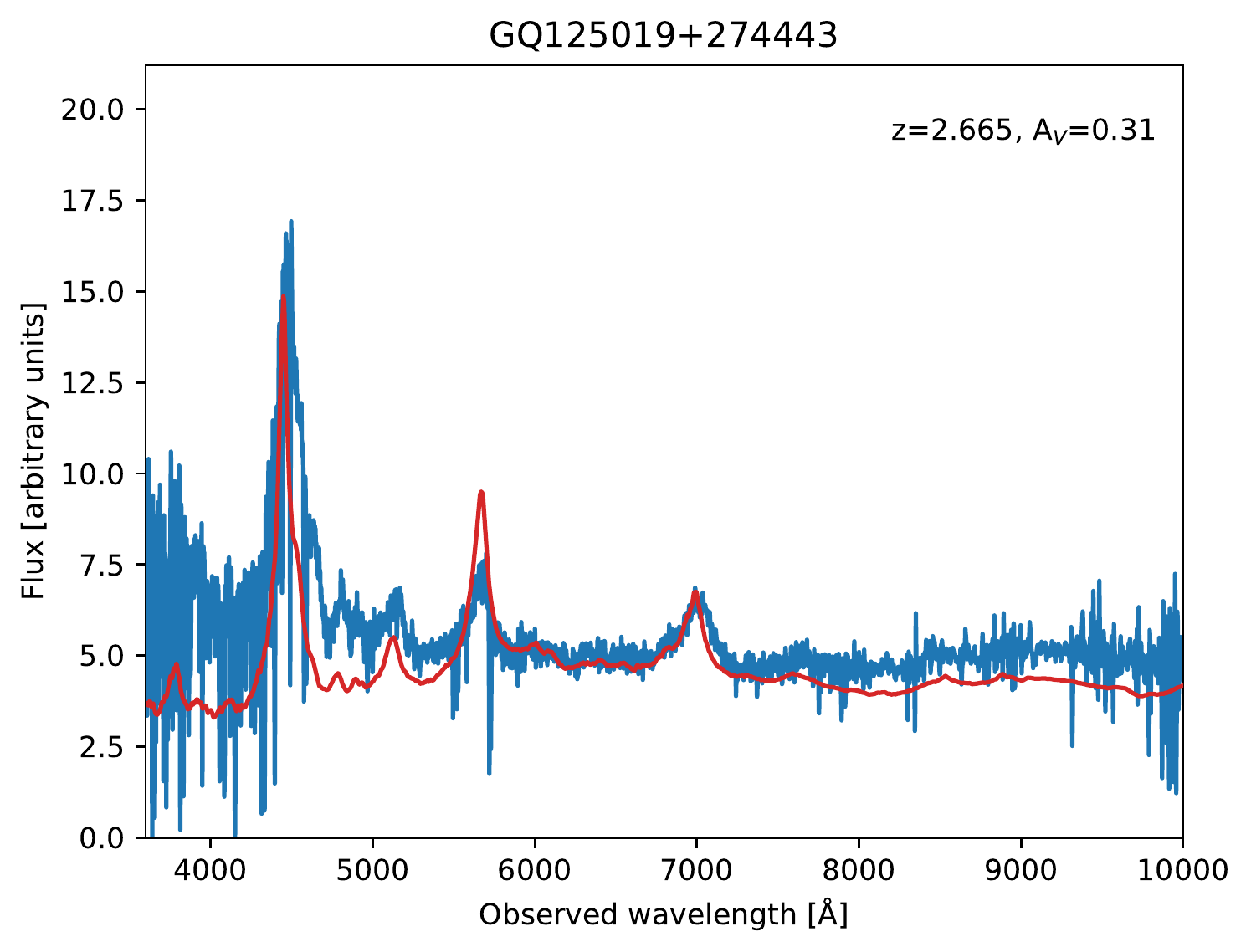,width=7.5cm}
\epsfig{file=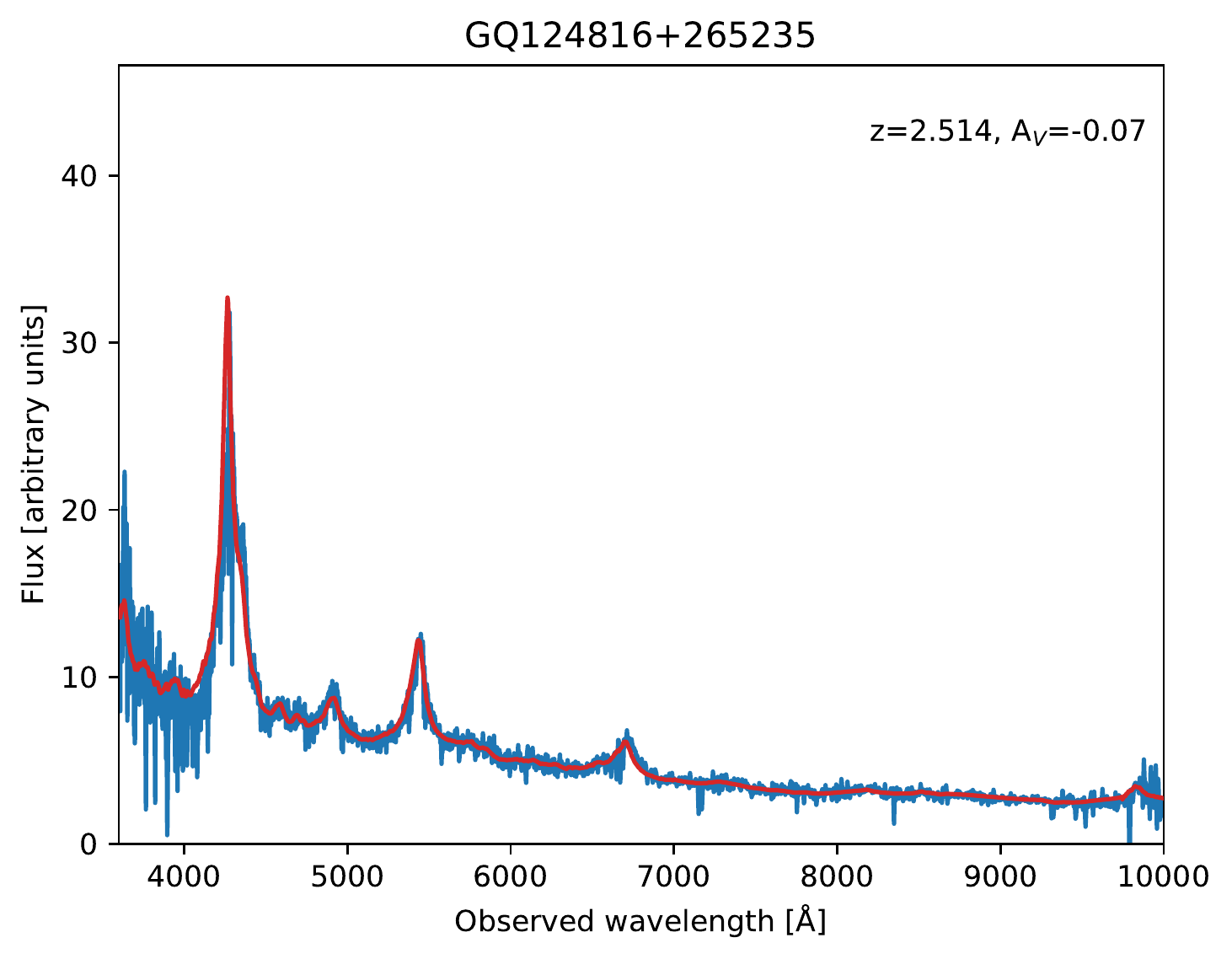,width=7.5cm}
\epsfig{file=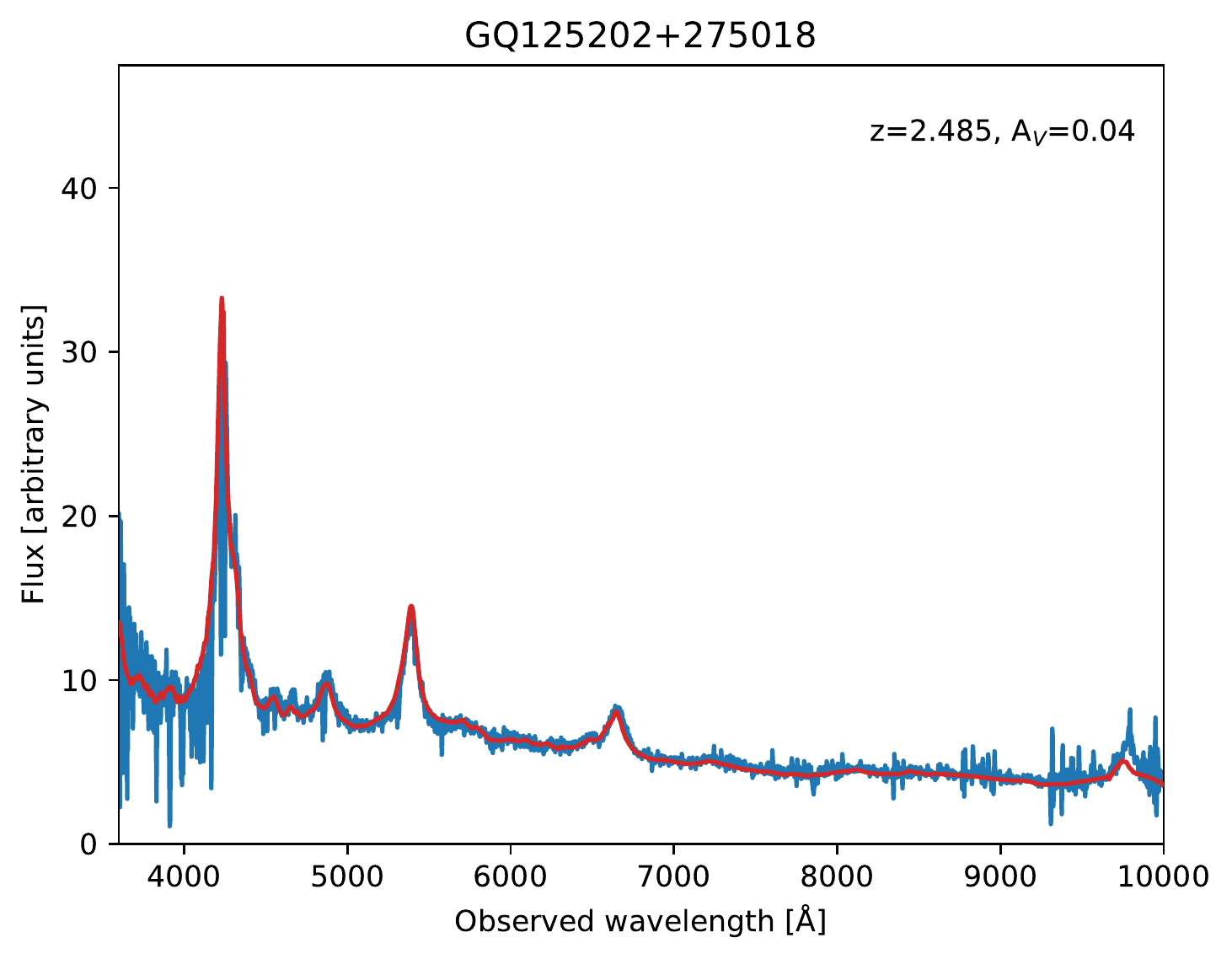,width=7.5cm}
\epsfig{file=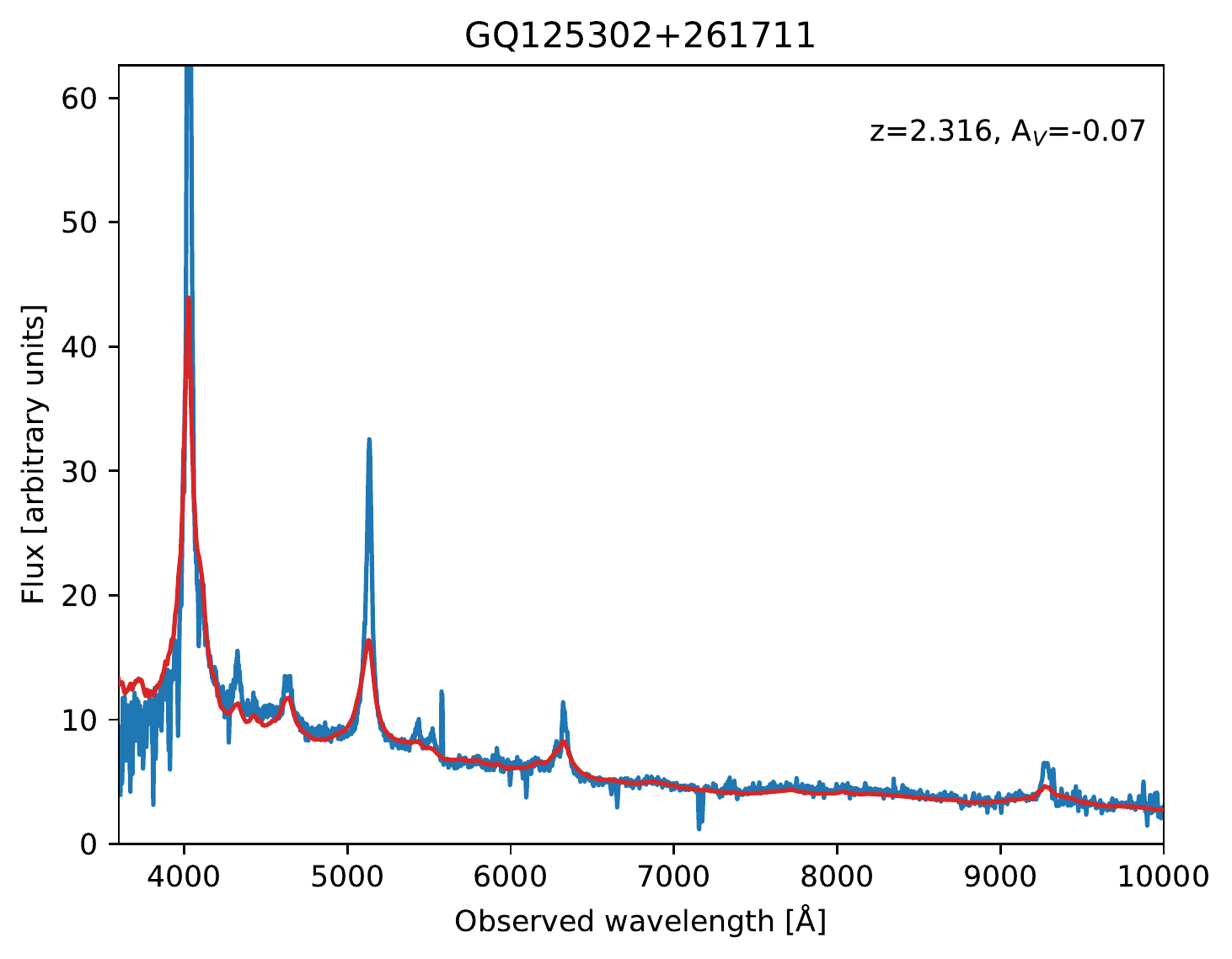,width=7.5cm}
\epsfig{file=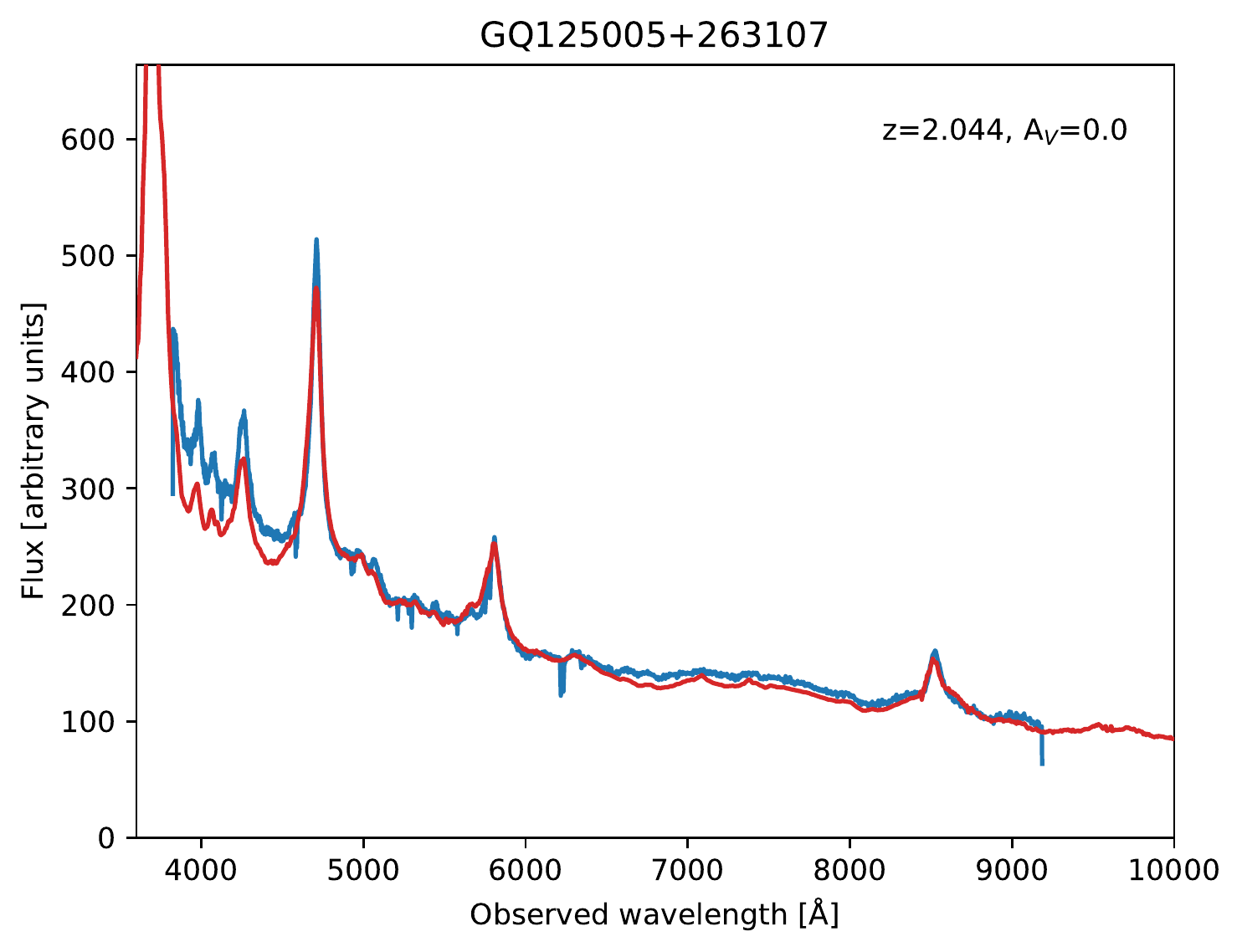,width=7.5cm}
\caption{Spectra of the SDSS quasars fulfilling our selection criteria.}
\label{fig:sdssspectra}
\end{figure*}

\begin{figure*} [!b]
\centering
\epsfig{file=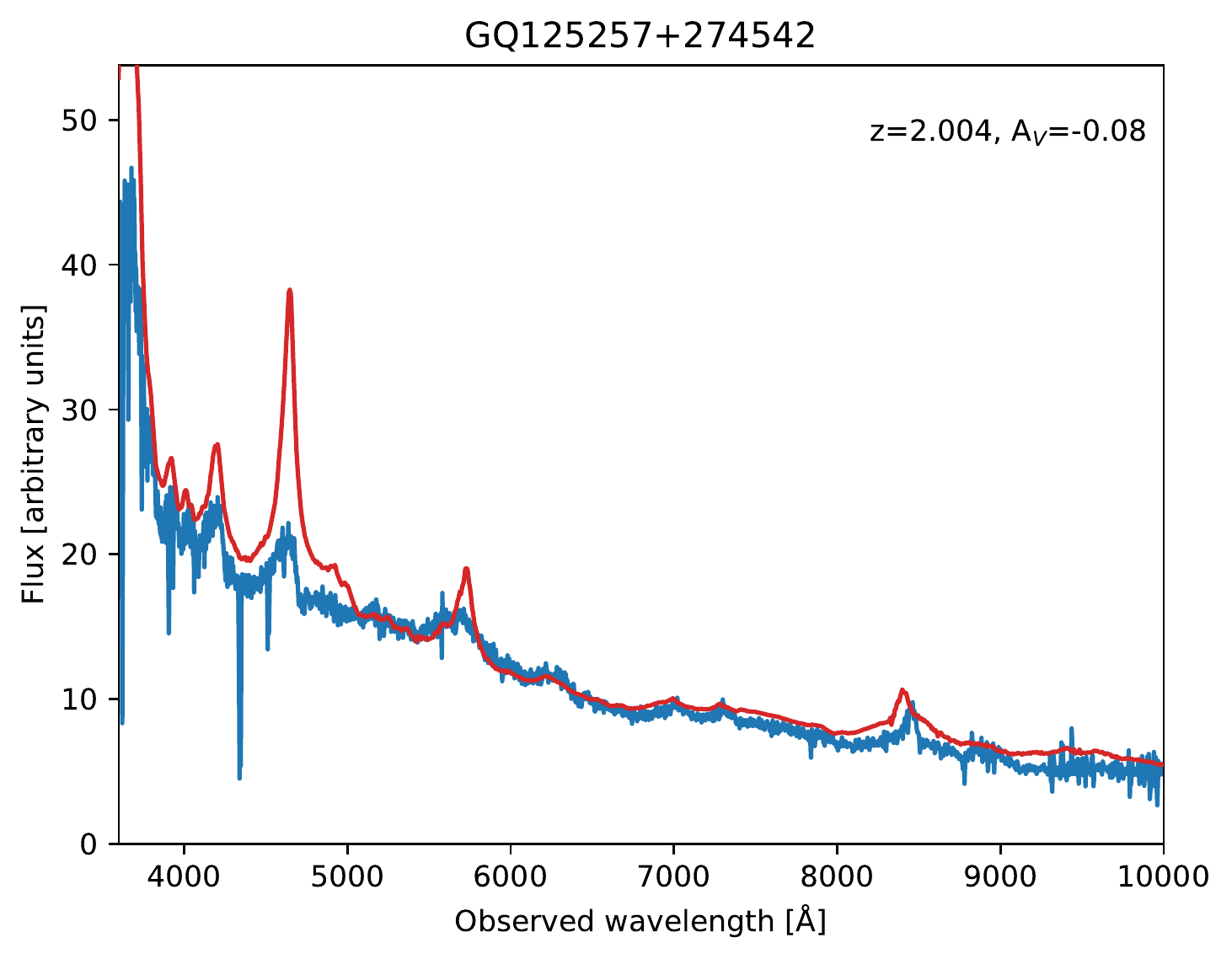,width=7.5cm}
\epsfig{file=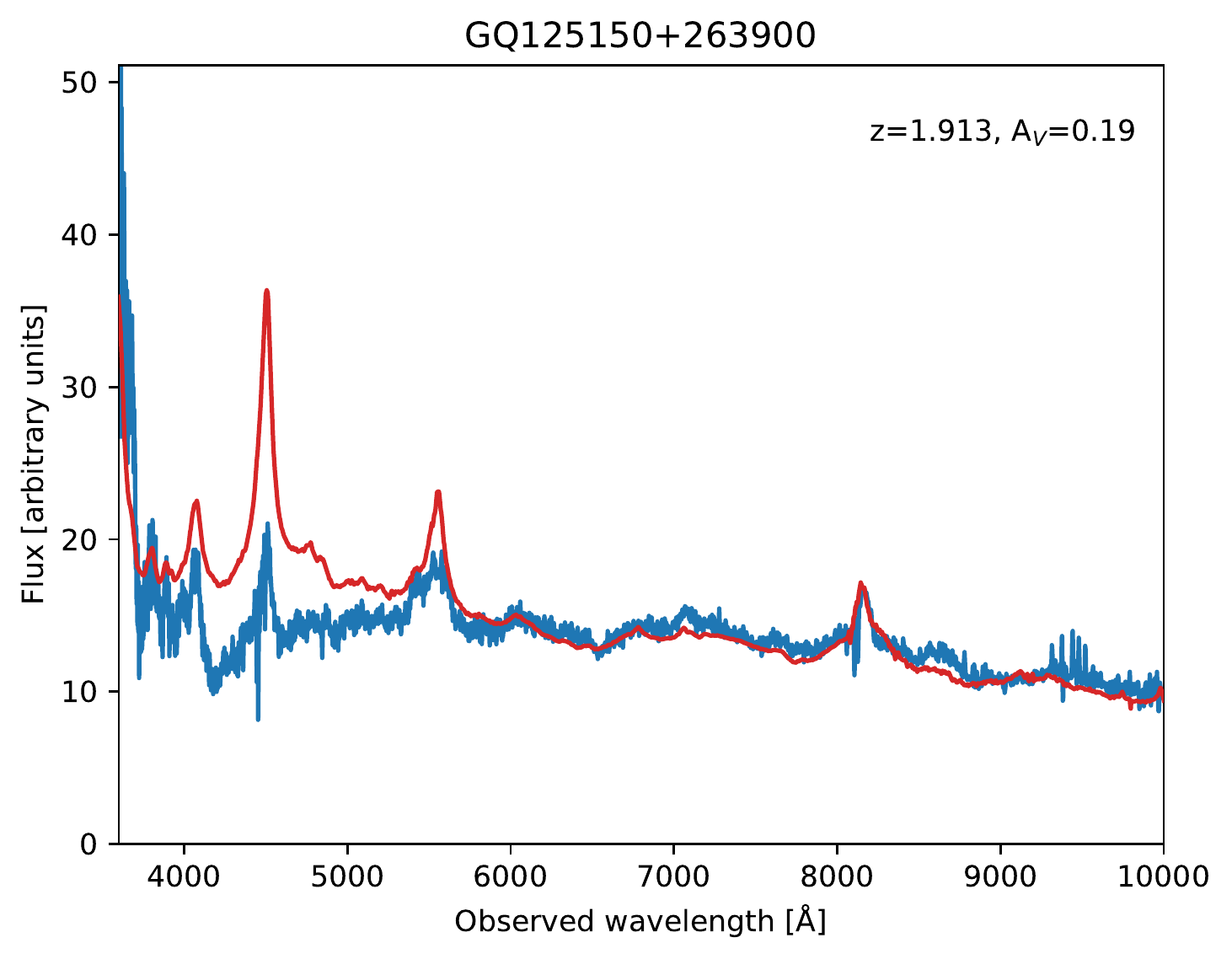,width=7.5cm}
\epsfig{file=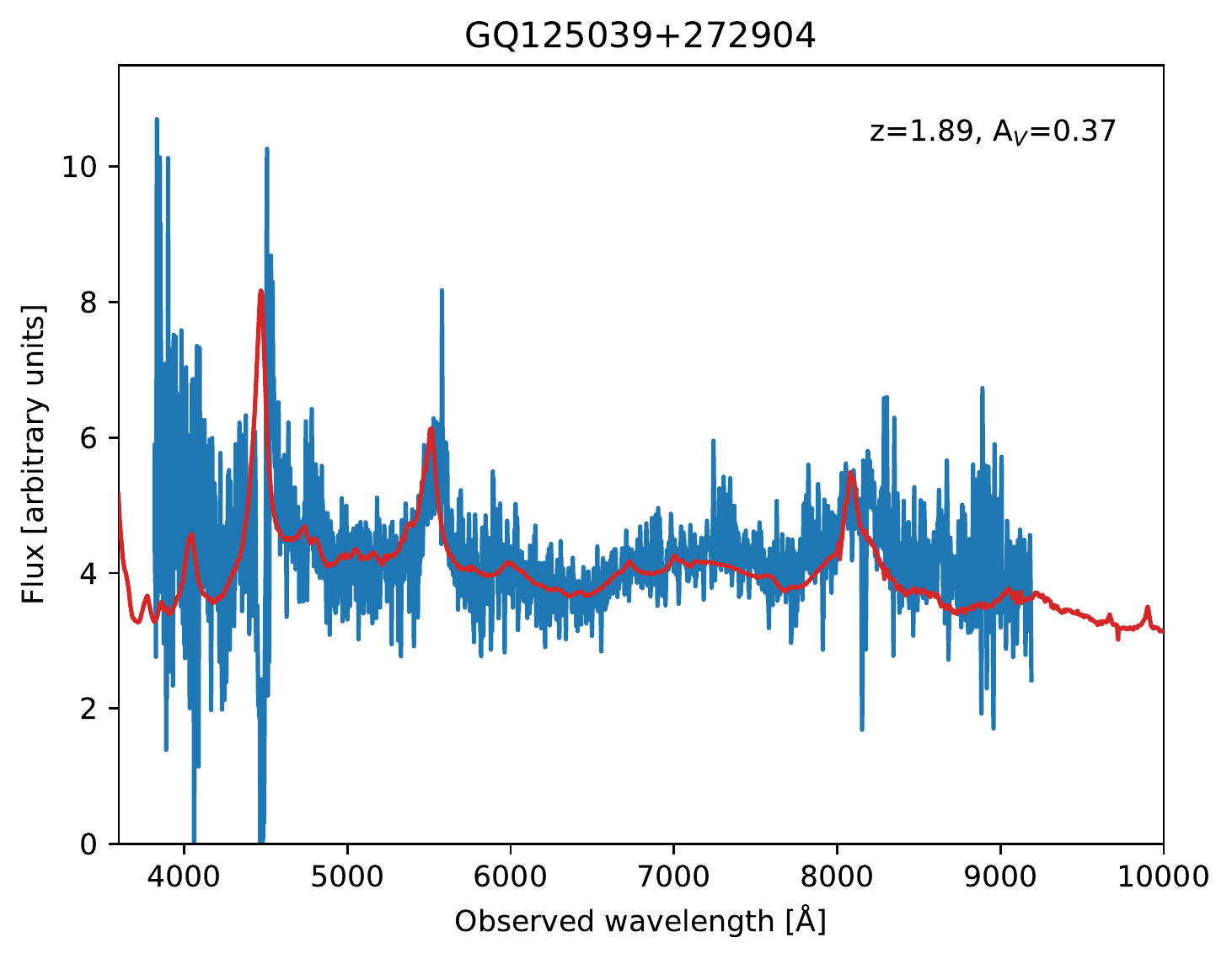,width=7.5cm}
\epsfig{file=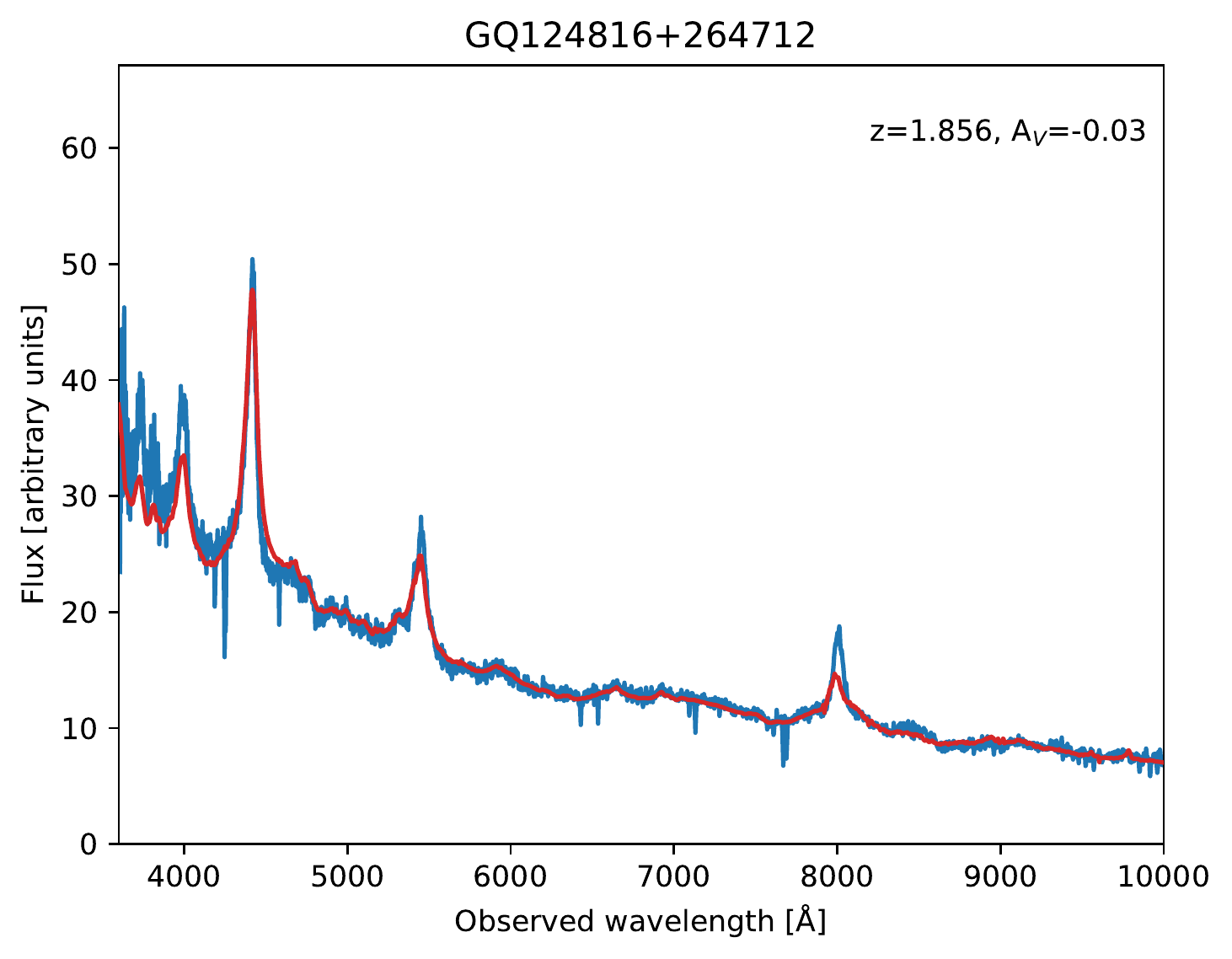,width=7.5cm}
\epsfig{file=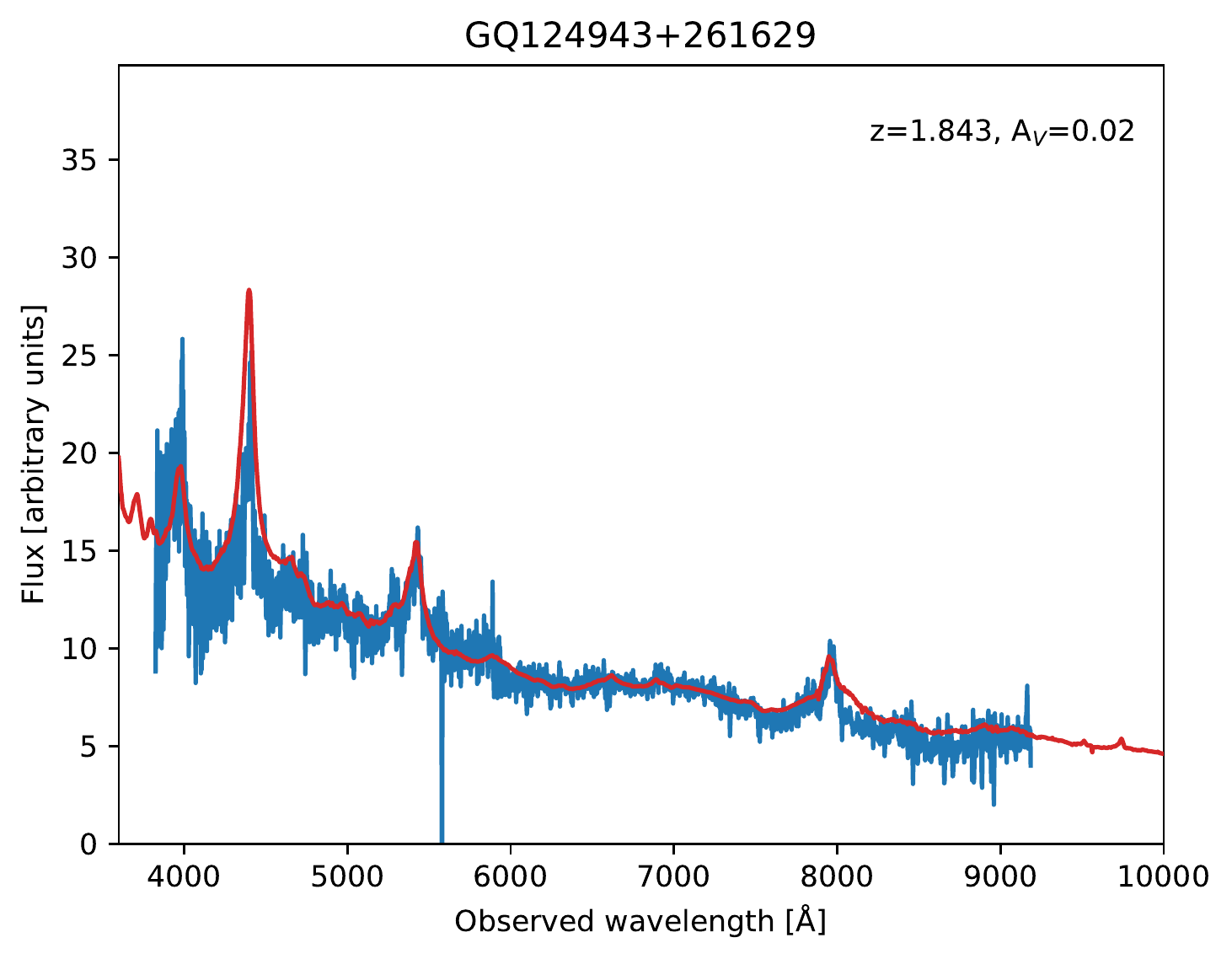,width=7.5cm}
\epsfig{file=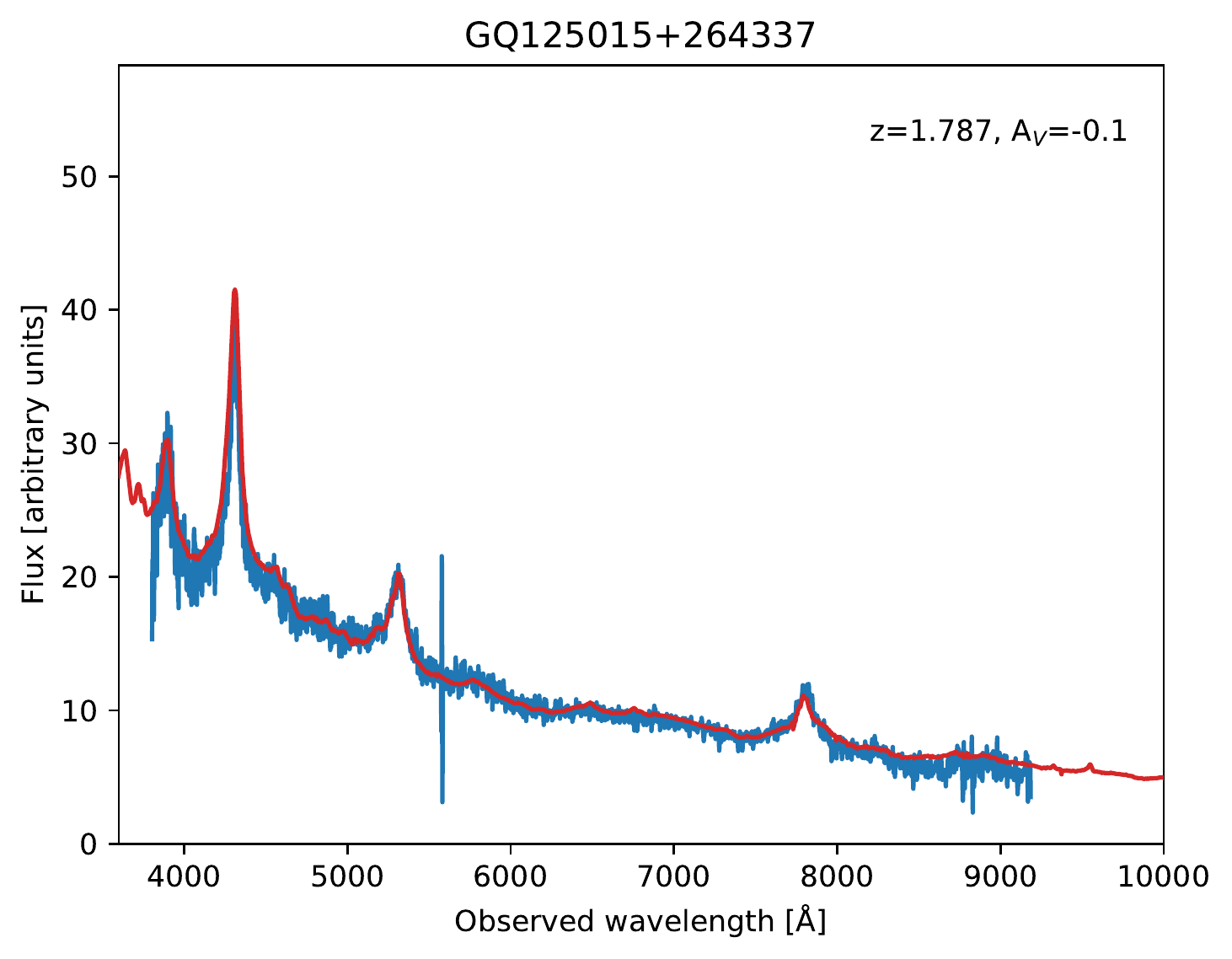,width=7.5cm}
\epsfig{file=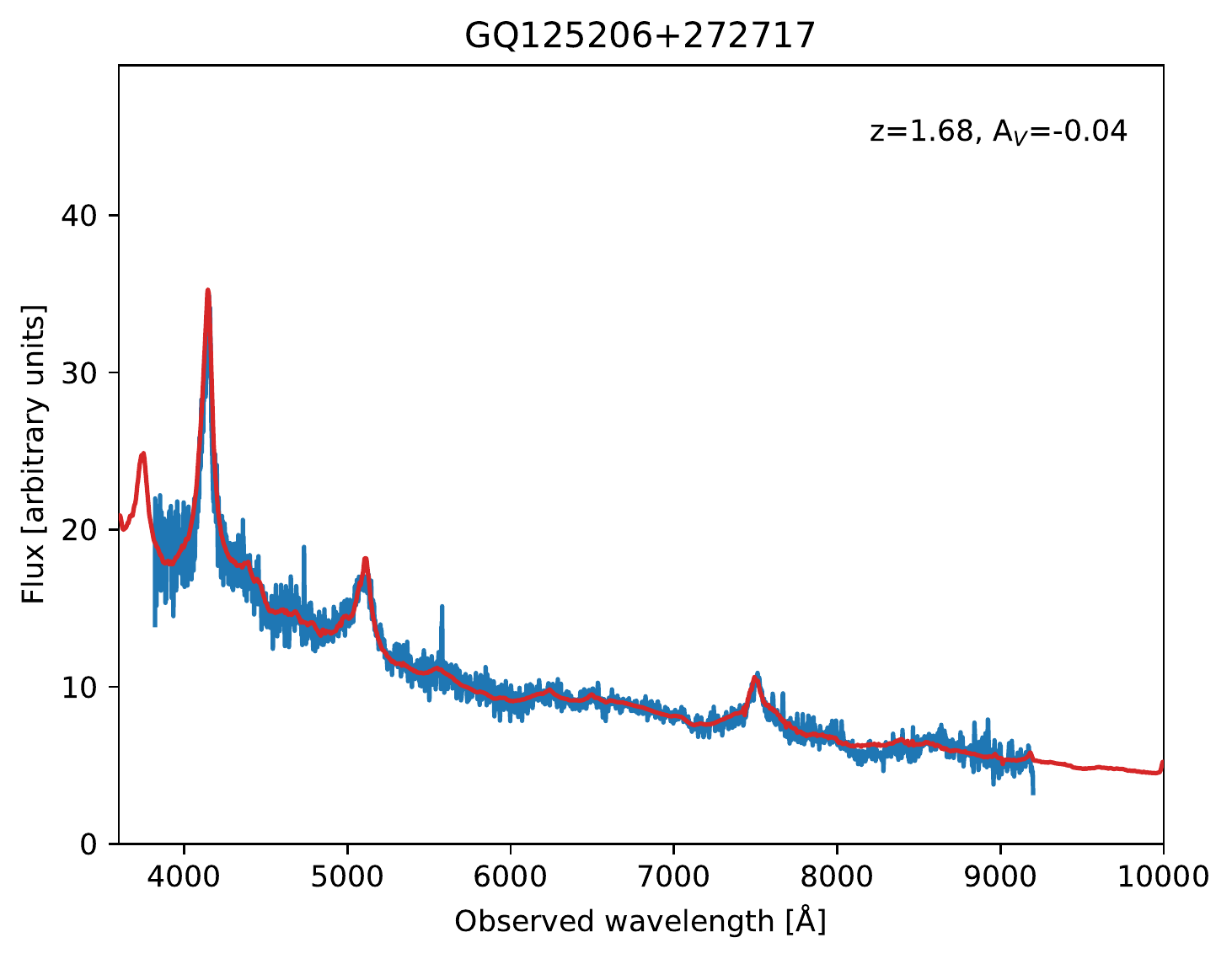,width=7.5cm}
\epsfig{file=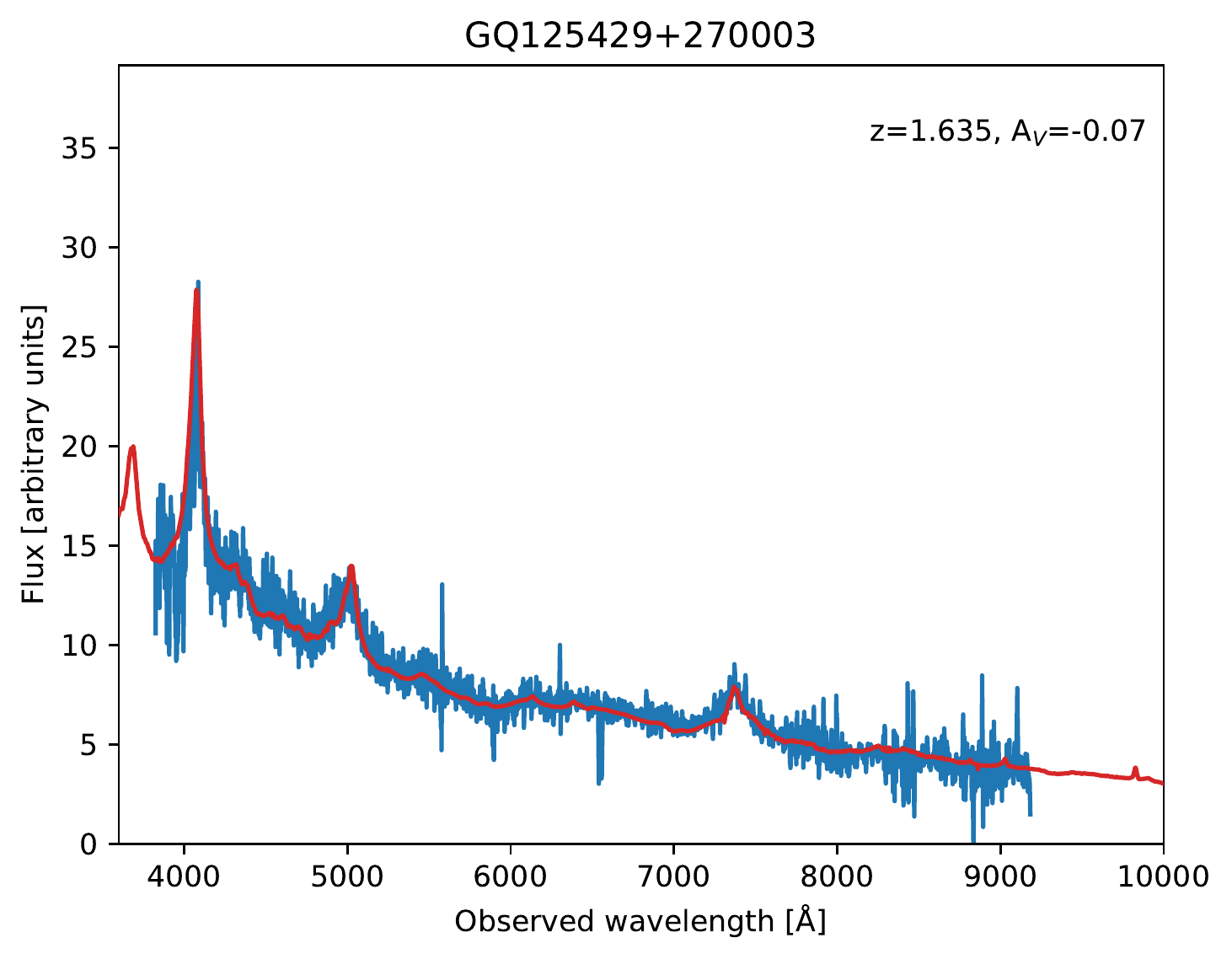,width=7.5cm}
\end{figure*}

\begin{figure*} [!b]
\centering
\epsfig{file=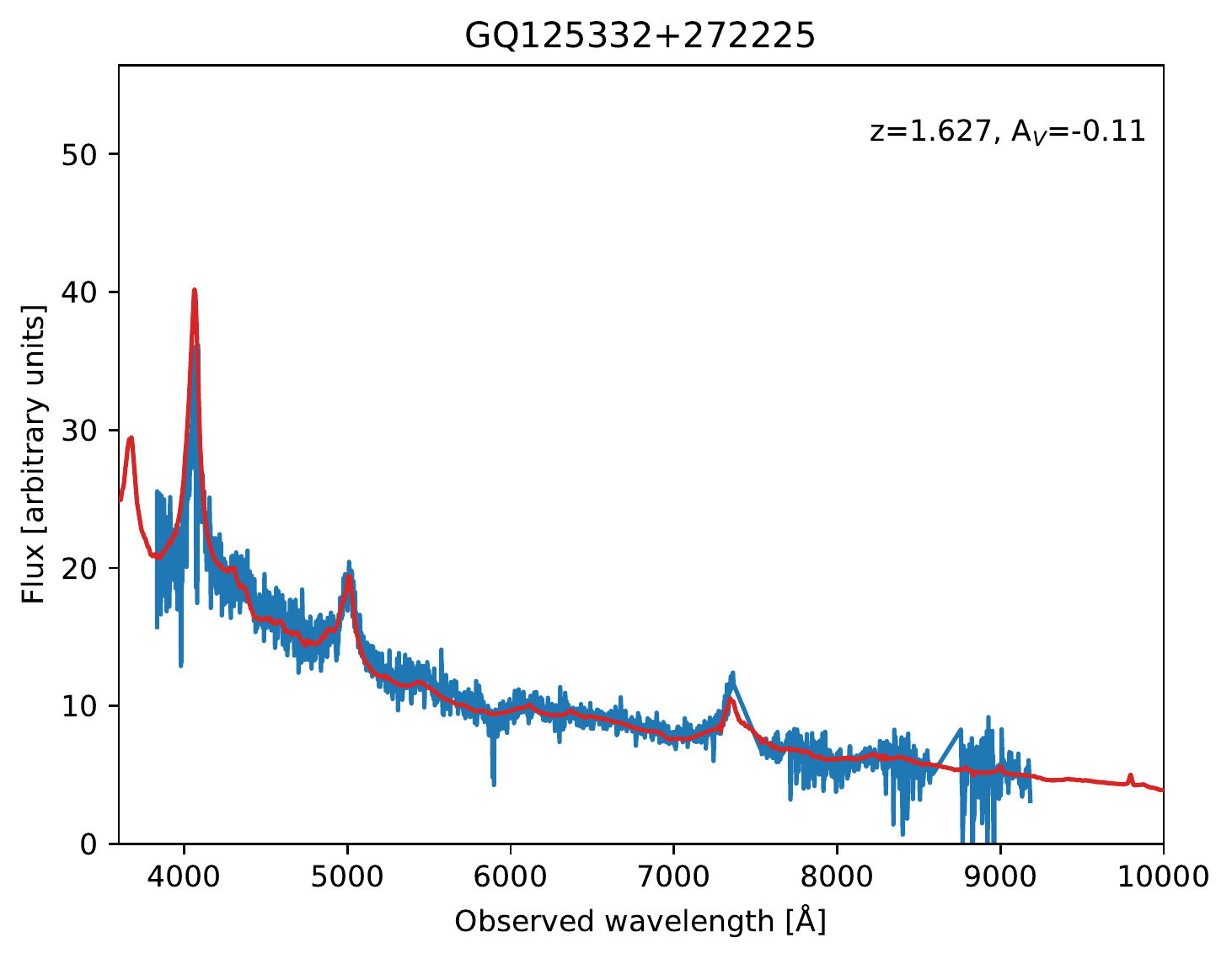,width=7.5cm}
\epsfig{file=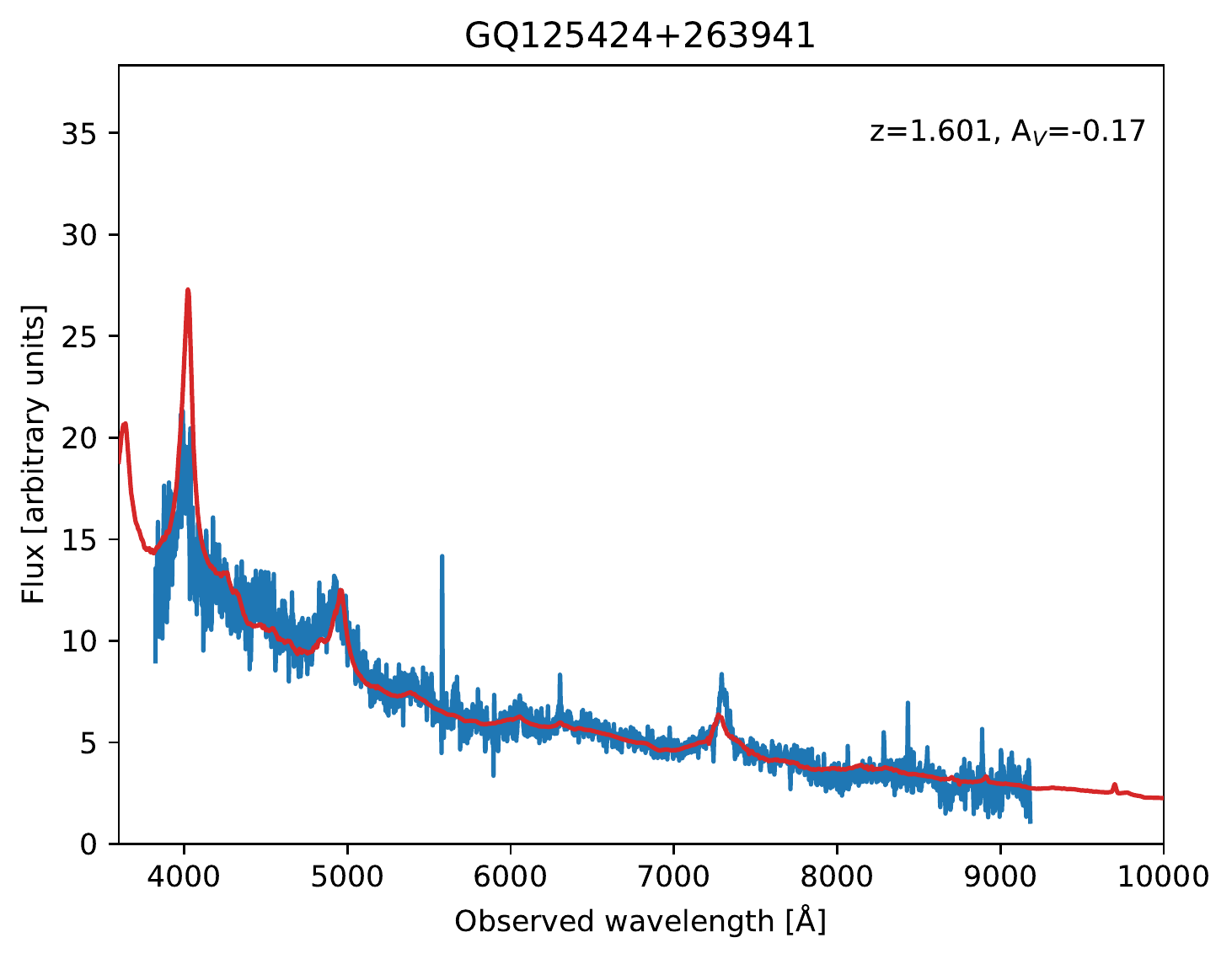,width=7.5cm}
\epsfig{file=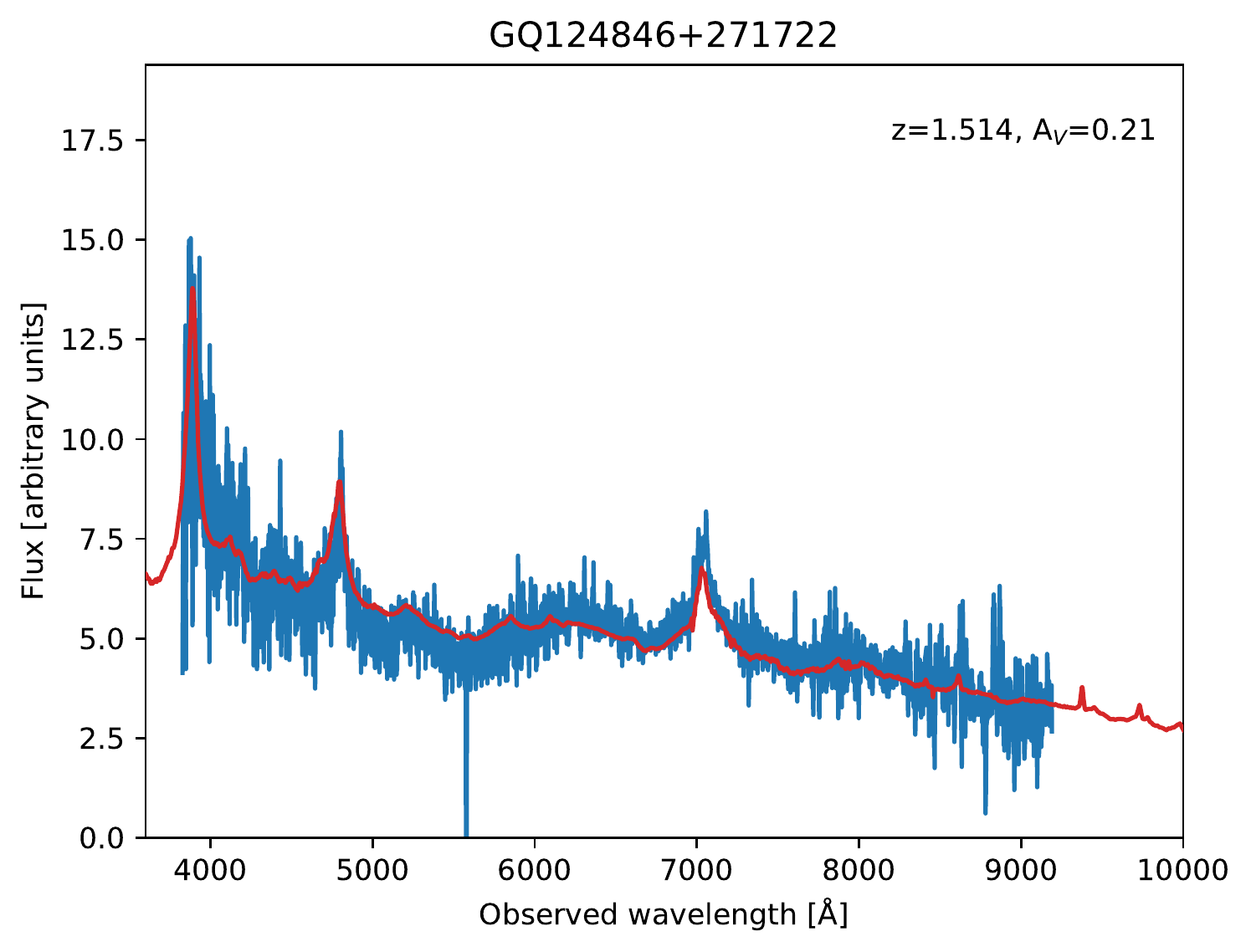,width=7.5cm}
\epsfig{file=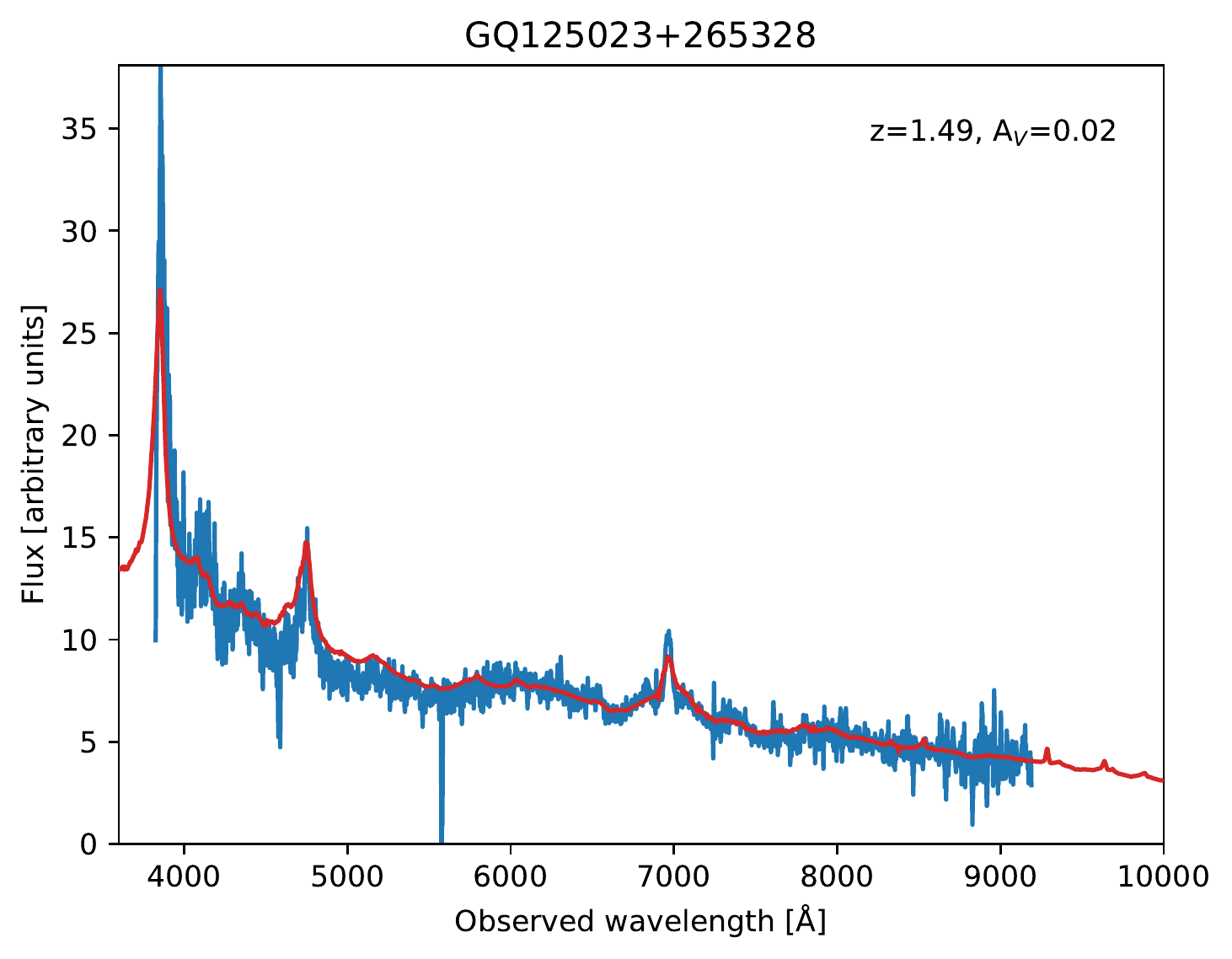,width=7.5cm}
\epsfig{file=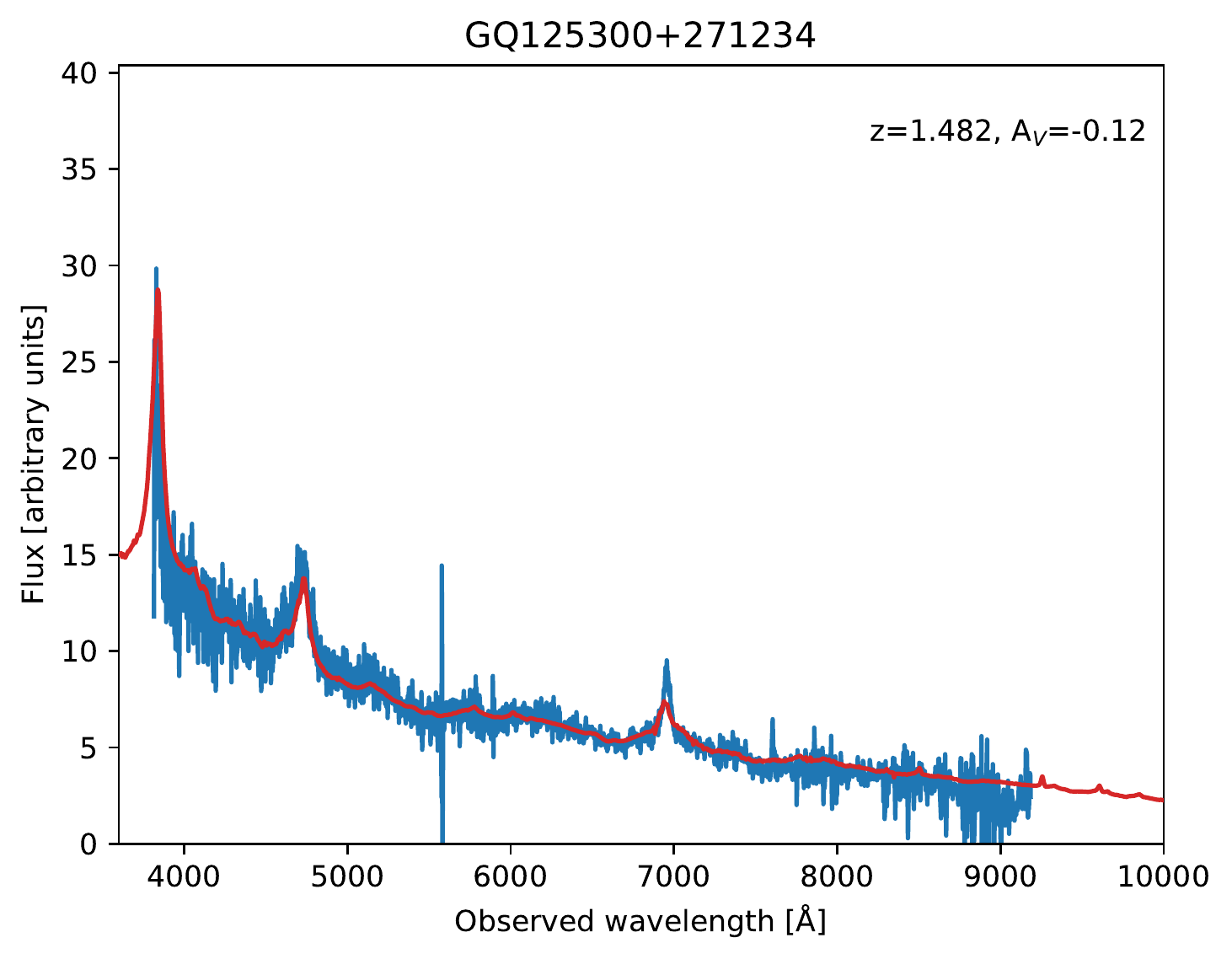,width=7.5cm}
\epsfig{file=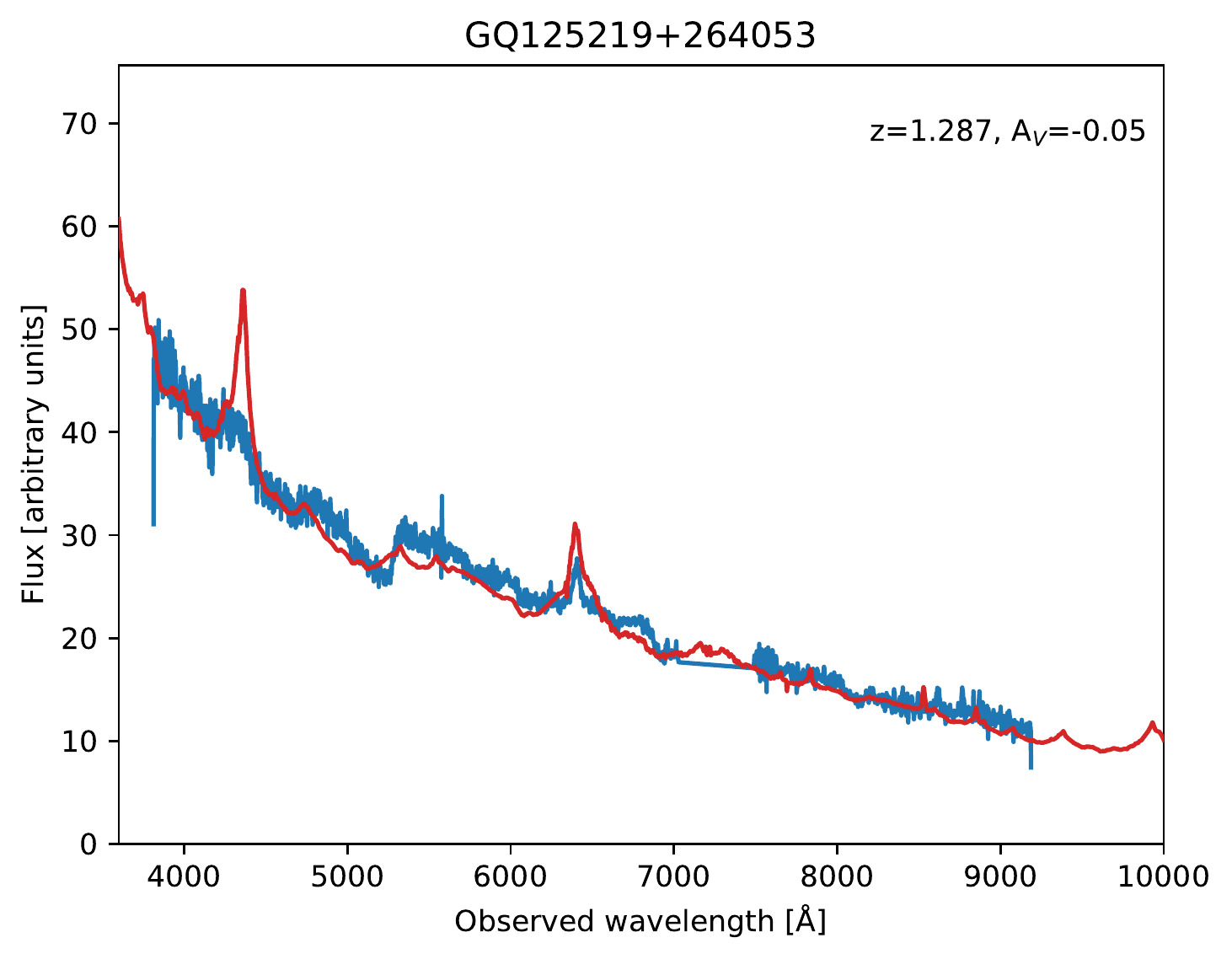,width=7.5cm}
\epsfig{file=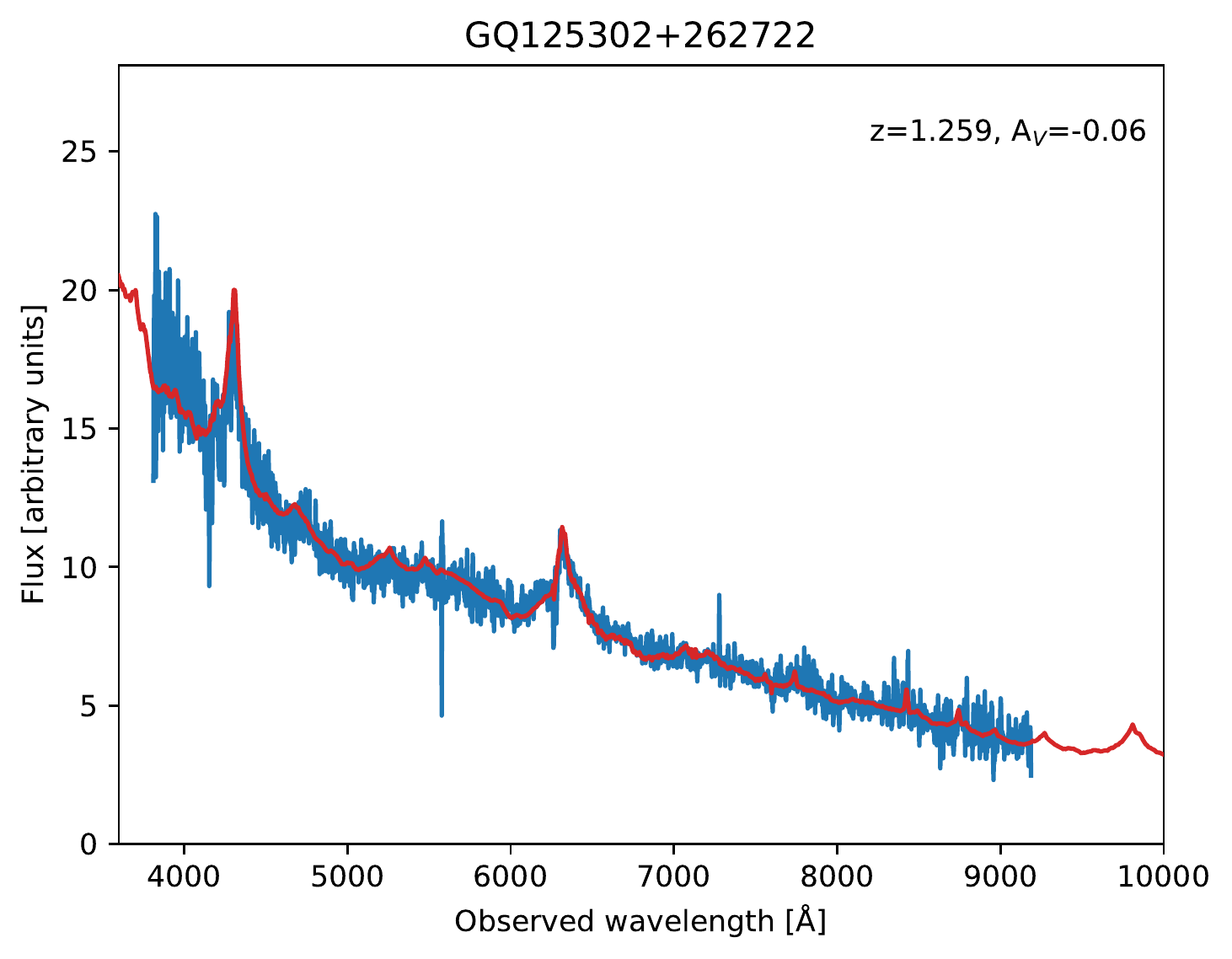,width=7.5cm}
\epsfig{file=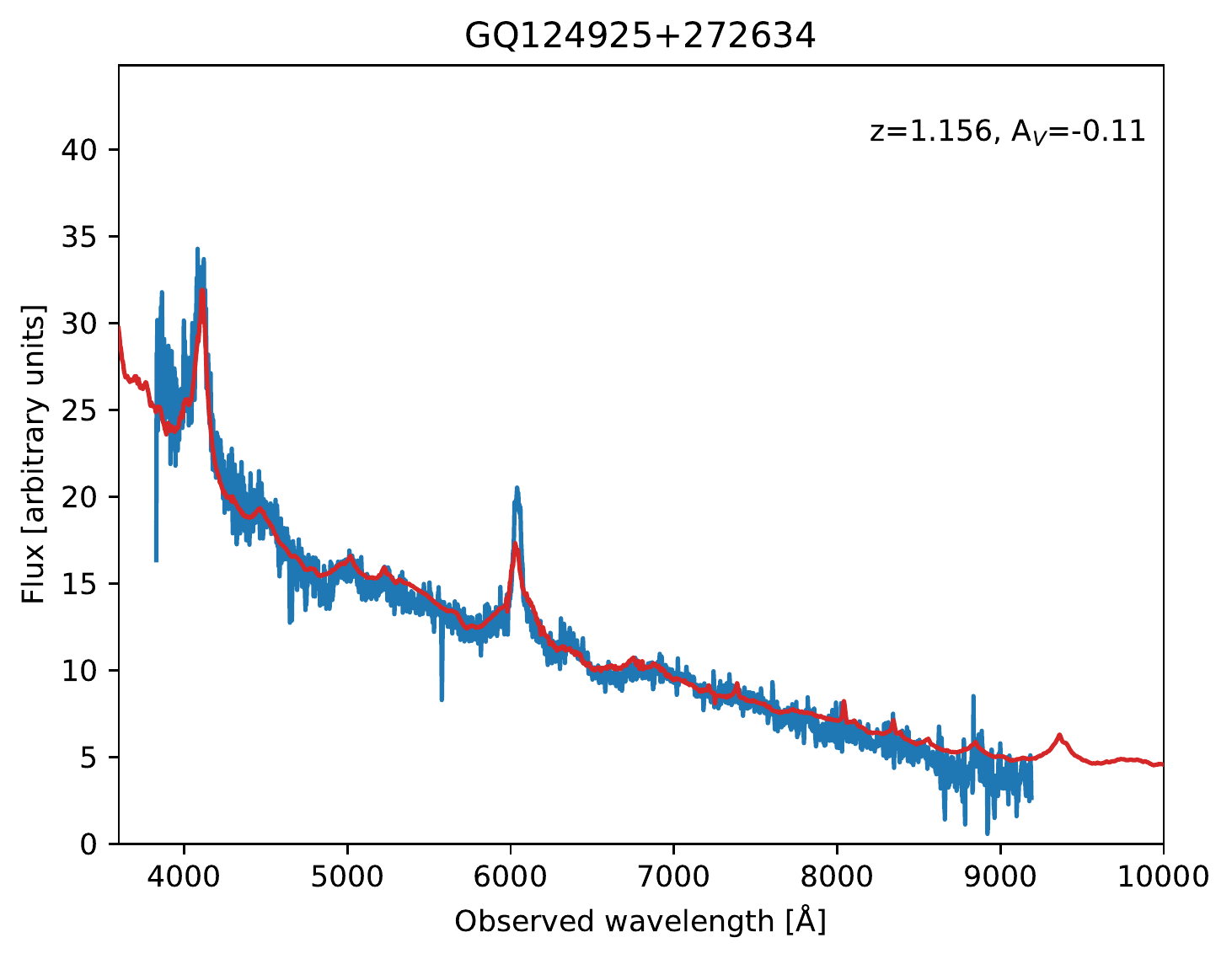,width=7.5cm}
\end{figure*}

\begin{figure*} [!b]
\centering
\epsfig{file=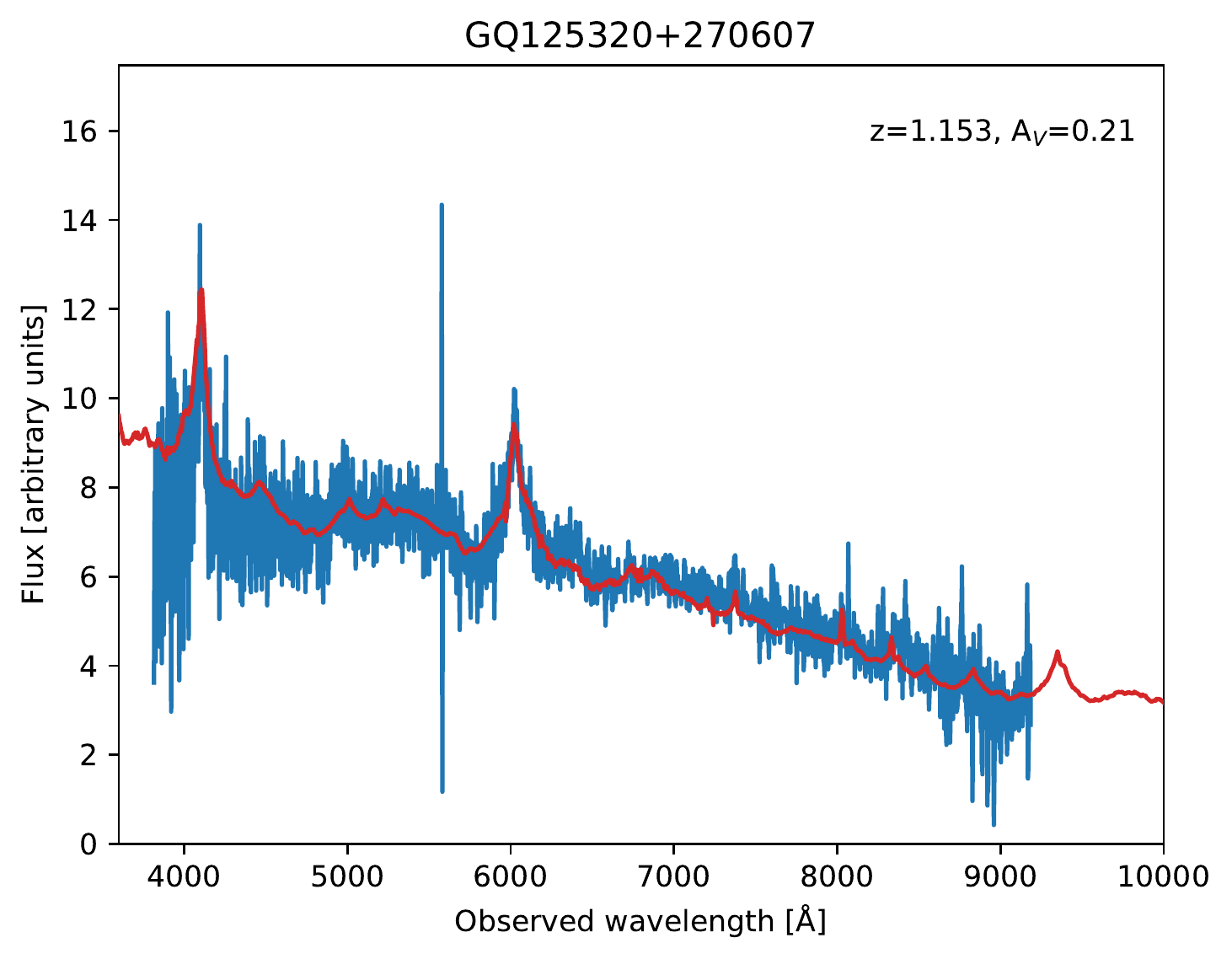,width=7.5cm}
\epsfig{file=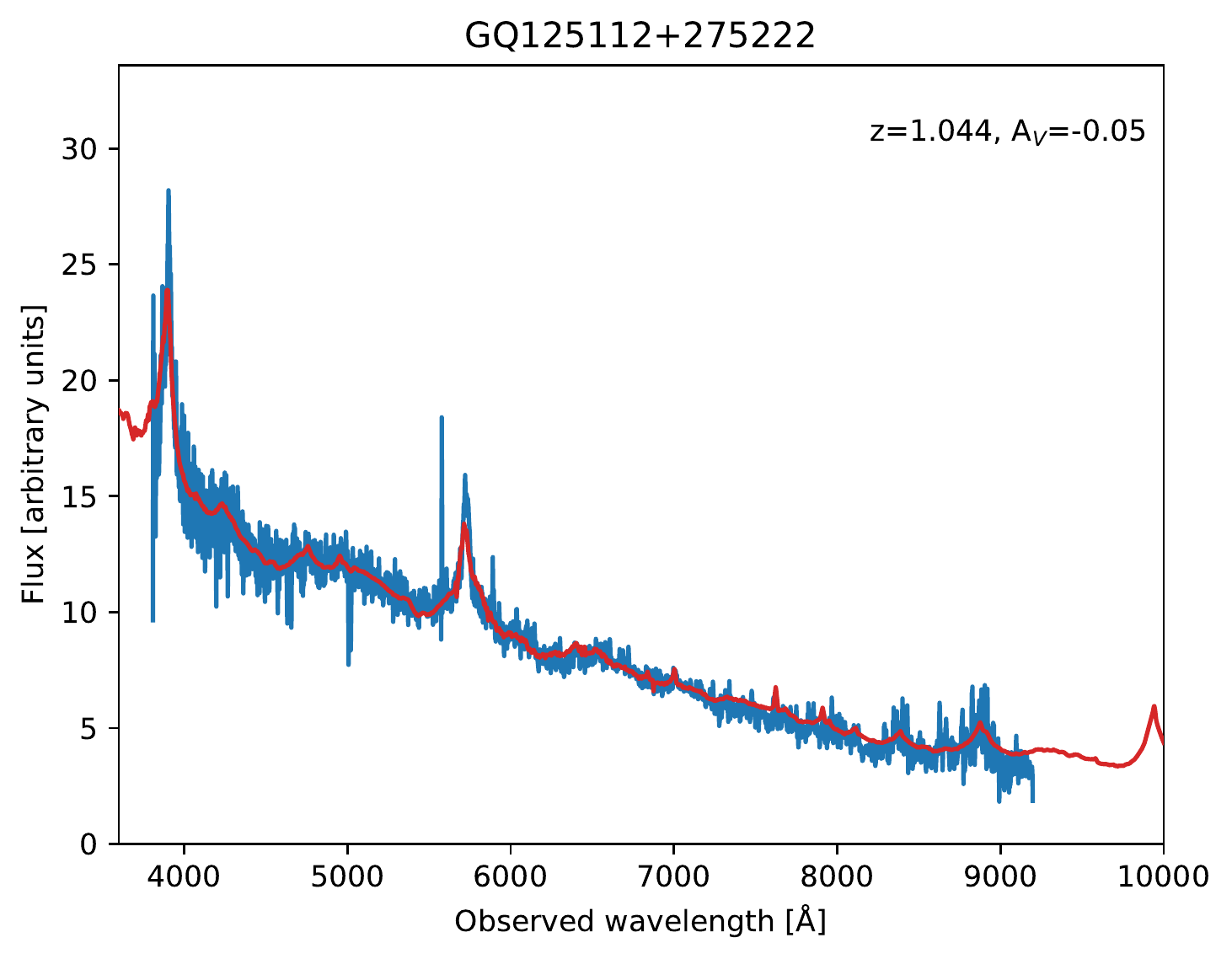,width=7.5cm}
\epsfig{file=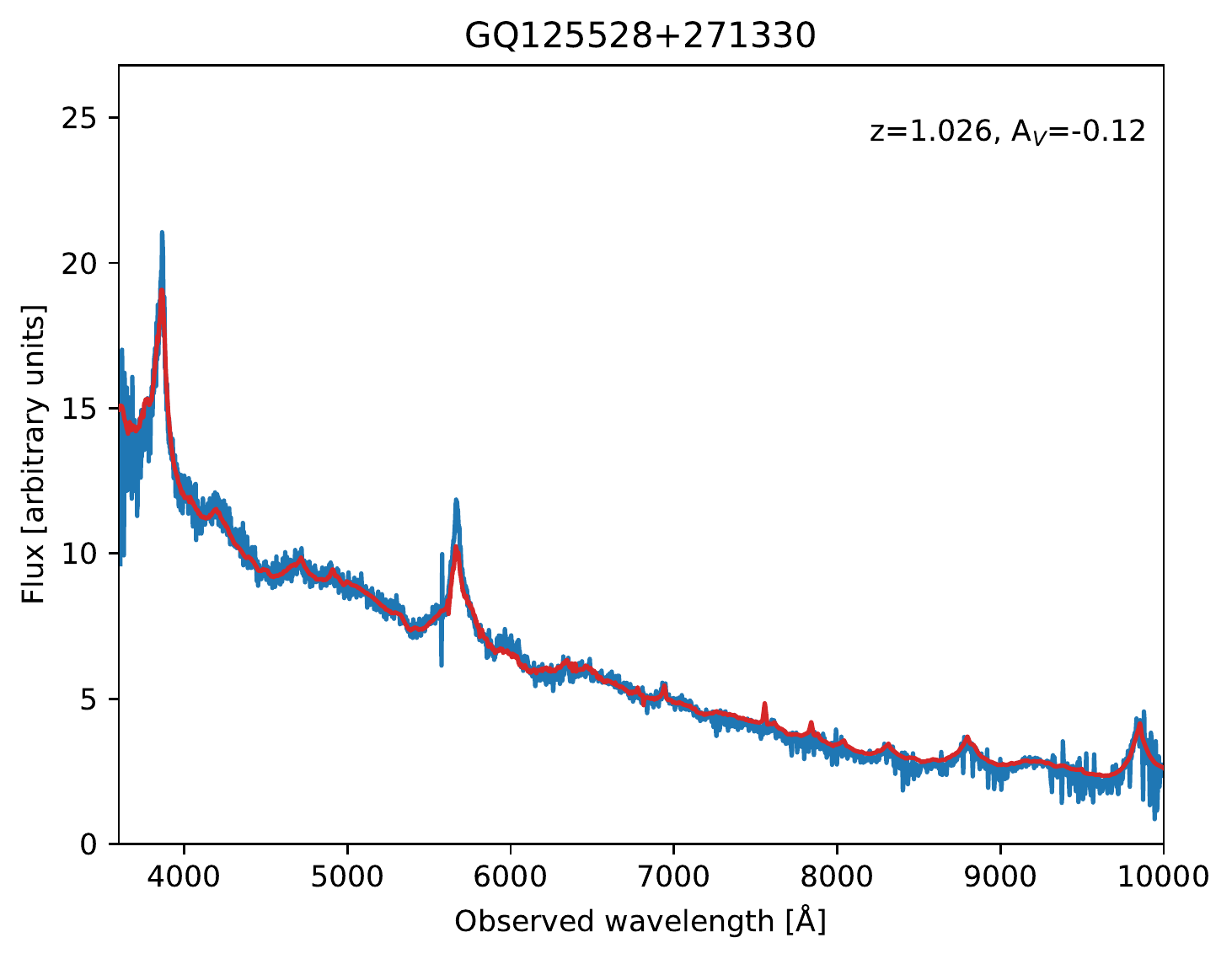,width=7.5cm}
\epsfig{file=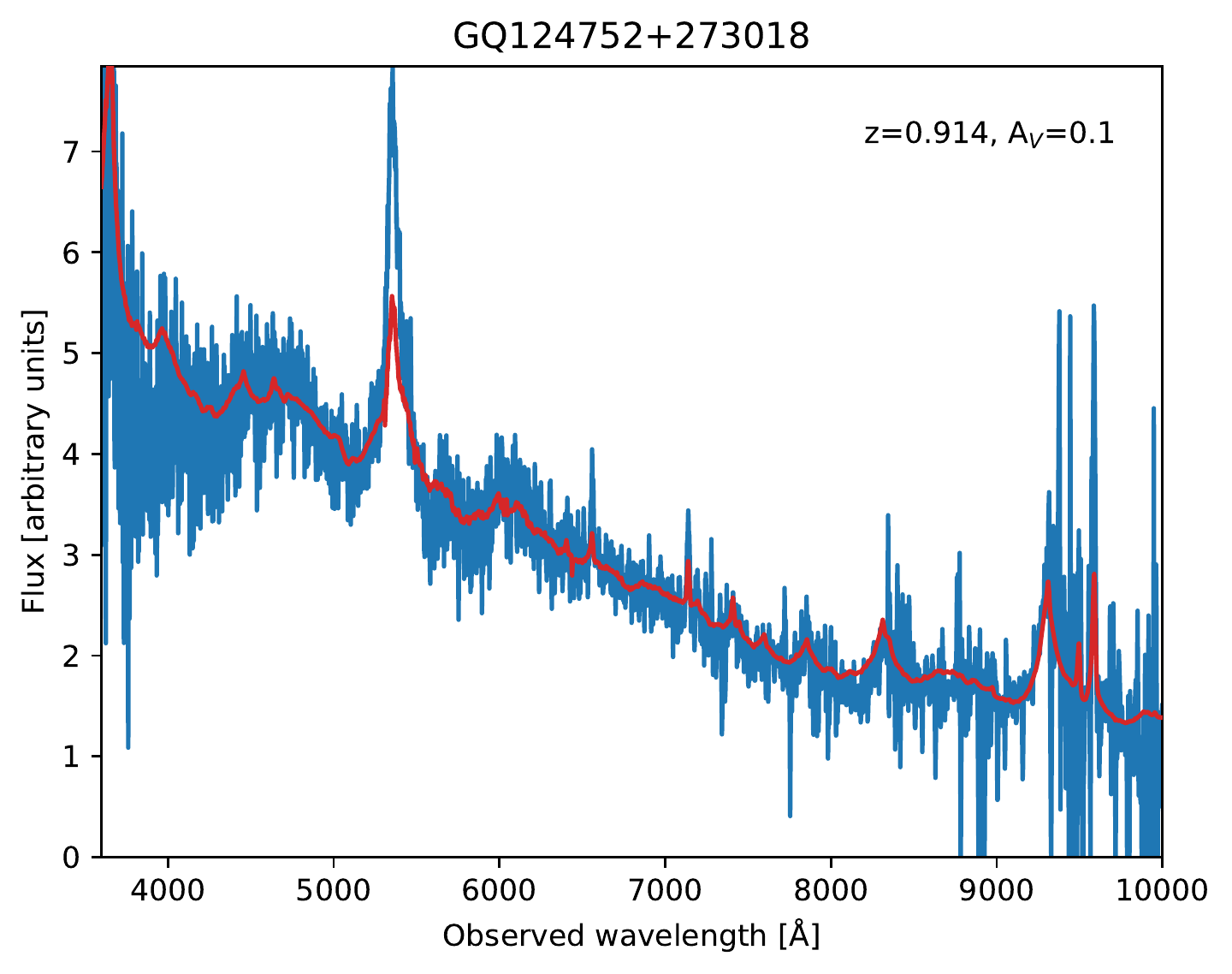,width=7.5cm}
\epsfig{file=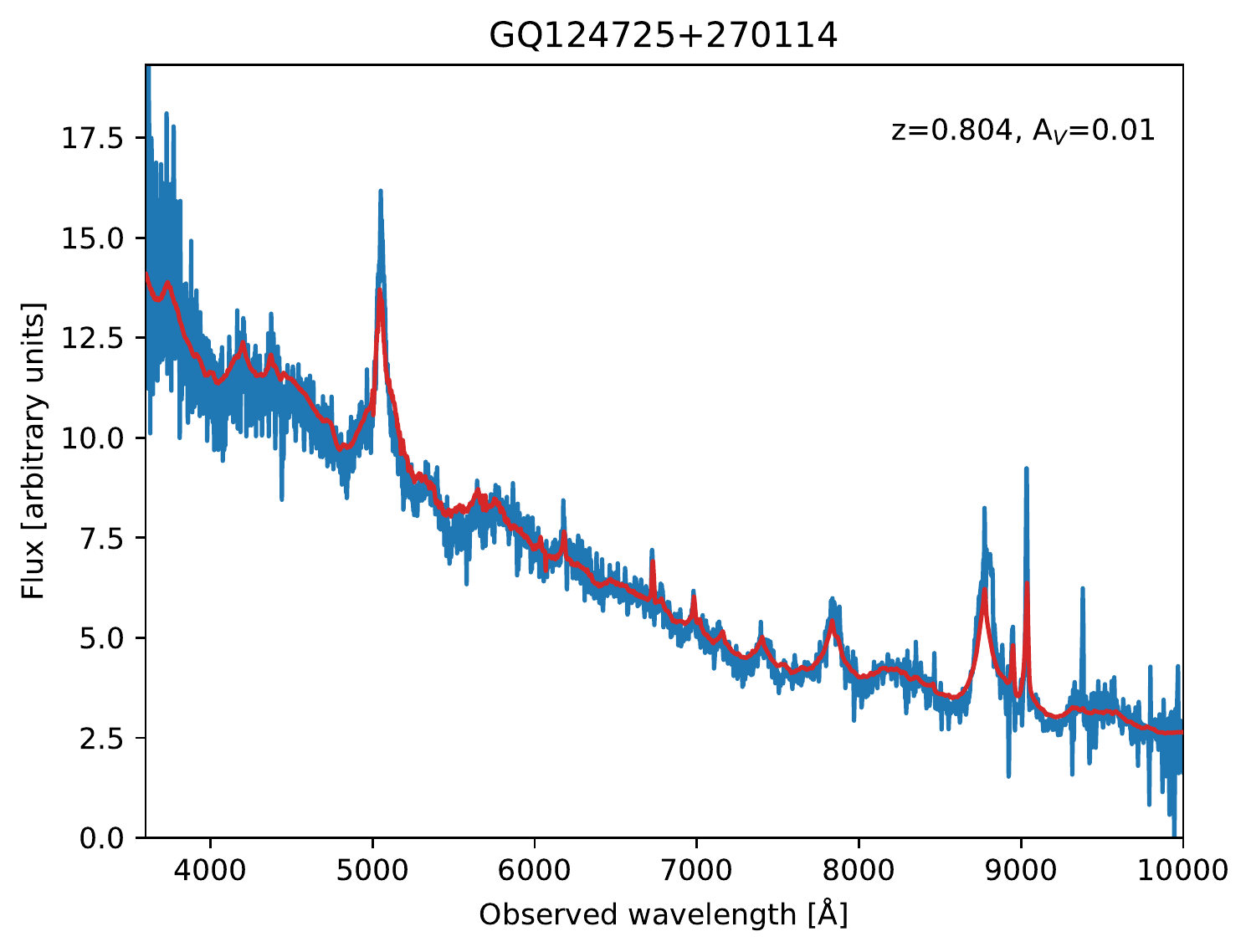,width=7.5cm}
\epsfig{file=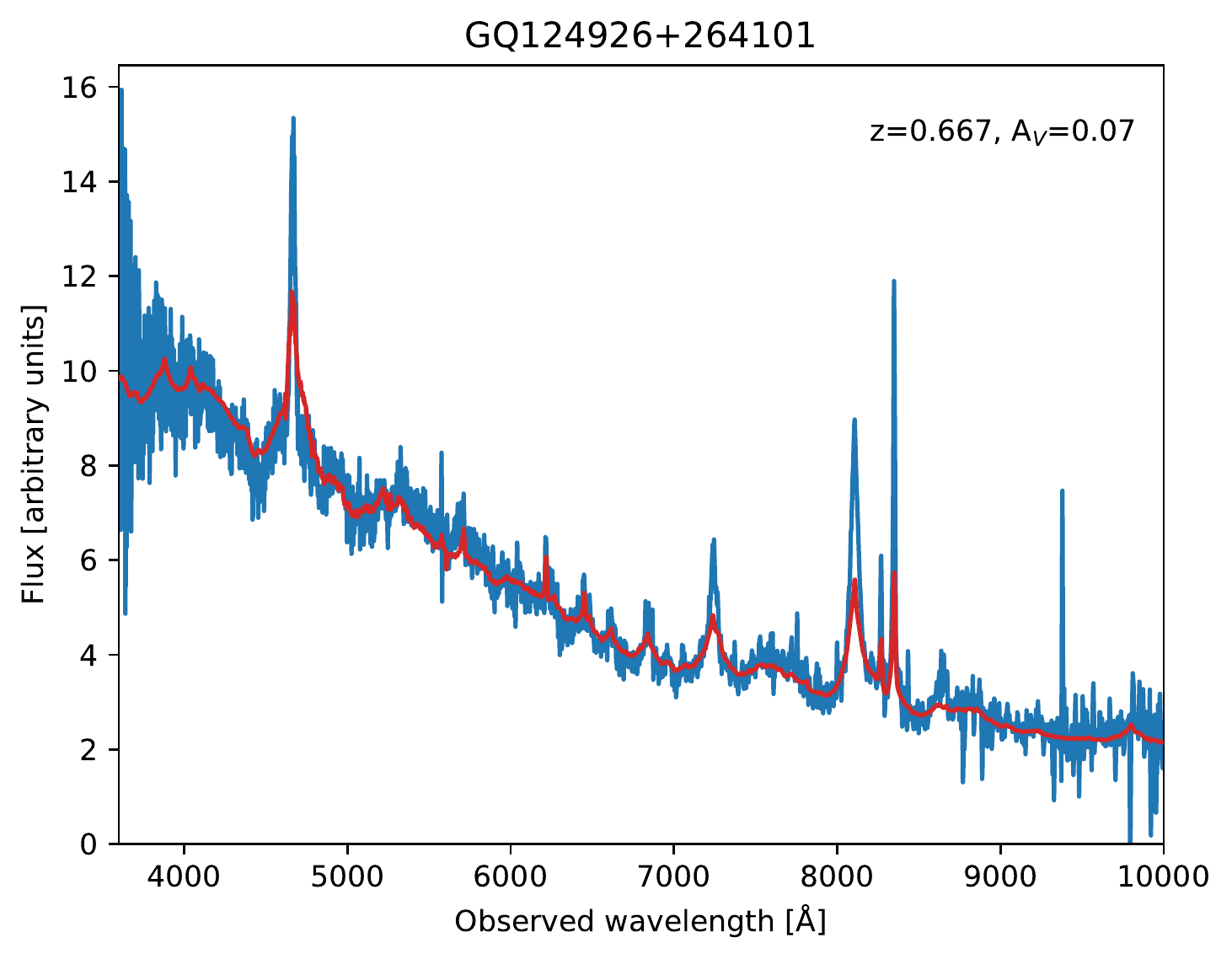,width=7.5cm}
\epsfig{file=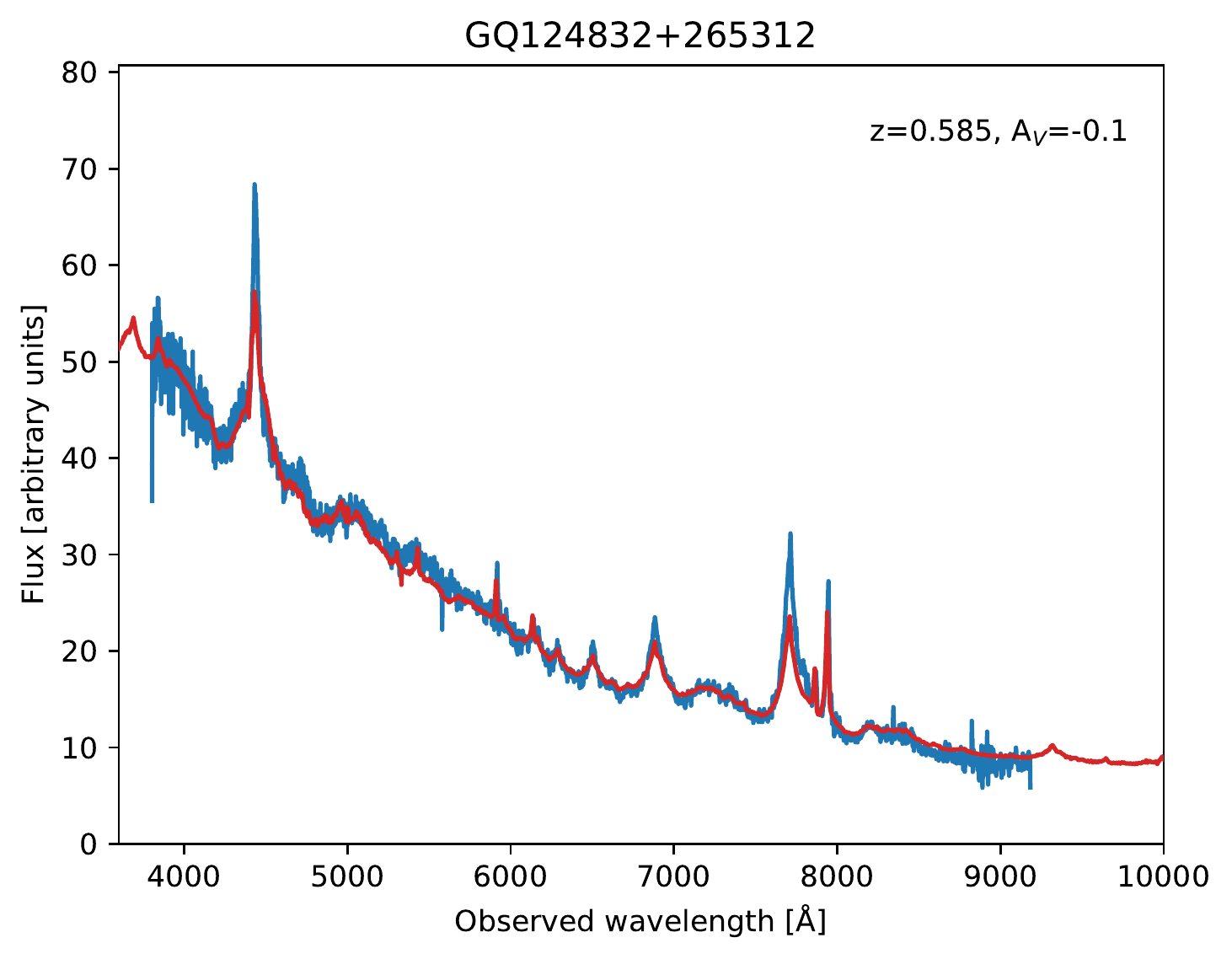,width=7.5cm}
\end{figure*}

\begin{figure*} [!b]
\centering
\epsfig{file=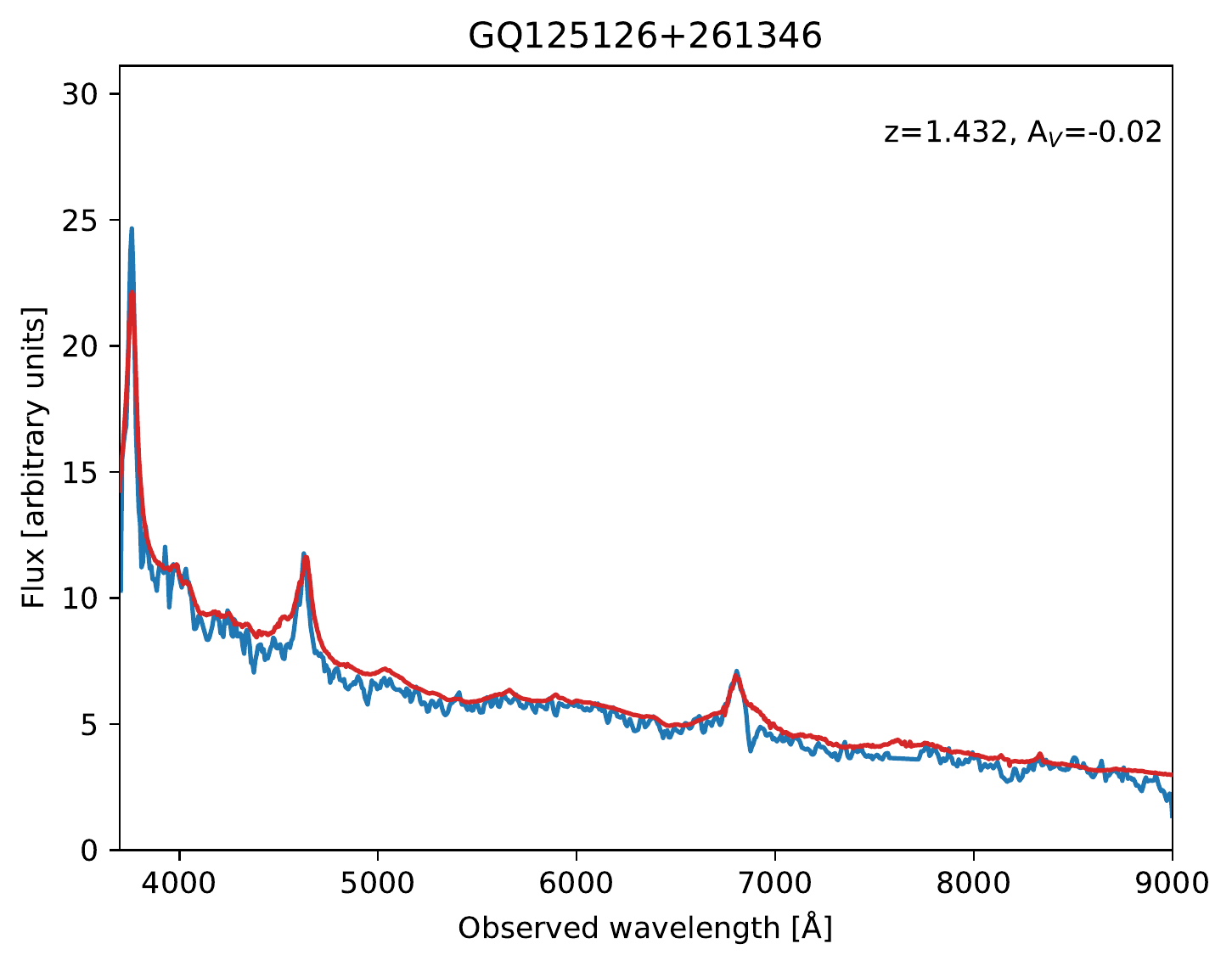,width=7.5cm}
\epsfig{file=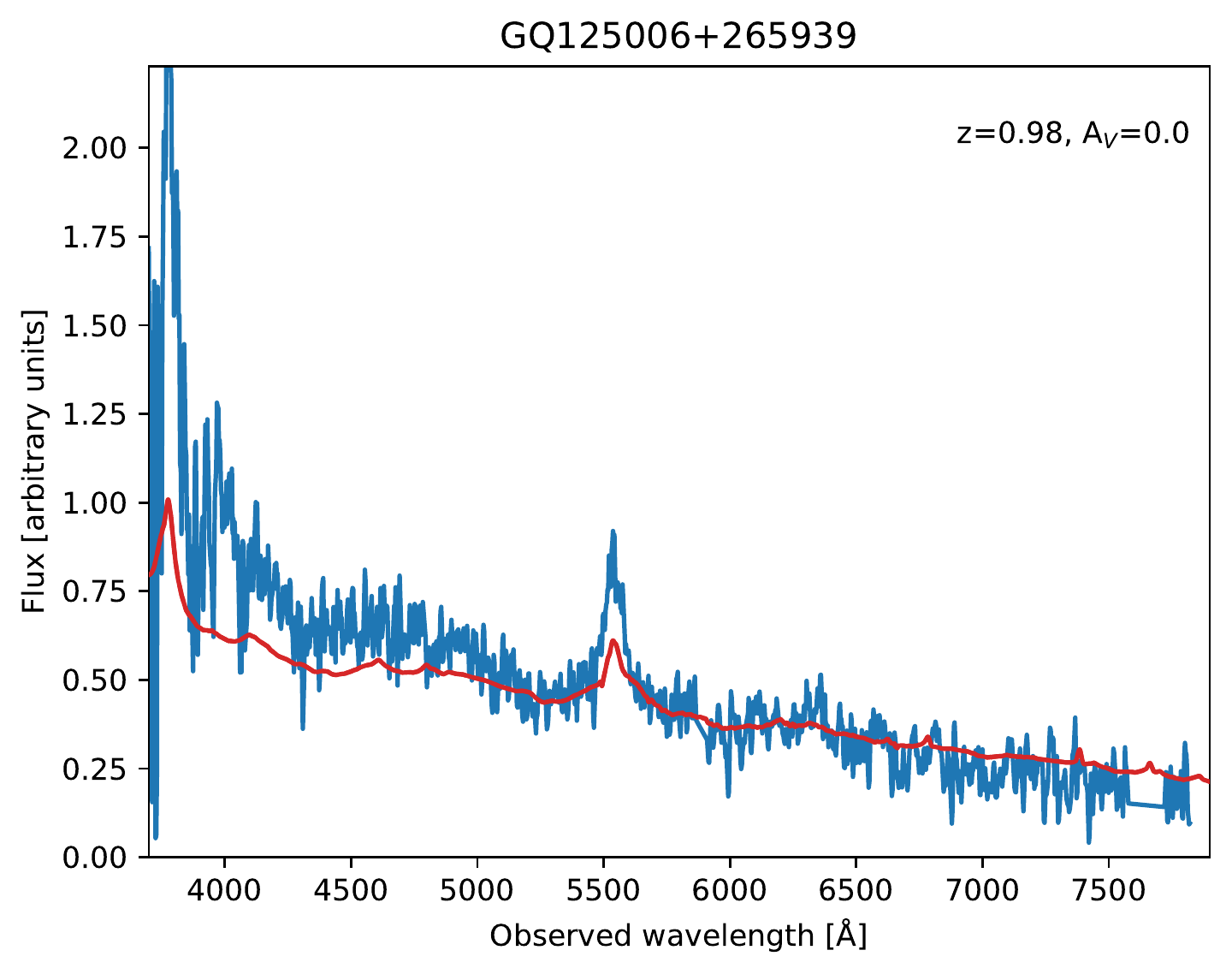,width=7.5cm}
\caption{Spectra of two previously known quasars \citep[redshifts can be found in][]{Crampton1987} for which spectra could not be located in the literature.}
\label{fig:otherspectra}
\end{figure*}

\end{appendix}

\clearpage

\onecolumn

\begin{longtable}{ccccccc}
\caption{\label{tab:log}Observing log ordered by telescope and date.}\\
\hline
Source & R.A. & Decl. & $r_{\rm AB}$ & Telescope/ & Observation Date & Exp. time \\
& (J2000) & (J2000) & (mag) & Instrument && (s) \\
\hline
\hline
\endfirsthead
\caption{continued.}\\
\hline\hline
Source & R.A. & Decl. & $r_{\rm AB}$ & Telescope/ & Observation Date & Exp. time \\
& (J2000) & (J2000) & (mag) & Instrument && (s) \\
\hline
\hline
\endhead
GQ125518+270034 & 12:55:18.6 & +27:00:34.3 & 19.7 & GTC/OSIRIS & 2019-02-06 & $2\times 500$ \\
GQ124728+272742 & 12:47:28.6 & +27:27:42.5 & 20.3 & GTC/OSIRIS & 2019-12-01 & $4\times 300$ \\
GQ124935+265906 & 12:49:35.2 & +26:59:06.4 & 19.9 & GTC/OSIRIS & 2019-12-02 & $2\times 720$ \\
GQ125312+265355 & 12:53:12.1 & +26:53:55.7 & 20.0 & GTC/OSIRIS & 2019-12-02 & $2\times 900$ \\
GQ124922+265938 & 12:49:22.2 & +26:59:38.9 & 19.7 & GTC/OSIRIS & 2019-12-02 & $1\times 720$ \\
GQ124902+262014 & 12:49:02.2 & +26:20:14.7 & 19.7 & GTC/OSIRIS & 2019-12-04 & $2\times 400$ \\
GQ124731+270622 & 12:47:31.5 & +27:06:22.2 & 19.8 & GTC/OSIRIS & 2019-12-19 & $3\times 400$ \\
GQ124734+270615 & 12:47:34.4 & +27:06:15.2 & 20.3 & GTC/OSIRIS & 2019-12-19 & $3\times 900$ \\
GQ125007+274709 & 12:50:07.9 & +27:47:09.4 & 19.8 & GTC/OSIRIS & 2019-12-20 & $2\times 700$  \\
GQ125159+280446 & 12:51:59.1 & +28:04:46.3 & 19.9 & GTC/OSIRIS & 2019-12-27 & $2\times 720$ \\
GQ125516+265049 & 12:55:16.1 & +26:50:49.0 & 19.8 & GTC/OSIRIS & 2019-12-28 & $2\times 600$ \\
GQ124804+265836 & 12:48:05.0 & +26:58:36.1 & 20.7 & GTC/OSIRIS & 2019-12-30 & $2\times 1600$ \\
GQ125420+274609 & 12:54:20.0 & +27:46:09.6 & 20.0 & GTC/OSIRIS & 2020-01-02 & $2\times 720$ \\
GQ125050+272448 & 12:50:50.2 & +27:24:48.4 & 20.1 & GTC/OSIRIS & 2020-01-02 & $2\times 900$ \\
GQ125248+263805 & 12:52:48.3 & +26:38:05.6 & 20.1 & GTC/OSIRIS & 2020-01-03 & $2\times 1000$ \\
GQ125308+273331 & 12:53:08.5 & +27:33:31.2 & 20.4 & GTC/OSIRIS & 2020-01-03 & $3\times 900$ \\
GQ124945+264953 & 12:49:45.5 & +26:49:53.3 & 19.8 & GTC/OSIRIS & 2020-02-02 & $2\times 750$ \\
GQ125107+271451 & 12:51:08.0 & +27:14:51.5 & 19.4 & GTC/OSIRIS & 2020-02-05 & $2\times 480$ \\
GQ125043+271934 & 12:50:44.0 & +27:19:34.3 & 20.1 & GTC/OSIRIS & 2020-02-06 & $2\times 600$ \\
GQ124728+272742 & 12:47:28.6 & +27:27:42.5 & 20.3 & GTC/OSIRIS & 2020-02-15 & $4\times 300$ \\
GQ125241+275942 & 12:52:42.0 & +27:59:42.7 & 20.6 & GTC/OSIRIS & 2020-02-15 & $3\times 1200$ \\
GQ125239+280127 & 12:52:39.2 & +28:01:27.2 & 19.5 & GTC/OSIRIS & 2020-02-15 & $3\times 1200$ \\
GQ125158+271304 & 12:51:58.6 & +27:13:04.1 & 19.7 & GTC/OSIRIS & 2020-03-07 & $1\times 750$ \\
GQ125149+265349 & 12:51:49.2 & +26:53:50.0 & 19.8 & GTC/OSIRIS & 2020-04-28 & $2\times 450$ \\
GQ125543+270714 & 12:55:43.7 & +27:07:14.1 & 19.7 & GTC/OSIRIS & 2020-04-28 & $2\times 500$ \\
GQ124814+271603 & 12:48:14.5 & +27:16:03.6 & 19.2 & GTC/OSIRIS & 2020-05-15 & $2\times 350$ \\
GQ124728+272742 & 12:47:28.6 & +27:27:42.5 & 20.3 & GTC/OSIRIS & 2020-05-18 & $2\times 1000$ \\
GQ124835+263759 & 12:48:35.0 & +26:37:59.8 & 19.9 & WHT/ISIS & 2020-01-03 & $3\times 720$ \\
GQ125353+273405 & 12:53:53.7 & +27:34:05.9 & 19.8 & WHT/ISIS & 2020-01-03 & $3\times 1800$ \\
GQ125008+274001 & 12:50:08.6 & +27:40:01.5 & 19.9 & WHT/ISIS & 2020-01-03 & $3\times 1200$ \\
GQ125043+271934 & 12:50:44.0 & +27:19:34.3 & 20.1 & WHT/ISIS & 2020-01-04 & $2\times 1350$ \\
GQ125104+273341 & 12:51:04.1 & +27:33:41.2 & 19.8 & WHT/ISIS & 2020-01-04 & $2\times 1000$ \\
GQ125312+275524 & 12:53:12.1 & +27:55:24.6 & 19.3 & WHT/ISIS & 2020-01-04 & $2\times 1600$ \\
GQ125230+275410 & 12:52:30.1 & +27:54:10.0 & 19.0 & WHT/ISIS & 2020-01-04 & $2\times 1200$ \\
GQ124921+264446 & 12:49:21.2 & +26:44:46.8 & 19.6 & WHT/ISIS & 2020-01-05 & $2\times 1000$ \\
GQ125107+271451 & 12:51:08.0 & +27:14:51.5 & 19.4 & WHT/ISIS & 2020-01-05 & $2\times 1000$ \\
GQ125042+274012 & 12:50:42.4 & +27:40:13.0 & 19.7 & WHT/ISIS & 2020-01-05 & $2\times 1200$ \\
GQ125033+274519 & 12:50:33.6 & +27:45:19.1 & 19.7 & WHT/ISIS & 2020-01-05 & $2\times 1000$ \\
GQ124915+274554 & 12:49:15.9 & +27:45:54.8 & 19.3 & WHT/ISIS & 2020-01-05 & $2\times 900$ \\
GQ124955+275657 & 12:49:55.4 & +27:56:57.7 & 19.3 & WHT/ISIS & 2020-01-05 & $2\times 1250$ \\
GQ124958+275703 & 12:49:58.1 & +27:57:03.7 & 19.4 & WHT/ISIS & 2020-01-05 & $2\times 1250$ \\
GQ125456+263623 & 12:54:56.8 & +26:36:23.8 & 19.6 & NOT/ALFOSC & 2019-01-28 & $2\times 600$ \\
GQ125510+263724 & 12:55:10.1 & +26:37:25.0 & 19.9 & NOT/ALFOSC & 2019-01-28 & $2\times 600$ \\
GQ125424+264833 & 12:54:24.6 & +26:48:33.8 & 19.7 & NOT/ALFOSC & 2019-01-28 & $2\times 600$ \\
GQ125453+263721 & 12:54:53.0 & +26:37:21.8 & 19.5 & NOT/ALFOSC & 2019-01-28 & $2\times 600$ \\
GQ125133+264550 & 12:51:33.7 & +26:45:50.4 & 19.4 & NOT/ALFOSC & 2019-01-28 & $2\times 600$ \\
GQ125008+263658 & 12:50:09.0 & +26:36:58.6 & 19.3 & NOT/ALFOSC & 2019-01-28 & $2\times 450$ \\
GQ125054+264353 & 12:50:55.0 & +26:43:53.6 & 19.6 & NOT/ALFOSC & 2019-01-28 & $2\times 720$ \\
GQ124941+265314 & 12:49:41.0 & +26:53:14.8 & 19.7 & NOT/ALFOSC & 2019-01-28 & $2\times 600$ \\
GQ125359+272014 & 12:53:59.8 & +27:20:14.8 & 20.3 & NOT/ALFOSC & 2019-01-28 & $2\times 900$ \\
GQ125327+273733 & 12:53:27.1 & +27:37:33.1 & 19.8 & NOT/ALFOSC & 2019-01-28 & $2\times 600$ \\
GQ125241+272550 & 12:52:41.1 & +27:25:50.8 & 20.7 & NOT/ALFOSC & 2019-01-29 & $2\times 1800$ \\
GQ125211+272803 & 12:52:11.9 & +27:28:03.8 & 20.4 & NOT/ALFOSC & 2019-01-29 & $2\times 1200$ \\
GQ125031+261953 & 12:50:31.6 & +26:19:53.7 & 19.3 & NOT/ALFOSC & 2019-04-01 & $2\times 300$ \\
GQ124715+265528 & 12:47:15.9 & +26:55:28.1 & 18.9 & NOT/ALFOSC & 2019-04-01 & $4\times 200$ \\
GQ125027+263048 & 12:50:27.3 & +26:30:48.1 & 19.0 & NOT/ALFOSC & 2020-01-21 & $2\times 1000$ \\
GQ125357+271630 & 12:53:57.5 & +27:16:30.6 & 18.9 & NOT/ALFOSC & 2020-01-21 & $2\times 1000$ \\
GQ124955+271335 & 12:49:55.7 & +27:13:35.3 & 19.3 & NOT/ALFOSC & 2019-01-29 & $2\times 300$ \\
GQ125005+280041 & 12:50:05.2 & +28:00:41.9 & 20.0 & NOT/ALFOSC & 2020-03-20 & $2\times 990$ \\
GQ124959+261949 & 12:49:59.0 & +26:19:49.3 & 19.9 & NOT/ALFOSC & 2020-03-20 & $2\times 1050$ \\
GQ125421+265454 & 12:54:21.0 & +26:54:54.7 & 19.6 & NOT/ALFOSC & 2020-03-20 & $2\times 900$ \\
GQ124746+264709 & 12:47:46.5 & +26:47:09.8 & 20.0 & NOT/ALFOSC & 2020-03-20 & $2\times 1080$ \\
GQ124721+265728 & 12:47:21.3 & +26:57:28.9 & 19.7 & NOT/ALFOSC & 2020-03-20 & $2\times 1125$ \\
GQ124814+271603 & 12:48:14.5 & +27:16:03.6 & 19.2 & NOT/ALFOSC & 2020-03-20 & $2\times 1800$ \\
GQ124958+273317 & 12:49:58.2 & +27:33:17.5 & 19.8 & NOT/ALFOSC & 2020-03-20 & $2\times 550$ \\
GQ124901+274330 & 12:49:01.8 & +27:43:30.2 & 19.8 & NOT/ALFOSC & 2020-03-20 & $2\times 1050$ \\
GQ125005+280041 & 12:50:05.2 & +28:00:41.9 & 20.0 & NOT/ALFOSC & 2020-03-20 & $2\times 990$ \\
GQ125503+264502 & 12:55:03.8 & +26:45:02.7 & 19.6 & NOT/ALFOSC & 2020-03-21 & $2\times 900$ \\
GQ125249+280356 & 12:52:49.1 & +28:03:56.4 & 19.2 & NOT/ALFOSC & 2020-03-21 & $1\times 673$ \\
GQ125441+264637 & 12:54:41.3 & +26:46:37.8 & 19.3 & NOT/ALFOSC & 2020-03-28 & $2\times 900$ \\
GQ124820+264225 & 12:48:20.6 & +26:42:25.1 & 19.4 & NOT/ALFOSC & 2020-03-28 & $2\times 1000$ \\
GQ125134+272000 & 12:51:34.4 & +27:20:00.2 & 19.7 & NOT/ALFOSC & 2020-03-28 & $2\times 9100$ \\
GQ124809+271650 & 12:48:09.4 & +27:16:50.1 & 19.7 & NOT/ALFOSC & 2020-03-28 & $2\times 800$ \\
GQ125013+273256 & 12:50:13.7 & +27:32:56.4 & 19.6 & NOT/ALFOSC & 2020-03-28 & $2\times 900$  \\
GQ125126+261346 & 12:51:26.6 & +26:13:46.9 & 19.4 & NOT/ALFOSC & 2020-03-28 & $2\times 600$ \\
GQ124758+272556 & 12:47:58.7 & +27:25:56.2 & 17.1 & WHT/DOLORES & 2020-04-28 & $5\times 100$ \\
GQ125109+263505 & 12:51:09.8 & +26:35:05.3 & 17.5 & WHT/DOLORES & 2020-04-28 & $3\times 100$ \\
GQ125006+265939 & 12:50:06.3 & +26:59:39.9 & 19.5 & WHT/DOLORES & 2020-04-29 & $5\times 300$ \\
\hline\noalign{\smallskip}
\end{longtable}

\begin{longtable}{cccccccccc}
\caption{\label{qsotab}Classification of the stationary sources identified as quasars. We have ordered the sources by redshift and separated the quasars that were observed spectroscopically by previous surveys and this work. The redshifts from the SDSS archive is from the most recent DR14Q compilation \citep{Paris18} and the redshifts from the CFHT survey are from \citet{Crampton1987}. For the objects identified as quasars as part of this work, we also provide the photometric redshifts from \citet{Richards2009}. We include measurements of the signal-to-noise of the proper motions and parallaxes as reported in the {\it Gaia}-DR2 catalogue and the typical optical, NIR, and MIR colours used to identify quasars.} \\
\hline\noalign{\smallskip}\hline\noalign{\smallskip}
Source & (S / N)$_{\rm \mu}$ & (S / N)$_{\rm plx}$ & $u-g$ & $J-K$ & $W1-W2$ & $z_{\mathrm{prev.}}$ & $z_{\mathrm{this\,work}}$ & $A_V$ (mag) & Survey \\
\noalign{\smallskip}\hline\hline\noalign{\smallskip}
\endfirsthead
\caption{continued.}\\
\noalign{\smallskip}\hline\hline\noalign{\smallskip}
Source & (S / N)$_{\rm PM}$ & (S / N)$_{\rm Plx}$ & $u-g$ & $J-K$ & $W1-W2$ & $z_{\mathrm{SDSS}}$ & $z_{\mathrm{this work}}$ & $A_V$ (mag) & Survey \\
\noalign{\smallskip}\hline\hline\noalign{\smallskip}
\endhead
GQ125005+263107 & 0.63 & 0.19 & -0.04 & 0.98 & 1.06 & 2.044 & - & 0.00 & SDSS \\
GQ125219+264053 & 1.07 & 1.56 & - & - & - & 1.287 & - & -0.05 & SDSS \\
GQ125209+265018 & 1.17 & -0.96 & 1.51 & - & 0.59 & 3.441 & - & 0.03 & SDSS \\
GQ125358+262453 & 0.49 & -1.48 & 0.75 & - & 0.87 & 3.100 & - & -0.02 & SDSS \\
GQ125302+270519 & 1.67 & -1.15 & 2.18 & 1.00 & 0.83 & 2.988 & - & -0.05 & SDSS \\
GQ125019+274443 & 1.75 & 0.02 & 0.84 & 1.47 & 1.03 & 2.665 & - & 0.31 & SDSS \\
GQ124816+265235 & 1.05 & -1.78 & 0.62 & 1.19 & 0.72 & 2.514 & - & -0.07 & SDSS \\
GQ125202+275018 & 0.66 & 0.80 & 0.53 & 1.32 & 1.16 & 2.485 & - & 0.04 & SDSS \\
GQ125302+261711 & 0.46 & -0.05 & 0.59 & 1.50 & 0.44 & 2.316 & - & -0.07 & SDSS \\
GQ125257+274542 & 1.61 & 1.26 & 0.31 & 0.89 & 1.37 & 2.004 & - & -0.08 & SDSS \\
GQ125150+263900 & 0.86 & 0.39 & 0.12 & 0.92 & 0.92 & 1.913 & - & 0.19 & SDSS \\
GQ125039+272904 & 1.16 & -0.85 & 0.60 & 1.23 & 1.19 & 1.890 & - & 0.37 & SDSS \\
GQ124816+264712 & 0.60 & 1.08 & -0.04 & 0.94 & 1.37 & 1.856 & - & -0.03 & SDSS \\
GQ124943+261629 & 0.53 & 0.36 & 0.05 & 0.90 & 0.11 & 1.843 & - & 0.02 & SDSS \\
GQ125015+264337 & 0.21 & -0.85 & 0.06 & 1.09 & 1.46 & 1.787 & - & -0.10 & SDSS \\
GQ125206+272717 & 1.69 & -1.63 & 0.07 & 0.84 & 1.23 & 1.680 & - & -0.04 & SDSS \\
GQ125429+270003 & 1.25 & 1.61 & 0.19 & 1.11 & 1.25 & 1.635 & - & -0.07 & SDSS \\
GQ125332+272225 & 1.69 & 0.87 & 0.27 & - & 1.32 & 1.627 & - & -0.11 & SDSS \\
GQ125424+263941 & 1.19 & -0.64 & 0.23 & 1.08 & 1.57 & 1.601 & - & -0.17 & SDSS \\
GQ124846+271722 & 0.79 & -1.43 & 0.36 & 1.07 & 1.35 & 1.514 & - & 0.21 & SDSS \\
GQ125023+265328 & 0.21 & -0.23 & 0.36 & 1.09 & 1.10 & 1.490 & - & 0.02 & SDSS \\
GQ125300+271234 & 1.16 & -0.31 & 0.26 & 1.03 & 1.44 & 1.482 & - & -0.12 & SDSS \\
GQ125302+262722 & 1.26 & 3.30 & 0.21 & 1.36 & 1.25 & 1.259 & - & -0.06 & SDSS \\
GQ124925+272634 & 0.55 & -0.55 & 0.13 & 1.41 & 1.41 & 1.156 & - & -0.11 & SDSS \\
GQ125320+270607 & 0.40 & -0.24 & 0.42 & 1.36 & 1.39 & 1.153 & - & 0.21 & SDSS \\
GQ125112+275222 & 0.38 & -0.26 & 0.19 & 1.12 & 1.30 & 1.044 & - & -0.05 & SDSS \\
GQ125528+271330 & 0.73 & -1.45 & 0.15 & 1.42 & 1.30 & 1.026 & - & -0.12 & SDSS \\
GQ124752+273018 & 1.37 & 0.63 & 0.39 & 1.23 & 0.70 & 0.914 & - & 0.10 & SDSS \\
GQ124725+270114 & 0.63 & -0.36 & 0.45 & 1.23 & 1.04 & 0.804 & - & 0.01 & SDSS \\
GQ124926+264101 & 0.64 & 0.09 & 0.43 & 1.28 & 1.09 & 0.667 & - & 0.07 & SDSS \\
GQ124832+265312 & 1.14 & 0.72 & 0.41 & 1.92 & 1.09 & 0.585 & - & -0.10 & SDSS \\
GQ125320+272116 & 0.59 & 0.89 & 0.25 & 1.50 & 1.03 & 0.512 & - & -0.14 & SDSS \\
GQ125126+261346 & 1.14 & -1.63 & -0.03 & 0.79 & 1.23 & 1.430 & 1.432 & -0.03 & CFHT \\
GQ125006+265939 & 0.63 & 0.82 & 0.11 & 1.15 & 1.20 & 0.970 & 0.980 & 0.00 & CFHT \\
GQ124915+274554 & 0.95 & 0.40 & 0.24 & 1.21 & 1.55 & 2.16P & 2.160 & -0.03 & This work \\
GQ125239+280127 & 1.96 & 1.63 & 0.06 & 1.03 & 1.31 & 2.04P & 2.070 & 0.03 & This work \\
GQ125211+272803 & 0.63 & -1.18 & 0.08 & 1.39 & 1.06 & 1.87P & 2.044 & -0.08 & This work \\
GQ125241+272550 & 1.87 & -1.36 & 0.16 & 1.16 & 1.37 & 0.22P & 2.000 & 0.28 & This work \\
GQ124958+275703 & 0.49 & 0.66 & 0.23 & 0.99 & 1.33 & 1.91P & 1.960 & 0.03 & This work \\
GQ124958+273317 & 0.48 & -0.48 & 0.15 & 0.69 & 1.51 & 1.72P & 1.950 & -0.05 & This work \\
GQ124728+272742 & 1.28 & 1.60 & 0.58 & 1.15 & - & 0.46P & 1.840 & 0.01 & This work \\
GQ125054+264353 & 1.67 & 0.85 & 0.39 & 2.17 & 1.36 & 1.59P & 1.760 & 0.04 & This work \\
GQ125327+273733 & 1.32 & 1.65 & 0.07 & 0.24 & 1.03 & 1.96P & 1.700 & -0.11 & This work \\
GQ125359+272014 & 1.09 & 1.72 & 0.26 & - & 1.55 & 0.08P & 1.700 & -0.09 & This work \\
GQ124955+271335 & 0.92 & -0.18 & 0.05 & 1.11 & 1.46 & 1.47P & 1.500 & -0.18 & This work \\
GQ125033+274519 & 0.55 & -1.24 & 0.12 & 1.72 & 1.44 & 1.48P & 1.485 & -0.07 & This work \\
GQ125424+264833 & 0.84 & 0.46 & 0.20 & - & 1.36 & 1.47P & 1.470 & -0.17 & This work \\
GQ125042+274012 & 0.16 & -1.19 & 0.21 & 1.04 & 1.44 & 1.48P & 1.455 & 0.01 & This work \\
GQ125013+273256 & 0.85 & -2.01 & 0.09 & 1.28 & 1.32 & 1.15P & 1.423 & -0.02 & This work \\
GQ125104+273341 & 0.46 & 0.38 & 0.02 & 1.48 & 1.28 & 1.15P & 1.320 & 0.00 & This work \\
GQ125133+264550 & 1.16 & -0.19 & 0.12 & 1.62 & 1.43 & 1.28P & 1.280 & -0.04 & This work \\
GQ125421+265454 & 0.15 & -2.05 & 0.21 & 1.53 & 1.22 & 1.30P & 1.266 & 0.07 & This work \\
GQ125230+275410 & 1.66 & -0.73 & 0.11 & 1.87 & 1.05 & 1.04P & 1.250 & -0.06 & This work \\
GQ124835+263759 & 1.29 & -1.23 & 0.03 & 1.26 & 1.18 & 1.15P & 1.225 & -0.03 & This work \\
GQ124809+271650 & 0.32 & -0.69 & 0.24 & 1.33 & 1.16 & 0.95P & 1.017 & 0.01 & This work \\
GQ125312+275524 & 1.80 & 2.45 & 0.23 & 0.95 & 1.26 & 1.00P & 0.980 & -0.01 & This work \\
GQ125249+280356 & 0.15 & 0.62 & 0.11 & 1.16 & 1.07 & 0.93P & 0.944 & -0.10 & This work \\
GQ125043+271934 & 1.29 & -0.81 & 0.12 & 1.62 & 0.84 & 0.93P & 0.924 & -0.17 & This work \\
GQ125107+271451 & 0.90 & -0.55 & 0.40 & 1.04 & 0.72 & 0.38P & 0.930 & 0.42 & This work \\
GQ125031+261953 & 0.28 & -1.37 & 0.22 & 1.22 & 1.06 & 0.88P & 0.860 & -0.14 & This work \\
GQ124941+265314 & 0.45 & 0.05 & 0.62 & 1.49 & 0.69 & 2.46P & 0.750 & -0.02 & This work \\
GQ125008+263658 & 0.93 & 0.32 & 0.39 & 1.44 & 1.04 & 0.68P & 0.678 & -0.07 & This work  \\
GQ124921+264446 & 0.33 & -1.08 & 0.01 & 1.89 & 1.02 & 0.31P & 0.300 & 0.19 & This work \\
\hline\noalign{\smallskip}
\end{longtable}

\begin{longtable}{ccccccc}
\caption{\label{startab}Classification of the stationary sources identified as stars. We have ordered the objects by the stellar classifications. We include measurements of the signal-to-noise of the proper motions and parallaxes as reported in the {\it Gaia}-DR2 catalogue and the typical optical, NIR, and MIR colours used to identify quasars.} \\
\noalign{\smallskip}\hline\hline\noalign{\smallskip}
Source & (S / N)$_{\rm PM}$ & (S / N)$_{\rm Plx}$ & $u-g$ & $J-K$ & $W1-W2$ & Type \\
\noalign{\smallskip}\hline\hline\noalign{\smallskip}
\endfirsthead
\caption{continued.}\\
\noalign{\smallskip}\hline\hline\noalign{\smallskip}
Source & (S / N)$_{\rm PM}$ & (S / N)$_{\rm Plx}$ & $u-g$ & $J-K$ & $W1-W2$ & Type \\
\noalign{\smallskip}\hline\hline\noalign{\smallskip}
\endhead
GQ124804+265836 & 1.83 & 3.48 & 2.59 & 0.66 & 0.15 & M5 star \\
GQ125241+275942 & 1.48 & 2.18 & 2.17 & 0.81 & 0.18 & M5 star \\
GQ125543+270714 & 1.81 & 4.94 & 2.72 & 0.77 & -0.11 & M5 star \\
GQ125518+270034 & 1.91 & 1.49 & 2.72 & 0.87 & 0.18 & M4 star \\
GQ124734+270615 & 0.89 & 3.86 & 1.87 & 0.73 & -0.31 & M4 star \\
GQ125050+272448 & 1.06 & -0.62 & 3.10 & 0.89 & -0.02 & M4 star \\
GQ125159+280446 & 1.20 & 3.39 & 1.67 & 0.78 & -0.09 & M3 star \\
GQ125312+265355 & 1.88 & 0.76 & 3.07 & 0.81 & 0.04 & M3 star \\
GQ125248+263805 & 0.69 & 2.20 & 3.21 & 0.88 & -0.04 & M3 star \\
GQ125308+273331 & 1.44 & 1.27 & 2.14 & 0.87 & -0.50 &  M3 star \\
GQ125007+274709 & 0.99 & 2.46 & 2.27 & 0.74 & 0.34 & M2 star \\
GQ125420+274609 & 0.76 & 0.96 & 1.85 & 0.78 & 0.10 & M1 star\\
GQ124814+271603 & 0.22 & 1.20 & 2.35 & 0.63 & -0.05 & M1 star \\
GQ125516+265049 & 1.33 & -0.10 & 3.92 & 0.76 & - & M0 star \\
GQ124945+264953 & 0.63 & 1.39 & 3.78 & 0.73 & 0.20 & M0 star \\
GQ124935+265906 & 1.06 & 1.07 & 3.65 & 0.96 & -0.19 & M0 star \\
GQ125149+265349 & 1.78 & 0.97 & 2.17 & - & 0.51 & M0 star \\
GQ124715+265528 & 1.81 & 4.05 & 2.29 & 0.77 & 0.23 & K9 star \\
GQ124955+275657 & 1.20 & -1.76 & 1.32 & 0.60 & 0.88 & K6 star \\
GQ124902+262014 & 1.44 & 0.28 & 1.72 & 0.66 & 0.46 & K5 star \\
GQ125357+271630 & 1.75 & 0.69 & 2.25 & - & -0.28 & K5 star \\
GQ125456+263623 & 0.49 & -0.07 & 2.41 & 0.73 & -0.02 & K5 star \\
GQ125353+273405 & 1.72 & -1.75 & 1.05 & 0.95 & -0.03 & K4 star \\
GQ124758+272556 & 1.87 & 6.19 & 1.89 & 0.52 & 0.00 & G8 star \\
GQ125158+271304 & 1.53 & 0.67 & 1.38 & - & - & G6 star \\
GQ125027+263048 & 1.05 & -0.45 & 1.16 & 0.43 & - & G6 star \\
GQ124721+265728 & 1.19 & -1.59 & 1.06 & 0.51 & - & G6 star \\
GQ124922+265938 & 0.83 & -0.86 & 1.20 & - & - & G4 star \\
GQ125441+264637 & 1.80 & -0.16 & 1.15 & 0.52 & - & G4 star \\
GQ125503+264502 & 0.61 & -1.31 & 1.10 & 0.62 & - & G4 star \\
GQ124901+274330 & 0.91 & -2.50 & 0.89 & 0.50 & - & G0 star \\
GQ124959+261949 & 1.80 & -0.31 & 0.94 & - & - & F8 star \\
GQ125008+274001 & 1.81 & -2.43 & 0.97 & 0.93 & 0.68 & F7 star \\
GQ124731+270622 & 1.69 & -1.69 & 0.85 & - & - & F6 star \\
GQ124746+264709 & 1.87 & 0.52 & 1.01 & - & - & F6 star \\
GQ125453+263721 & 1.68 & -1.24 & 1.09 & 0.64 & - & F4 star \\
GQ125005+280041 & 1.46 & 0.51 & 0.81 & - & - & F4 star \\
GQ125134+272000 & 0.35 & -0.78 & 0.94 & - & - & F4 star \\
GQ125109+263505 & 1.59 & -3.75 & 1.09 & 0.81 & 0.13 & F1 star \\
GQ125510+263724 & 1.25 & 0.51 & 0.89 & - & - & F0 star \\
GQ124820+264225 & 1.64 & -1.26 & 1.06 & - & - & F0 star \\
\hline\noalign{\smallskip}
\end{longtable}

\end{document}